\documentclass[12pt,preprint]{aastex}
\begin{document}

\title{Simultaneous {\it Chandra} and {\it VLA} Observations of  
Young Stars and Protostars in $\rho$~Ophiuchus Cloud Core A}
\author{Marc Gagn\'{e}}
\affil{Department of Geology and Astronomy, West Chester University, West Chester, PA 19383}
\email{mgagne@wcupa.edu}
\author{Stephen L. Skinner}
\affil{Center for Astrophysics and Space Astronomy, 
University of Colorado, Boulder, CO 80309-0389}
\email{skinners@casa.colorado.edu}

\and

\author{Kathryne J. Daniel\altaffilmark{1}}
\affil{Department of Geology and Astronomy, West Chester University, West Chester, PA 19383}
\altaffiltext{1}{Department of Physics and Astronomy, Johns Hopkins University, Baltimore, MD 21218}
\email{kdaniel@pha.jhu.edu}

\newcommand{\ltsimeq}{\raisebox{-0.6ex}{$\,\stackrel{\raisebox{-.2ex}%
{$\textstyle<$}}{\sim}\,$}}
\newcommand{\gtsimeq}{\raisebox{-0.6ex}{$\,\stackrel{\raisebox{-.2ex}%
{$\textstyle>$}}{\sim}\,$}}

\begin{abstract}
A 96-ks {\it Chandra} X-ray  observation of $\rho$~Ophiuchus 
cloud core~A  detected 87 sources, of which 60 were identified
with counterparts at other wavelengths. The X-ray detections
include 12 of 14 known classical T Tauri stars (CTTS) in the 
field, 15 of 17 known weak-lined TTS (WTTS), and 4 of 15 brown dwarf
candidates. The X-ray detections are characterized by hard,
heavily absorbed emission. Most X-ray detections have visual
extinctions in the range A$_{V}$ $\approx$ 10 - 20 mag, but
several sources with visual absorptions as high as 
A$_{V}$ $\approx$ 40 - 56 mag were detected. The mean photon
energy of a typical source is  $\langle$E$\rangle$ $\approx$ 3 keV,
and more than half of the detections are variable. Prominent
X-ray flares were detected in the unusual close binary system
Oph S1, the X-ray bright WTTS DoAr~21, and the brown dwarf
candidate GY 31 (M5.5). Time-resolved spectroscopic analysis
of the DoAr~21 flare clearly reveals a sequence of secondary
flares during the decay phase which may have reheated the plasma.
We find that the X-ray luminosity distributions and spectral
hardnesses of CTTS and WTTS are similar. We also conclude that
the X-ray emission of detected brown-dwarf candidates is less
luminous than T Tauri stars, but spectroscopically similar.

Simultaneous multifrequency {\it VLA} observations detected
31 radio sources at 6 cm, of which ten were also detected by
{\em Chandra}. We report new radio detections of the optically
invisible IR source WLY 2-11 and the faint H$\alpha$ emission line 
star Elias 24 (class II). We confirm circular polarization in
Oph S1 and report a new detection of circular polarization in 
DoAr~21. We find no evidence that X-ray and radio luminosities
are correlated in the small sample of TTS detected 
simultaneously with {\em Chandra} and the {\em VLA}.

We describe a new non-parametric method for estimating X-ray
spectral properties from unbinned photon event
lists that is applicable to both faint and bright
X-ray sources. The method is used to generate $f_{\rm X}$,
$\log T$ and $L_{\rm X}$ light curves.
In addition, we provide a publically-available
electronic database containing multi-wavelength data for
345 known X-ray, IR, and radio sources in the core~A region.  
\end{abstract}


\maketitle
\clearpage

\section{Introduction}
The $\rho$~Ophiuchus cloud is one of the nearest sites of 
on-going star formation. The optical and infrared study of
\citet{Chi81} gave a mean distance $d$ = 165 $\pm$ 20 pc,
but more recent work based on {\em Hipparcos} and
{\em Tycho} data for off-core sources \citep{Knu98} suggests 
the value may be as low as 120 pc. We use $d=165$~pc in this paper.
Despite its proximity,
very few stars are optically revealed due to high extinction.
Even so, previous observations at X-ray, near-infrared,
far-infrared, and radio wavelengths show a diverse collection of
objects including protostars, T Tauri stars, and brown dwarf 
candidates (BDCs). Near the center of Cloud Core A
(as designated by \citet{Lor90}) are the magnetic B4 star Oph~S1
\citep{And88,And91}, the Class~0 radio source LFAM~5 \citep{Leo91}, and the
class~I protostars GSS~30 IRS-1  and GSS IRS-3, first identified by 
\citet{Gra73}.

The first X-ray images of $\rho$~Oph obtained with the {\it Einstein}
Observatory revealed a large population \citep{Mon83} of flaring
young stellar objects (YSOs).  In subsequent observations with {\it ROSAT},
{\it ASCA}, and {\it Chandra}, Class~I protostars, Class~II classical T-Tauri
stars, and Class~III weak-line T-Tauri stars
\footnote{Class I, II, and III designations for YSOs are based
on the value of the spectral index
$\alpha$ = $d$~log($\lambda$F$_{\lambda}$)/$d$~log$\lambda$,
which is usually evaluated in the near to mid-IR spectral range.
As discussed by \citet{Hai01}, class I sources
have  $\alpha$ $>$ 0.3, flat spectrum sources have
$-$0.3 $\leq$ $\alpha$ $<$ 0.3, class II sources
have $-$1.6 $\leq$ $\alpha$ $<$ $-$0.3, and class III
sources have $\alpha$ $<$ $-$1.6. By comparison, class 
0 sources are usually not detected in the near-IR.}
in $\rho$~Oph have exhibited some
of the largest stellar flares ever observed
\citep{Koy94,Cas95,Gro97,Kam97,Gro00,Tsu00,Gro01,Ima01,Ima02}.  

The ubiquity of X-ray flaring and rapid
variability suggests that magnetic fields play a central role
in the X-ray emission of most young stars. However, the details
of this process and the geometry of the magnetic structures 
involved are still under intense study.   Evidence for 
extended magnetic structures around YSOs comes mainly from 
high-resolution radio observations. {\it VLA} observations have
detected moderately circularly polarized nonthermal emission 
indicative of ordered magnetic fields in WTTS such as Hubble 4
(e.g. \citet{Ski93}). At higher angular resolution, VLBI observations
have shown that the radio emission of WTTS typically arises in
compact regions of diameter $\leq$25 R$_{*}$ with high brightness
temperatures T$_{b}$ $\geq$ 10$^{7.5}$ K \citep[e.g.,][]{And92},
\citep{Phi91}. Of particular interest here is the strong nonthermal
radio source Oph~S1 in $\rho$~Oph core~A whose magnetospheric 
diameter $D\approx12R_\star$ has been directly measured using
VLBI techniques \citep{And91}. Oph~S1 consists of a magnetic
B4 star (Oph~S1 A) and a K $=$ 8.3 companion (Oph S1~B) at a separation of
20 mas discovered by lunar occultation \citep{Sim95}.
The B4 star is unusual, belonging to a small class of magnetic
OB stars which includes $\theta^1$~Ori~C (O7~V), the central
star of the Orion Nebula.

Although the existence of extended magnetic structures in young stars is 
now firmly established, it is not yet clear how these magnetic 
structures relate to the observed X-ray emission (and its variability)
nor is it even certain that the emission processes in YSOs are the same as in 
older late-type stars like the Sun. 
In the Sun, X-ray emission is ultimately linked to the presence
of dynamo-driven magnetic fields. X-ray emitting plasma is confined 
in magnetic loops and X-ray flares are an observational consequence
of rapid energy release during magnetic reconnection events 
in the corona.  However, in younger stars the situation is more
complex since magnetic fields may be of primordial origin or 
produced by dynamo mechanisms that are different from the Sun, and interaction
of the stellar magnetosphere and the surrounding disk may be possible.

\citet{Shu94} and \citet{AC96} have invoked extended magnetic fields that 
couple the photosphere to the disk in order to explain the slow rotation of 
classical T~Tauri stars.  If this basic model is correct,
then the magnetic fields must extend to the Keplerian co-rotation radius,
several stellar radii from the photosphere. 
Quasi-periodic  count rate variations in objects such as the Class~I
source YLW~15 in  $\rho$~Oph core-F \citep{Tsu00} have been interpreted
in terms of differential rotation of the protostar and its accretion disk,
leading to magnetic shearing and reconnection thereby producing
large-amplitude, long-duration, quasi-periodic X-ray flares \citep{Mon00}.
Also, high-resolution {\it Chandra} grating spectra of the CTTS
TW~Hya show weak forbidden-line emission that may be attributed
either to high electron densities \citep{Kas02} or to UV irradiation
of the X-ray emitting plasma \citep{Gag02}, possibly from a UV hot spot.
Either explanation suggests that material may be funneled inward
along field lines from the inner edge of the disk to the photosphere,
producing an X-ray emitting shock at or near the stellar surface.

The possible effects of circumstellar disks and/or accretion on
the X-ray emission of young stars is a subject of current 
interest and debate. The presence of a disk is usually inferred
from  non-photospheric excesses at near to mid-IR
wavelengths. In a {\it Chandra} study of the  Orion Nebula
Cluster (ONC), it was found that stars with and without IR
excesses have similar X-ray luminosity functions \citep{Fei02,Fei03}.
This suggests that IR disks themselves have little if any effect on
X-ray emission levels. 

However, it is not necessarily the case that all  stars surrounded
by IR disks are actively accreting, and there is now some evidence
that accretion itself (rather than IR disks) may have an effect on X-ray 
emission. Accreting sources are usually identified by UV excesses
or by optical diagnostics such as emission at H$\alpha$, O~I, or
Ca~II. Such optical diagnostics are not generally available in heavily-obscured
regions such as $\rho$ Oph, but are accessible in other star-forming regions.
The {\it ROSAT} surveys of Taurus-Auriga
\citep{Ste01}, NGC 2264 and Chameleon I \citep{Fla03b} and {\it Chandra}
observations of the Orion Nebula Cluster and IC 348 \citep{Fla03a,Pre02,Sta04}
find that stars with high accretion rates have, on average, 2-3 times lower
X-ray luminosity than stars with low accretion rates.
Moreover \citet{Fla03b} show that, in the Orion Nebula Cluster and in NGC 2264,
stars with infrared excesses have lower $L_{\rm X}$ and $L_{\rm X}/L_{\rm bol}$
than stars with no IR excess, although the effect is less pronounced.
\citet{Fla03a} argue that accretion and a disk affect the magnetic
geometry, thereby reducing the available surface area at the photosphere for
closed magnetic loops, thus reducing $L_{\rm X}$ and $L_{\rm X}/L_{\rm bol}$.
\citet{Fla03c} further suggest that disk dissipation and/or decreased 
mass accretion causes an increase in $L_{\rm X}/L_{\rm bol}$ during the
first few Myrs of PMS evolution. After 10 Myr, $L_{\rm X}/L_{\rm bol}$ 
declines with rotation.

We present simultaneous
observations of the $\rho$~Oph~A cloud with the {\it Chandra} ACIS-I camera
and the {\it VLA} that provide new information
on the cloud population and its X-ray and radio properties. Our 
primary objectives were (i) to obtain a sensitive 
high angular resolution X-ray and 
radio census of the young stellar population in  $\rho$~Oph~A and
quantify the X-ray and radio properties of the population, 
(ii) to accurately identify counterparts to X-ray sources at
other wavelengths, particularly in the near-infrared and radio,
(iii) to determine if significant differences exist in the X-ray properties 
of the different stellar subgroups (Class 0/I, II, III and brown dwarf
candidates), and (iv) to search for possible relationships (including correlated
variability) between the X-ray and radio emission of YSOs in 
$\rho$~Oph~A. These simultaneous high-angular resolution X-ray and radio
observations are so far unique and provide compelling evidence
for  strong magnetic activity in the embedded $\rho$~Oph~A population.

\section{A Database of Young Stellar Objects in  $\rho$~Oph A}

As part of this study, we  have compiled an electronic database 
consisting of 345 objects in the  {\it Chandra} ACIS-I FOV.
It includes previously  cataloged infrared, X-ray, and radio
data and visual extinction estimates, as well as new data 
presented here. All figures and tables herein are based on information
in this database. The database is
publically available and is described in more detail in Appendix A.

\section{{\it Chandra} Observation, Data Reduction, and Source Identification}

\subsection{{\it Chandra} Observation and Data Reduction}

The {\it Chandra} X-ray Observatory observed the
$\rho$~Ophiuchus Cloud A \citep[L1688;][]{Lyn62} continuously for 96.5~ks,
beginning at 2335 UT on 2000 May 15 and ending at 0319 UT on May 17.  
{\it Chandra} collected consecutive exposures of 3.24~s on the Advanced CCD 
Imaging Spectrometer (ACIS), using ACIS chips I0-I3 and S2-S3. The ACIS-S chips
were located far off-axis and were not used in our analysis,
 and will not be discussed in this paper.
The high-resolution mirror assembly (HRMA) and ACIS-I camera
are described in detail in the {\it Chandra Proposers' Observatory Guide}
\footnote{http://cxc.harvard.edu/udocs/docs/docs.html} and in
\citet{Wei02}.
The ACIS I3 aim-point was on Oph~S1 at (J2000) 
$\alpha = 16^{\rm h}26^{\rm m}35{\stackrel{\rm s}{\textstyle{.}}}2$, 
$\delta = -24^{\circ}23\arcmin13\arcsec$.  

We applied a standard data reduction using the
{\it Chandra Interactive 
Analysis of Observations} software
\footnote{http://asc.harvard.edu/ciao/} 
CIAO v2.2.1. The ACIS data
were obtained in VFAINT mode, allowing us to remove  afterglow events.
The event list was filtered to include events with standard ASCA grades
and with photon energies in the 0.5-7.0~keV band, thereby significantly
reducing the particle background.  We checked for background flares
and none were found.
Source identification (\S3.2) and light curve variability analysis
(\S5.3) are described below. Spectra of selected sources, along
with source-specific ARF and RMF response files, were extracted
using the CIAO tool PSEXTRACT.

\subsection{{\it Chandra} Images and Source Identification}

Two images which contain the entire 4-CCD ACIS-I FOV were generated to 
search for X-ray sources:
a full-resolution 2800$\times$2800 pixel image ($0.492\arcsec$ pixels)
and a binned 1400$\times$1400 pixel image ($0.984\arcsec$ per pixel).
Figure 1 shows the full-resolution 2800$\times$2800 pixel image.
To register the {{\it Chandra} image against infrared positions, 
we selected 43 2MASS sources with obvious counterparts on the {\it Chandra}
image.  The 2MASS J2000 positions and {\it Chandra} physical pixel positions
were then used to derive a 4-coefficient ACIS-I plate scale solution using the
Starlink program ASTROM. The ASTROM $-$ nominal offsets in R.A. and Decl.
were $-0.84\arcsec$ and $-3.68\arcsec$, respectively.
These offsets were applied to the {\it Chandra} image header
keywords to obtain the coordinates listed in Table 1.

We detected X-ray sources on each image using CIAO's wavelet-based tool, 
WAVDETECT.  We ran WAVDETECT using scaling 
factors 2, 4, and 8, with a false alarm probability of $10^{-5}$.  
In some cases, regions near the edge of
the field of view were detected as sources by WAVDETECT because the off-field
background is zero.  After visual inspection, these spurious detections were
omitted.  Two X-ray sources in Table~1 were not detected by WAVDETECT:
SKS~3-12=J162622.2-242447 and BKLT J162636-241902=J162636.8-241900.
They have been added because they contain more than 7 counts and correspond to
known IR sources.

WAVDETECT computes the 1-$\sigma$ major- and minor-axis radii of each source
ellipse.  Events within a 3-$\sigma$ major- and minor-axis radii ellipse
were extracted for each source.  In a few cases, the extraction ellipses were
shrunk to avoid region overlap with nearby sources.
The net counts given in Table 1 are the raw counts in the elliptical extraction region
minus the estimated background in the same region. To determine
the number of background counts we used the area of the extraction
region and assumed a constant 0.5-7 keV background rate over the CCD of 
0.17 counts~s$^{-1}$ per ACIS-I CCD (see the {\it Chandra Proposers' 
Observatory Guide}). The typical background during our 96.5 ks exposure
is approximately 4 counts per
source. Table~1 lists the 87 sources in the 17$\arcmin\times17\arcmin$ 
ACIS-I FOV. 
Of the 87 {\it Chandra} sources, 60 are identified with catalogued IR or radio sources.
All but three of these identifications have X-ray $-$ IR offsets $< 1\arcsec$.

\section{{\it VLA} Observations and Data Reduction} 

Multifrequency radio observations of $\rho$~Oph core~A were obtained 
simultaneously with {\it Chandra} on 2000 May 16 using the 
NRAO\footnote{The National Radio Astronomy Observatory (NRAO) is a facility of 
the National Science Foundation, operated under cooperative agreement by 
Associated Universities Inc.} {\it Very Large Array}.  The observations 
provided continuous rise-to-set monitoring of $\rho$~Oph over $\sim$7 
consecutive hours, as summarized in Table~2. 
The VLA was in C-configuration and all 26 operational antennas 
were used at each observing frequency (no subarrays). The primary frequency
was 4.86~GHz (6 cm), but briefer coverage was also obtained at 1.42 and 8.46~GHz
to search for additional radio sources that might have been missed at 4.86~GHz.
Since none were found, the discussion below focuses mainly on the 4.86~GHz results, 
which provide the best sensitivity and most complete spatial coverage. 

Nine separate 6 cm pointings were distributed on a grid centered on
Oph~S1 with a $\sim4.5\arcmin$ grid spacing in order to provide nearly
uniform radio sensitivity across the full {\it Chandra} FOV. Data at
each grid point were obtained in two  $\sim$15 minute scans, with each
scan bracketed by phase calibrator observations. In addition, we obtained 
a single 1.4~GHz pointing centered on Oph~S1 and four 8.4~GHz pointings
centered on Oph~S1 and the known radio sources LFAM~2,~5,~and~15 
identified in the previous VLA survey of \citet[hereafter LFAM]{Leo91}.

We edited and calibrated the data using the AIPS\footnote{Astronomical Image 
Processing System (AIPS) is a software package developed by NRAO.} 
software package.  The data at each grid pointing position were 
combined to produce separate cleaned maps at each frequency
using the AIPS task IMAGR with natural weighting. 
In those cases where the field contained a bright radio source, we used 
phase-only self-calibration to improve the dynamic range.  Typical rms noise 
levels at phase center in Stokes I cleaned maps were 36~$\mu$Jy/beam (6~cm), 
40~$\mu$Jy/beam (8~GHz), and 220~$\mu$Jy/beam (1.4~GHz).  The higher noise 
level at 1.4~GHz is due mainly to the presence of a bright extragalactic source (BZ6) 
located $\sim7\arcmin$ southeast of Oph~S1. 

\subsection{{\it VLA} Source Identification}

We identified radio sources using cleaned maps at a 5$\sigma$ detection threshold, 
yielding the 31 detections listed in Table~3.  We measured peak fluxes 
with the AIPS task IMEAN, and we measured total fluxes using IMFIT 
(Gaussian source model) and TVSTAT (pixel summation inside of the 
2$\sigma$ flux contour). Fluxes for off-axis sources were corrected for primary 
beam attenuation using PBCOR.

Using our database, we attempted to identify known counterparts for 
each of the radio detections in Table~3. Previous detections at one or
more wavelengths were found for 18 of the 31 VLA sources (17 of 28 in the
{\it Chandra} FOV). Sixteen of the radio sources are confirmations
of previous VLA detections obtained by LFAM and \citet[hereafter SFAM]{Sti88}.
We note that the VLA beam did not resolve Oph S1~A and S1~B.  As a result,
we assign the radio emission to the binary Oph S1.

We report new radio detections of the IR source WLY~2-11 
(J162556.1-243014) and the emission line star Elias~24
($=$WSB~31$=$J162624.0-241613). 
WLY~2-11 is an embedded IR source with no visible counterpart and an 
estimated extinction of $A_{\rm V} = 7.3$ \citep{Wil89}.  
Elias~24 is a faint ($R = 18.5$) star whose H$\alpha$ emission is barely
detectable above the continuum \citep{Wil87}. Its spectral type
based on near-IR spectra is  M2  \citep{Dop03}.
 
Two relatively bright 6-cm radio sources observed by LFAM were not detected in 
our VLA observations: LFAM~20 and LFAM~22 in Table~1 of LFAM, whose 
respective flux densities in April 1988 were 824~$\mu$Jy and 599~$\mu$Jy.  Our 
upper limits of $\sim$150~$\mu$Jy (3$\sigma$) for these two objects indicate 
that they have weakened considerably. 

Thirteen of the {\it VLA} detections have no known counterparts and were
not detected in previous VLA surveys of comparable sensitivity (LFAM, SFAM).
One of these, J162615.5-243428, is relatively bright with $S_{\rm 6cm}=4.09$~mJy
and is offset by $\approx6\farcs7$ from the HII region Oph 11.
Given that these unidentified sources were not detected previously, some
may be variable emitters associated with deeply embedded objects.

\section{X-ray Source Properties}

\subsection{Summary of X-ray Properties}

{\it Chandra} detected X-ray sources with luminosities
spanning more than four orders of magnitude from $\log L_{\rm X}=27.67$
- 31.82~ergs~s$^{-1}$ (\S5.2). The mean photon energy
of detections is typically $\langle$E$\rangle$ $\approx$ 3 keV,
but the two hardest sources have 
$\langle$E$\rangle$ $\approx$ 5 keV (Table 1).
Variability is detected in
45 of 87 sources (52\%) as judged by a value of 
the Kolmogorov-Smirnov statistic KS $>$ 1 (\S5.3).
For those detections with known IR counterparts (Table 4), visual absorption
ranges from $A_V$ = 1.8 - 56.0 mag. The faintest IR source detected
by {\it Chandra} was GY~144 (K = 13.46). 
  
The list of X-ray detections in Tables 1 and 4 includes 12 of 14 known CTTS 
and 15 of 17 known WTTS within the ACIS-I FOV. The X-ray
luminosity distributions of CTTS and WTTS are 
similar (\S7). Four of 15 BD candidates
were detected. Apart from their lower X-ray luminosities,
the X-ray properties of BD candidates are otherwise 
similar to TTS (\S9). 
The class 0 radio source LFAM~5 and the class I protostars
GSS 30 IRS-1/IRS-3 were not detected by {\it Chandra}, but 
LFAM~5 and GSS 30 IRS-3 are confirmed as  radio sources.  

\subsection{X-ray Fluxes and Spectral Parameters}

The absorbed fluxes in column 8 of Table~1 were computed
from the event list of each source, which contains the time-tagged
energy of each detected photon. An auxiliary response file
(ARF) was generated for each source, which provides the effective 
area corrected for quantum efficiency (cm$^{2}$ counts photon$^{-1}$)
versus energy (keV) at the source's position on the detector.
The absorbed X-ray flux at Earth is,
\begin{equation}
   f_{\rm X} = \frac{1}{t} \sum_{i=1}^{n} \frac{E_i}{a_i(E)} 
   - f_{\rm bkg}A,
\end{equation}
where $E_i$ is the energy of each photon, $a_i(E)$ is the corresponding
effective area, $t=96439.96$~s is the livetime of the observation,
A is the area of the source extraction region, and
$f_{\rm bkg}=3.605\times10^{-18}$~ergs~cm$^{-2}$~s$^{-1}$~pixel$^{-1}$
is the 0.5-7.0~keV background flux as determined from a deep, source-subtracted
observation from the {\it Chandra} calibration database. 

Traditionally, extraction and model fitting of a background-subtracted
source spectrum are used to determine spectral parameters such as
the absorption column density N$_{\rm H}$, mean plasma energy kT,
and unabsorbed X-ray luminosity L$_{X}$. Obviously, an accurate
determination of unabsorbed L$_{X}$ requires knowledge of N$_{\rm H}$,
as discussed by \citet{Cas95}. For some sources in $\rho$ Oph,
reliable estimates of the visual extinction are available 
from previous studies, and we are able to constrain the 
absorption column density using the relation 
$N_{\rm H}/A_{\rm V}=1.57\times10^{21}~{\rm cm}^{-2}$ mag$^{-1}$
\citep{Vuo03} (\S2).
In the absence of  A$_{V}$ information,
the X-ray data were used to estimate N$_{\rm H}$.

Although the traditional spectral fitting approach gives satisfactory
results for bright sources ($\gtsimeq$100 counts), it does not work
well for fainter sources because of issues related to binning of the
spectrum when only a few counts are available. Since $\approx$60\%
of the X-ray detections in $\rho$ Oph~A have  $\leq$100 counts, we
have developed a non-parametric method for estimating spectral   
parameters that operates on unbinned photon event data, as described
in more detail in Appendix B.  This method
is based on a large number of simulated isothermal ACIS-I spectra spanning  
a range of ($\log N_{\rm H}$, $\log T$) values, utilizing the APEC emission model
\citep{Smi01}, the Wisconsin absorption model \citep{Mor83}, and the MARX 
ray-tracing {\em Chandra} simulation software
\footnote{http://space.mit.edu/CXC/MARX/}. The method was used to 
determine the X-ray temperatures and unabsorbed luminosities 
for all {\em Chandra} detections given in Table 4, as well
as N$_{\rm H}$ values for those detections where a previous determination 
of A$_{V}$ was not available. The X-ray temperature and luminosity estimates
are based on the assumption that the plasma is isothermal. The inferred values
thus provide only a rough temperature estimate that can be used to 
compare a large number of sources. In reality, the plasma is 
likely to be more complex, time variable and involve multiple temperature
components. 

\subsection{X-ray Variability and Light Curves}
 
To quantify X-ray variability, the event list of each source was used
to calculate the KS statistic, given by 
\begin{equation}
   {\rm KS} = \sqrt{n}\,\sup|f_{i}(t)-f_{0}(t)|.
\end{equation}
where $n$ is the number of events,
$f_i(t)$ is the normalized observed cumulative distribution
and $f_0(t)$ is the normalized model cumulative distribution,
assuming a constant flux.  
If $z$ is the value of the KS statistic then the null hypothesis
probability P($z$), or probability of constant flux, is given by
a convergent infinite series \citep{Pre92}: \\

\begin{equation}
{\rm P}(z) = 2 \sum_{j=1}^{\infty} (-1)^{j-1}e^{-2j^{2}z^{2}} 
\end{equation}

Truncating the series at six terms gives a few representative
values:   P(0.5) = 0.96, P(1.0) = 0.27,
P(1.5) = 0.02,  P(1.7) = 0.006,
and P (2.0) = 6.7 $\times$ 10$^{-4}$.

Sources with $z$ = ${\rm KS}>1$ generally show some
light curve variability and sources with ${\rm KS}>2$ typically reveal obvious 
signs of flaring.
The source event files were used to generate light curves for all 87 X-ray
sources. Figures~\ref{f2}-\ref{f8} show light curves for seven sources showing
interesting variability. The time bin sizes in each light curve are varied so as
to include approximately the same number of counts per bin.

The top panel in each light curve shows $f_{\rm X}$, which as already 
noted is the the background subtracted flux at Earth (Sec. 5.2). 
In the simplest approximation, $\log T$ and $\log N_{\rm H}$ determine
the shape of the X-ray spectrum and the flux conversion factor, the
ratio of unabsorbed-to-absorbed flux. At each time bin, $\log T$ and $\log N_{\rm H}$ 
and their associated random errors are estimated by the non-parametric method described
in Appendix B.  The two lower panels in Figs.~\ref{f2}-\ref{f8} 
show $\log T$ and unabsorbed luminosity $L_{\rm X}$ assuming a distance $d=165$~pc.

Fig.~\ref{f2} reveals that the optically bright WTTS DoAr 21 was in the 
decay phase of a large flare throughout the observation, and the 
spectral changes that occurred during this decay are discussed below (Sec. 7.2.3).
As apparent in Fig.~\ref{f2}, the decay was punctuated by frequent hard-energy 
spikes, which can be interpreted as reheating events. The peak in mean energy often 
coincides with or immediately precedes the luminosity peak. The time-resolved spectral
fits suggest that the hard-energy spikes are not consistent with sudden
increases in column density. As a result, it may necessary to explicitly
consider reheating when modeling flare decay with loop models.
The relatively short rise and decay times suggest moderate-sized magnetic loops.

\subsection{Extragalactic X-ray Sources}

The KS statistic, count rate, and $\langle{\rm E}\rangle$ can be used to
isolate possible extragalactic X-ray sources \citep*{Dan02},
as demonstrated in Figure~\ref{f9}a-c. Of the 27 {\it Chandra} sources
with no IR or radio counterparts, 26 show $\langle{\rm E}\rangle>3$~keV,
count rates $<0.001$~cts~s$^{-1}$, and KS$\lesssim1$.
XSPEC simulations of faint, highly absorbed ($25<A_V<50$) power-law
($2<\alpha<4$) sources sources give rise to a similar distribution of
$\langle{\rm E}\rangle$ and count rate. Furthermore, extragalactic X-ray
sources are expected to have low KS statistic because they generally
do not show large-amplitude variations on short time scales \citep{Zam84,CM97}.

Using equations (2) and (3) of \citet{Cow02} to estimate the expected number
of extragalactic sources in a 96~ksec exposure of $\rho$~Oph, a limiting flux
$f_{\rm X}=9\times10^{-16}$~ergs~cm$^{-2}$~s$^{-1}$ yields approximately
16 extragalactic sources in the $16\farcm9\times16\farcm9$ ACIS-I FOV.
We note, though, that the predicted extragalactic source count depends on the assumed
extragalactic source spectrum and the spatially varying absorbing cloud column,
both of which are difficult to estimate. Nonetheless, it seems likely that
the majority of the 26 {\it Chandra} sources with no infrared or radio
counterparts are extragalactic. However, a few of the 26 unidentified
sources may be new cloud members. In particular, the unidentified
{\it Chandra} source J162627.4-242418 is a  candidate for 
cloud membership
since it is variable in X-rays (KS = 2.77) on a timescale of less
than one day.

\subsection{X-ray vs. Infrared Properties}

Figure~\ref{f10} shows a $J-K$ versus $K$ color-magnitude diagram for those
IR sources in the ACIS-I FOV with measured JHK magnitudes, including 
a few fainter sources which have only lower magnitude limits at J.
The faintest K-band source detected by {\em Chandra} is 
J162651.9-243039 (= GY 144) at $K=13.46$, although its $J-K$ color
is not well-determined.

In Figure~\ref{f11} we show a color-color diagram for those IR sources 
having measured J, H, and K magnitudes.  As Fig.~\ref{f11} shows,
most of the {\it Chandra} detections are confined to a strip
oriented roughly in the direction 
of the $A_{V}=10$ reddening vector, consistent with normal
reddening for cluster members. The colors of the CTTS GY 81 
and the WTTSs SKS~1-11 and GY 84 are consistent
with cluster membership but they were not detected with {\it Chandra}.
We note that the source  GY 101 has unusual colors for its assigned
class III type, but its J-H value is only a lower limit.
We used J $>$ 17.0 (Table 4) to compute its J-H in Fig.~\ref{f11}, 
but 2MASS gives a more stringent limit J $>$ 18.67.

\subsubsection{Unclassified Infrared Sources}

{\em Chandra} detected X-ray emission from more than two dozen
IR sources whose SED classifications are not yet known (Table 4). 
Are these unclassified IR sources cloud members, such as T Tauri
stars? We argue below that most of these X-ray/IR sources are 
{\em not} X-ray active AGNs, and conclude that many are likely
to be cloud members.

To substantiate the above conclusion, we consider the ratio of
X-ray to K-band fluxes. To estimate the unabsorbed IR fluxes, we
calculated the interstellar extinction for the J, H, and K bands using
\begin{equation}
   \frac{A_{\lambda}}{{\rm E}(J-K)} = 2.4 \lambda^{1.75}\,{\rm \mu m},
\end{equation}
as derived by \citep{Dra89}.  In conjunction with the fact that,
\begin{equation}
   \frac{N_{\rm H}}{{\rm E}(J-K)} = 1.1 \times 10^{22}\,{\rm cm^{-2}\,mag^{-1}},
\end{equation}
\citep{Dra89}, we derived $A_\lambda$ as a function of known $A_V$
for each source.  We found that for a typical source ($kT$ $\approx$ 2.0~keV and 
$N_{\rm H}=3.27\times10^{22}$~cm$^{-2}$), the K band 
extinction ($f_{\rm K}/f_{\rm K_\circ}=0.15$) is similar to the X-ray extinction
($f_{\rm X}/f_{\rm X_\circ}=0.25$).  Therefore, we used $f_{\rm K}$ to analyze
the relationship between IR and X-ray emission.
Figure~\ref{f12} illustrates that $\log(f_{\rm X}/f_{\rm K})$ is  
slightly lower for unclassified X-ray/IR sources 
(median $\log{f_{\rm X}/f_{\rm K}}=-3.0$) than YSOs with measured SEDs
(median $\log{f_{\rm X}/f_{\rm K}}=-2.6$).
This suggests that most of the unclassified X-ray/IR sources are probably
{\em not} X-ray active AGN because AGN typically show higher
X-ray-to-optical flux ratios than normal stars \citep{Mac88,Sto91}.

\section{Radio Source Properties}

\subsection{Radio Variability} 

Each of the {\it VLA} detections in Table~3 was observed in 
two or more $\sim$15 minute scans 
separated in time by anywhere from $\sim$0.5-3~hours.  To search for 
short-term ($\sim$hours) variability, we compared peak fluxes in the individual 
scans for the brightest radio detections (S/N$\geq$10).

Definite variability was detected in LFAM~2 (J162622.4-242252)  and 
Oph~S1 (J162634.2-242328). The peak 6-cm flux of LFAM~2 increased 
by a factor of $\sim$5 during the time interval 0629 -- 0931 UT, indicative 
of a radio flare.  This radio source is associated with the class III IR 
source GSS30 IRS-2 \citep{Bar97}.  No large amplitude flares were detected 
from Oph~S1, but its peak fluxes showed significant scan-to-scan variations 
of up to $\sim$30\% (30$\sigma$) over time intervals of $\sim$3 hours. 
For example, a 6 cm scan centered on Oph S1 in the time interval 0629 -- 0644 UT
gave a peak flux $S_{6\rm cm}^{\rm peak}$ = 7.94 $\pm$ 0.06 mJy ($\pm$1$\sigma$)
while a second scan at the same pointing position from 0921 -- 0931 UT gave
$S_{6\rm cm}^{\rm peak}$ = 5.86 $\pm$ 0.08 mJy. Four additional scans with
Oph S1 positioned $6\farcm4$ off-axis also confirm a trend of decreasing 
radio flux during the observation. Thus,
we find for the first time that radio variability is present in this 
unusual close binary system.   
We note little or no direct correlation between the VLA and {\it Chandra}
variability of Oph~S1.  For example the X-ray flux peak from 0820 -- 1005 UT 
does not coincide with the peak 6-cm emission interval from 0629 -- 0644 UT.
The apparent decoupling  between the X-ray and radio emission
is not unique to Oph~S1, and has been observed in other PMS objects
such as the multiple sytem V773 Tau (Sec. 7.3). 
If the radio emission is produced in an extended  magnetosphere around the B4 star
\citep{And91} and the X-ray flares are produced in a corona around the
K-type secondary, then one would not necessarily expect to see a 
correlation between the X-ray and radio emission. 

Given that the WTTS DoAr~21 was in the decay phase of a large X-ray flare,
it is noteworthy that its 6-cm radio flux appeared rather stable.
A 15-minute scan centered at 0619 UT with DoAr~21 located $2\farcm7$ from phase 
center gave a peak flux of $S^{\rm peak}_{\rm 6cm}$ = 12.04 $\pm$ 0.05 mJy,
and a second 10-minute scan at the same pointing position centered at 0939 UT
gave $S^{\rm peak}_{\rm 6cm}$ = 11.82 $\pm$ 0.04 mJy.
Two additional scans centered at 0542 UT and 1007 UT with DoAr~21 located
$7\farcm7$ from phase center also gave peak fluxes differing by less
than 3$\sigma$.  Thus, we find no compelling evidence for radio
variability in DoAr~21 up to the end of the last {\em VLA} 6 cm scan  at
1013 UT, but no simultaneous radio coverage is available during the
last $\sim$16 hours of the {\em Chandra} observation when several
X-ray temperature spikes occurred (Fig. 2). Even though no 
6 cm  variability was detected, DoAr~21 is clearly variable
on longer timescales. The 6 cm fluxes quoted above  are between
the peak flare value of 46.1~mJy on 1983 February 18 and the
much lower level of 3.7~mJy measured on 1983 June 18 \citep{Sti88}.

\subsection{Circular Polarization} 

Circularly polarized radio emission indicative of nonthermal processes was
clearly detected at 6~cm in three sources.  These are the optically visible
G-K type star DoAr~21 (J162603.0-242336), 
Oph~S1 (J162634.2-242328), and the extragalactic 
source LFAM~21$=$BZ6 (J162700.0-242640).  The fractional circular polarization 
($\pi_{c}=$V/I) and detection significance (S/N ratio) for these sources are 
given in the notes to Table~3. 

Previous detections of circular polarization in Oph~S1 at both 6~cm 
and 2~cm (with an upper limit at 20~cm) were reported by \citet{And88}.  
We confirm the presence of circular polarization at 6~cm and also provide a 
new detection at 3.5~cm, but likewise fail to detect circular polarization 
at 20~cm.

\section{Discussion: The Young Stellar Population in $\rho$ Oph A}

\subsection{Class 0/I Protostars}

The core-A region contains three well-known protostars, 
namely the class 0 source LFAM 5 and the class I sources GSS30-IRS1
and GSS30-IRS3. In many respects, LFAM 5 is the most unusual of
the three. It is a centimeter radio source that is thought to
be  driving a molecular outflow \citep{And90},
but has so far not been detected at millimeter, IR, or X-ray wavelengths.
Our {\it VLA} observation confirms radio emission from LFAM 5
(Table 3) at a 6 cm flux density similar to that reported by 
LFAM.

GSS30 is an asymmetric infrared bipolar reflection nebula with at
least 3 IR sources seen in projection \citep{Gra73,Cas85,Wei93}.
The class I source GSS30-IRS1  is thought to be the
illuminating star \citep{Cas85}. It has not been previously detected at
centimeter radio wavelengths and we likewise obtain only an
upper limit of S$_{6cm}$ $\leq$ 0.12 mJy (3$\sigma$) and an
identical upper limit at 3.5 cm.  Observations at 2.7 mm
have also failed to detect GSS30-IRS1 \citep{Zha97}.
There is no confirmed detectable outflow in this source.
However, line emission in recent VLT IR spectra was interpreted
as possible evidence for dense gas in an accretion shock \citep{Pon02}.

The class I source GSS30-IRS3 has K = 12.85 but no emission 
peak is seen at J or H \citep{Wei93}, and is thus much fainter  than 
IRS1. A 2.7 mm continuum source has been detected toward
GSS30-IRS3 at S$_{2.7mm}$ = 29 mJy \citep{Zha97}.
Centimeter radio emission is also detected from the radio
source LFAM 1, which is identified with GSS30-IRS3.
We confirm radio emission from LFAM 1 (Table 3) at a 6 cm flux near
the value reported by LFAM.

None of the three class 0/I sources discussed above was detected 
in our {\it Chandra} observation. Upper limits (Table 5) suggest
that their luminosities are below $\log L_{\rm X}\approx28.5$~ergs~s$^{-1}$.
However, these upper limits are based on uncertain $A_V$
and $kT$. In the absence of specific information, we have assumed
$kT\approx2.5$ keV and $\log N_{\rm H}\approx 22.5$~cm$^{-2}$,
typical of X-ray detected TTS in core~A. An extinction
estimate $A_V$ $\approx$ 38 is available for GSS30-IRS1 
\citep{Cas85},
and using this value along with $kT \approx 2.5$~keV and
a 7 count {\it Chandra} detection threshold we obtain
$\log L_{\rm X}\leq28.4$~ergs~s$^{-1}$ (0.5 - 7 keV).
Assuming a higher temperature
$kT \approx 6$~keV, as measured for two class I sources in
the cluster IC 348 \citep{Pre02}, then
the upper limit for GSS30-IRS1 becomes $\log L_{\rm X}\leq28.2$~ergs~s$^{-1}$.

We thus conclude that the X-ray luminosities of GSS30-IRS1 (and also
very likely GSS30-IRS3) are well 
below the  value of $L_{\rm X}\approx 10^{30}$~ergs~s$^{-1}$,
typical of class I sources detected by {\it Chandra} 
such as those in IC 348 \citep{Pre02} and
$\rho$ Oph core-F \citep{Ima01}. 
However, the {\it radio} luminosities of both LFAM 5 and 
GSS30-IRS3 are comparable to those of CTTS detected in 
core~A (Sec. 7.2.2).
Several factors may influence the X-ray detectability of
class 0/I sources. Most importantly, variability seems to be a
factor since several of the class I objects detected
by \citet{Ima01} in core~F were caught during 
X-ray flares. Extinction also plays a role, and class 0
objects in particular are believed to have very high absorption that
efficiently masks any soft X-ray emission that is present. 
Extinction will also make it more difficult to arrive at a 
complete picture of X-ray production mechanisms in protostars.
GSS30 IRS1 is a case in point. The suggestion that an
accretion shock is present \citep{Pon02} raises
the possibility that soft X-rays may be produced in the 
shock region, similar to what may be occurring in the
CTTS TW Hya \citep{Kas02}. Since shock-induced
X-rays are expected to have characteristic temperatures below
$\sim$1 keV, such emission (if present) would be 
heavily absorbed and difficult to detect in protostars
with typical extinctions $A_V\approx 40$.

\subsection{T Tauri Stars}

\subsubsection{X-ray Luminosity Functions}

Using the unabsorbed $L_{\rm X}$ values of X-ray 
detections (Table 4), the median
X-ray luminosity of the 12 detected CTTS is 
$\log L_{\rm X}=29.92$~ergs~s$^{-1}$ which is
similar to the median for 15 detected WTTS of
$\log L_{\rm X}=29.79$~ergs~s$^{-1}$. To make further
comparisons between the X-ray luminosities of CTTS
and WTTS, we take X-ray upper limits of the undetected
TTS into account (Table 5).
Including both detections and non-detections,
all four versions of the generalized
Wilcoxon test in the ASURV statistical 
analysis package \citep{Lav92} give a probability $p$ = 0.53 that
the CTTS and WTTS X-ray luminosities are drawn from
the same distribution. Thus, the Wilcoxon test is 
inconclusive for this sample. Even so, the
the  Kaplan-Meier estimators for the CTTS
and WTTS samples shown in Figure~\ref{f13}
appear very similar, agreeing to within the 
error bars.

We also find that the  spectral hardnesses in 
CTTS and WTTS are nearly identical. Using the mean photon energies
$\langle$E$\rangle$ in Table 1, the CTTS
detections have median $\langle$E$\rangle$ =
2.57 and WTTS detections have median 
$\langle$E$\rangle$ = 2.53.

On the basis of the shapes
of their IR spectral energy distributions, CTTS (class II sources)
are thought to be surrounded by circumstellar disks but WTTS (class III
sources ) are not. Given that the X-ray luminosities and 
spectral hardnesses of the CTTS and WTTS in our sample are quite similar,
there is no compelling evidence that the presence or absence
of infrared disks has any significant effect on their X-ray
emission. However, it should be emphasized that 
{\em accretion} diagnostics (like H$\alpha$ or Ca~II emission)
are not generally  available for sources in $\rho$~Oph cloud core A. 
Thus, our results cannot be used to determine whether accretion itself
affects the X-ray properties of this small sample.

\subsubsection{Radio vs. X-ray Luminosities}

Figure~\ref{f14} shows the 5~GHz radio luminosities ($L_{\rm 6cm}$) and
X-ray luminosities ($L_{\rm X}$) for those 6 TTS (class II and
III) that were simultaneously detected with the {\it VLA} and 
{\it Chandra}, along with {\em Chandra} upper limits for
class 0/I sources. We have also 
included Oph~S1, whose IR classification
is uncertain because of the close binary separation. 
Even though the sample is small, we have checked for a correlation
between $L_{\rm 6cm}$ and $L_{\rm X}$ amongst TTS. Excluding Oph~S1,
the generalized
Kendall's tau test in the ASURV software package gives a 
correlation probability $p$ = 0.15 and the Cox proportional
hazard model gives $p$ = 0.64. If  Oph~S1 is included (on the suspicion
that its faint companion Oph S1B is a TTS), then
these values increase slightly to $p$ = 0.35 and $p$ = 0.81, respectively.
Thus, we find that radio and X-ray luminosity
are not strongly correlated in this small sample of 
detected TTS.

We do note however that in our small sample there is a tendency for
WTTS to be more luminous radio sources than CTTS, even when comparing
objects with similar $L_{\rm X}$. In Table 3, all four WTTS (class III)
have larger $L_{\rm 6cm}$ values than the two CTTS (class II).
This may reflect different radio emission processes in WTTS
and CTTS, with nonthermal emission being more common in
WTTS.  It can also be seen from Figure~\ref{f14} that Oph~S1 is 
clearly situated in the region of higher $L_{\rm 6cm}$ occupied
by WTTS.

Two objects of some interest are the class II sources 
Elias 24 (J162624.0-241613) and GY51  (J162630.5-242256),
both of which were detected simultaneously  with {\it Chandra}
and the {\em VLA}. As  Figure~\ref{f14} shows, they have 
similar  $L_{\rm X}$, but GY 51 is nearly an order of magnitude
more luminous in the radio.  GY~51 shows a weak X-ray flare
near 1115 UT on May 16, while Elias~24 experienced a moderately strong
impulsive X-ray flare near 2400 UT.  The 3.6 cm and 6 cm radio fluxes
of GY 51 are nearly identical, suggesting a flat (or non-rising)
spectral energy distribution and possible non-thermal radio
emission. 

The faint M2 star Elias 24 was only detected at 
6 cm, so no definitive estimate of its radio spectral
index is available. However, there is no significant
3.6 cm emission in a  scan with Elias 24 
positioned 6.6$'$ from phase center, and we obtain
an upper limit S$_{3.6cm}$ $\leq$ 0.2 mJy (5$\sigma$).
This upper limit is approximately equal to the measured
flux at 6 cm (Table 3), suggesting a non-positive
spectral index $\alpha$ $\leq$ 0 (S$_{\nu}$ $\propto$
$\nu^{\alpha}$). This may be an indication of nonthermal
emission from magnetically-trapped particles. Such 
emission would be consistent with the impulsive 
X-ray variability, which is an indicator of magnetic
activity. However, the radio interpretion is not
yet clear since \citet{Gom03} have identified 
Elias 24 as the possible
driving source of a near-IR knot located 144$''$ away. If
this identification is correct, then the 6 cm radio emission 
could be thermal emission associated with an outflow, as
is already known to occur in other sources associated with
Herbig-Haro objects \citep{Bie84}. Higher sensitivity 
multi-wavelength observations will be needed to determine 
if the newly-detected
radio emission of Elias 24 is thermal or nonthermal.

\subsubsection{Time-Resolved X-ray Spectroscopy of a Large Flare on DoAr~21}

DoAr~21=GSS~23=V2246~Oph (K1, WTTS) is the brightest X-ray source in the 
ACIS-I FOV.  As is evident in Fig.~\ref{f2}, DoAr~21 was in
the decay phase of a large flare throughout the {\it Chandra} observation.
DoAr~21 was placed far off-axis to migitate pile-up in the ACIS~I3 
CCD. Although the total count rate of DoAr 21 exceeded 1 counts~s$^{-1}$ at 
the beginning of the observation, the source did not pile up 
significantly, contrary to a previous claim by \citet{Ima02}.
We find that the  pileup fraction never exceeded a few 
percent because the maximum count rate in the central pixel of the PSF was 
0.015~counts~s$^{-1}$~pixel$^{-1}$. Although the count rate decay in the top 
panel appears to be quite smooth up to a small flare around day 0.8, the 
$\log T$ light curve in the second panel is variable throughout.  We note that the 
temperature spikes appear to signal reheating events.  In many cases, 
a temperature spike immediately precedes a $f_{\rm X}$ maximum.
Based on its V-band absorption \citep{Wil01}, we expect DoAr~21 to have a 
column density $N_{\rm H}\approx 1.0\times10^{22}$~cm$^{-2}$.
$\log T$ and $L_{\rm X}$ in each time bin were estimated using the non-parametric
method in Appendix B.

In order to confirm the validity of the light curve analysis, 
the $\sim$81000 counts associated with DoAr~21 
were divided into 41 time segments each containing approximately 2000 counts.  
For each of the 41 ACIS-I spectra, background spectra and response files were 
generated using the CIAO tool PSEXTRACT.  Each of these were modeled in 
XSPEC V11 using a single-temperature VAPEC model.  We generally found good 
fits ($\chi^2 \lesssim 1$) using an Fe abundance of 0.5 solar and a column
density $N_{\rm H}\approx1.0\times10^{22}{\rm cm}^{-2}$. Because column-density 
variations have been reported during X-ray flares on YSOs \citep{Tsu98}, the column 
density, temperature, and normalization were free parameters.  After each fit, 
the emission measure and absorption-corrected X-ray luminosity were calculated 
assuming a distance of 165~pc \citep{Chi81}. Indeed, apparent variations
of 10--20\% in the best-fit column density values are seen throughout the
observation. The increases in $N_{\rm H}$ do not generally show corresponding
increases in hardness ratio or count rate.  The increases in best-fit
$N_{\rm H}$ are, however, associated with decreases in the best-fit temperature
and increases in $L_{\rm X}$ as a result of the increased correction due to
absorption. We believe that this result is an artifact of the fitting procedure:
moderate-resolution, moderate signal-to-noise spectra cannot distinguish small
increases in column density from small increases in temperature. Since we have
no reason to suspect large column-density changes on a WTTS like DoAr~21, the
column-density was fixed at the median best-fit value: 
$N_{\rm H}=1.08\times10^{22} {\rm cm}^{-2}$ and the spectra were then refit.
Note that this value of $N_{\rm H}$ determined from spectral fits differs
by only 6\% from the value determined by our non-parametric analysis
(Table 4).

Assuming $V=13.62$, ${\rm BC}=-0.36$, $d=165$~pc, 
$N_{\rm H}=1.08\times10^{22} {\rm cm}^{-2}$, 
and $N_{\rm H}/A_{\rm V}=1.57\times10^{21}~{\rm cm}^{-2}$ mag$^{-1}$,
the bolometric luminosity of DoAr~21 is 
$L_{\rm bol} \approx 2.3\times10^{35}$~ergs~s$^{-1}$.  During the 
{\it Chandra} observation, $L_{\rm X}/L_{\rm bol}$ decreases from 
$4.0\times 10^{-4}$ to $2.2\times 10^{-4}$.

The results of the time-resolved spectral analysis are shown in 
Fig.~2.  There is a steady overall decline in 
emission measure and $L_{\rm X}$ over the course of the observation and the 
plasma undergoes no fewer than five temperature spikes.  If the temperature 
spikes are ignored, then the plasma would appear to be cooling slowly from a 
giant flare which occurred before the start of the {\it Chandra} observation.  
The long decay time and the slow but steady decrease in temperature would 
indicate a large, low-density flaring loop.  For example, the two-ribbon flare
model of \citet{KP93} implies a loop radius in the range 
$1-3 R_{\odot}$. 

Previous observations of DoAr~21 with {\it ROSAT} and 
{\it ASCA} suggest approximately one large flare per day \citep{Ima02}.  If 
the flare decay in Fig.~2 is typical of flares on DoAr~21, then the corona 
is in a state of nearly continous flaring with low- and moderate-energy flares 
outnumbering large flares.

The frequent increases in $kT$ seen in Fig.~\ref{f2}
suggest that plasma is being intermittently reheated by moderate-energy flares.
If the reheating events occurred in the same loops as the primary decaying
flare then a simple cooling-loop model will significantly overestimate the
cooling time, thereby underestimating the density and overestimating the loop
radius. Or, the smaller flares may signal reconnection events in
separate magnetic loops.

Long, uninterrupted monitoring of DoAr~21 or simliar stars  may
provide a clue to the location of these smaller, shorter-duration flares. 
If short-duration flares during the decay phase of long-duration flares are
more frequent than short-duration flares during ``quiescent'' periods,
then the two types of flares probably occur in the same or adjoining arcades
of magnetic loops.

A final question concerns the geometrical relationship between the 
X-ray and radio emitting regions in DoAr~21. As already noted (Sec. 6.1),
two separate {\em VLA} scans of DoAr~21 at 0619 UT and 0939 UT (16 May)
gave nearly identical 6 cm fluxes differing by $<$2\%, while at the same time
its X-ray flux declined by about 15\%. This indicates that the 
circularly polarized nonthermal
radio emission originating in large magnetospheric structures is effectively
decoupled from the thermal X-ray emission, which  originates much
closer to the star - presumably in hot coronal plasma. Similar behavior
was noted by \citet{Fei94} in simultaneous  {\em ROSAT} and
{\em VLA}/{\em VLBI} observations of the unusual WTTS V773~Tau.
However, the radio flux of V773~Tau declined while its X-ray
emission remained steady, whereas the opposite behavior is 
observed in DoAr~21. The interpretation of the behavior of V773 Tau
is now complicated by the discovery that it is a quadruple system
consisting of a K2V + K5V spectroscopic binary with a nearby M-dwarf at
0.12$''$ \citep{Ghe95}, and an IR companion 0.21$''$ away 
\citep{Duc01}.  {\em ROSAT} could not spatially resolve these
four components, raising the obvious question of whether the (variable)
radio emission and (steady) X-ray emission originated in the same
star. Even though  DoAr~21 is not yet known to be such a closely-spaced
multiple system, this possibility should not yet be dismissed given
its similarities to V773~Tau. 

\subsection{The Unusual Close Binary System Oph S1}

Figure 15 shows the ACIS-I spectrum  of the 
unusual magnetically-active binary system Oph S1.
As we have noted,
this star consists of a B-type primary and a fainter secondary
at 20 mas separation \citep{Sim95}. At this close separation, 
the two components cannot be 
spatially resolved by {\it Chandra} or the {\it VLA}. 
The spectrum shows strong absorption with little or no 
detectable emission below  $\sim$1 keV. There are no clearly
discernible emission line features in the spectrum.
The light curve shown in Fig.~\ref{f4} shows $L_{\rm X}$ variations
from $2-4\times 10^{30}$~ergs~s$^{-1}$ with two small flare peaks 
at 0815 UT and 2453 UT.

We have analyzed the spectrum
using a variety of emission models
in XSPEC including  single-temperature (1T) and two-temperature (2T)
optically thin plasmas (VAPEC), bremsstrahlung, power-law (PL),
and the Chebyshev polynomial differential emission 
measure (DEM) model C6PVMKL. 
All models included an absorption component and the spectrum was
rebinned to a minimum of 15 counts per bin prior to fitting
in XSPEC. We compared spectral fits using spectra extracted 
with CIAO v. 2.2.1  and with the more recent CIAO v. 2.3/CALDB 2.0.
and CIAO v 3.0.2/CALDB 2.26 releases. Corrections for charge transfer
inefficiency (CTI) were applied to the spectra extracted with
CIAO v. 2.3 and  CIAO v 3.0.2. The best-fit values from 
spectral fits in Table 6 are from CIAO v. 2.3 with CTI corrections
applied. Generally, the three different versions of CIAO gave
similar best-fit values but the fit residuals as measured by
reduced $\chi^2$ were  15\% - 20\% smaller with CTI corrections
applied.

Because of the lack of strong emission features, several different
models can acceptably fit the Oph S1 spectrum. Table 6 gives
the derived spectral parameters for a power-law model as well as
1T and 2T VAPEC models. As can be seen, the fit statistics are
very similar for these three models. Thus, we cannot distinguish
between a power-law spectrum and optically thin  plasma emission
at the signal-to-noise ratio of the Oph S1 X-ray data. 

The 1T VAPEC 
model implies a mean plasma temperature 
kT $\approx$ 2.4 keV. A bremsstrahlung model gives a nearly 
identical temperature kT $\approx$ 2.1 keV and the C6PVMKL
model also shows a strong emission measure peak at 
kT $\approx$ 2.0 - 2.4 keV. Considered collectively, these models
indicate that most of the emission arises from plasma at 
or slightly above kT  $\approx$ 2 keV (T $\approx$ 2.3 MK), 
assuming a thermal origin.

A 1T VAPEC model using solar abundances gives a relatively poor
fit ($\chi^2$/dof = 165.4/135 = 1.23). The fit can be substantially
improved by allowing the Fe abundance to vary. As Table 6 shows,
the inferred Fe abundance is quite low, amounting to only a few tenths
of the solar value. Little or no further improvement was obtained
by allowing the abundances of other elements (Ne, Mg, Si, S) to
vary. We thus conclude that if the emission is thermal then
the iron abundance is well below solar.

All models give an absorbed  
flux  $\log f_{\rm X}=-12.52$~ergs~cm$^{-2}$~s$^{-1}$ 
(0.5 - 7 keV) and an absorption column density near
log N$_{\rm H}$ = 22.27 cm$^{-2}$. The 1T and 2T VAPEC models
(Table 6) give unabsorbed luminosities in the range
log L$_{\rm X}$ = 30.37 - 30.59 ergs~s$^{-1}$ (0.5 - 7 keV) 
at  d = 165 pc. The above values are  in very good
agreement with those derived using the independent non-parametric 
approach (Tables 1 and 4; Appendix B).

The temperature range deduced from  VAPEC models of Oph S1 
is similar to that 
seen in T Tauri stars. The X-ray luminosity of Oph S1 is somewhat
higher than the median value for TTS detected in core~A,
but still within the observed range for TTS. Thus, most of
the X-ray emission could be  produced by
the faint companion star Oph S1~B discovered by lunar occultation.

\subsection{Brown Dwarf Candidates}

{\it Chandra} detected 4 of 15 known brown dwarf candidates
(BDCs) in core~A, namely GY 5 (spectral type M7), GY 31 (M5.5),
GY 37 (M6) and GY 59 (M6) \citep{Wil99}. As a class, these
objects are very weak X-ray emitters. The median luminosity
of the four BDC detections is $\log L_{\rm X}=28.3$~ergs~s$^{-1}$,
which is a factor of $\sim$50 below the median $L_{\rm X}$ for
TTS detected in core~A. Assuming a detection threshold of
7 counts, the 11 undetected BDCs have upper limits
$\log L_{\rm X}\leq 27.80$ ergs~s$^{-1}$.
Using the bolometric luminosities of \citet{Wil99},
the four detected BDCs have $\log L_{\rm X}/L_{\rm bol}$ in the range
$-4.71$ to $-3.46$, where the latter value is for the
flare source GY 31. Brown dwarfs in other star-forming regions
exhibit similar X-ray activity levels \citep{Ima02,Pre02,Fei02,Mok02}.

In addition to their faint X-ray
emission, the BDCs in core~A are also undetected in
our {\it VLA} observations.
Typical radio upper limits for X-ray detected BDCs
are $S_{\rm 6cm}$ $\leq$ 0.16 mJy 
or $L_{\rm 6cm}$ $\leq$ 15.7 ergs~s$^{-1}$~Hz$^{-1}$.

Apart from their faint $L_{\rm X}$, the X-ray properties of 
detected BDCs appear to be similar to TTS. The median
value of mean photon energy for the detected BDCs is
$\langle$E$\rangle$ = 2.3 keV, only slightly lower
than the median of 2.5 keV for TTS (\S7.2.1).
Strong variability in the form of a flare was detected
in one BDC (GY 31), as discussed below. Thus, GY 31 has
a larger X-ray to K-band flux ratio $f_X$/$f_K$ than most TTS,
but the other 3 BDCs which did not flare have $f_X$/$f_K$
similar to TTS (Table 4).

Figure~\ref{f16} shows the flaring light curve of GY~31.
The flux declined
by a factor of $\sim5$ during the first 10 ksec of the
observation.  The hard-band 
light curve (not shown) closely tracks the broad-band behavior, but
there is very little emission in the soft-band below
2 keV due to strong absorption.

We attempted to fit the time-averaged spectrum of 
GY~31 (Figure~\ref{f17}) using optically
thin plasma models as well as bremsstrahlung and power-law
models. These different models give nearly identical 
reduced $\chi^2$ values and we conclude that there are
insufficient counts ($\approx$380 counts) to reliably distinguish between 
optically thin plasma emission and other alternatives.
All models give a high absorption column density
log N$_{H}$ (cm$^{-2}$) = 22.76 $\pm$ 0.08. Equating
this to a visual extinction \citep{Vuo03} yields A$_{V}$ $=$
37 mag. Thus, this object is viewed 
through very heavy absorption, as also determined
from infrared observations \citep{Wil99}.

If the emission is assumed to originate in an 
optically thin plasma as is usually the case for TTS, 
then 1T VAPEC fits using
a subsolar Fe abundance in the range Fe = 0.3 -
0.5 solar give $kT = 2.3$ [1.6 - 2.9] keV. 
Due to the high absorption, 2T models with 
a cooler component give no significant 
improvement over 1T models. This time-averaged temperature
structure is similar to that found in TTS in 
our sample.

\section{Summary}

We report new results based on simultaneous X-ray and 
radio continuum observations of $\rho$ Oph cloud core~A using
{\it Chandra} and the {\it VLA}. The most important findings
of this study are the following:

\begin{enumerate}

\item {\it Chandra} detected 87 X-ray sources, of which 60 have
      known infrared or radio counterparts.
      At least 16 of the 26 unidentified {\it Chandra} sources 
      are suspected to be extragalactic. More than one-half of 
      the X-ray detections showed variability, and the X-ray
      sources are typically hard and heavily absorbed 
      ($\langle$E$\rangle\approx 3$ keV, 
      $A_V$ $\approx$ 2 - 56 mag).

\item {\it Chandra} detected 12 of 14 known CTTSs and 15 of 17 WTTSs in
      $\rho$ Oph core~A. Their X-ray luminosities and
      spectral characteristics are similar, and we thus find
      no compelling evidence that the presence or absence of circumstellar
      disks (as discerned from IR excesses) has any significant effect 
      on their X-ray emission. The effects (if any) of accretion 
      on the X-ray emission in this small TTS sample  are not yet known
      due to the lack of suitable (e.g. optical) accretion 
      diagnostics. 

\item None of the three class 0/I protostellar sources in core~A was 
      detected by {\it Chandra} with upper limits at 
      $\log L_X\ltsimeq28$ ergs~s$^{-1}$. A comparison
      with X-ray detections of class I sources in other star-forming
      regions suggests that such protostars are preferentially detected
      during periods of enhanced X-ray emission (flares). Our data 
      indicate that in the absence of flares, X-ray luminosities 
      can be one to two orders of magnitude below the typical
      value $\log L_{\rm X}\sim 30$ ergs~s$^{-1}$ reported for
      flaring class 0/I sources.

\item {\it Chandra} detected 4 of 15 known brown dwarf candidates in
      core~A, including the flaring source GY 31. Their X-ray emission 
      is faint with a median $\log L_{\rm X} = 28.3$ ergs s$^{-1}$
      that is a factor of $\sim$50 lower than T Tauri stars in core~A.
      The mean X-ray energies and variability of BDCs
      appear to be similar to TTS. None of the X-ray emitting
      BDCs was detected with the {\em VLA}.

\item The WTTS DoAr 21 was in the decay phase of a large X-ray flare
      during the {\em Chandra} observation, but no radio variablity
      was detected. Time-resolved X-ray spectroscopy reveals multiple
      secondary flares during the decay. These secondary flares are
      associated with temperature increases, suggesting that the plasma 
      is being reheated if the secondary events occur in the primary
      flaring structure. Such reheating could contribute
      to the long decay times seen in some large YSO X-ray flares.

\item The X-ray light curve of the binary Oph~S1 shows hot, moderate-amplitude,
      short-duration flares typically seen on TTSs.  
      Much of the X-ray emission may be produced around the K-type secondary
      located 20 mas from the B4 primary. It remains to be determined
      whether the variable radio emission originates in a magnetosphere
      around the B4 star or is instead due to the nearby late-type
      companion.

\item Multifrequency {\it VLA} observations detected 31 radio sources in
      $\rho$ Oph core~A, of which 16 are confirmations of previous radio
      detections. New radio detections are reported
      for the emission-line star Elias 24 and the optically invisible IR
      source WLY 2-11, as well as the first detection of circular polarization
      in the radio-bright WTTS DoAr~21. There is no significant correlation
      between the X-ray and radio luminosities in the small sample of T Tauri
      stars detected simultaneously with {\it Chandra} and the {\em VLA}.

\end{enumerate}

\acknowledgments 

This work was supported by NASA grants NAS8-39073 and NAG5-3224 and SAO grant
GO0-1088C.  
This research made use of the SIMBAD astronomical database 
operated by the CDS at Strasbourg, France and the ASURV statistical
software package maintained at Penn State.
We thank the referee, Eric Feigelson, for many helpful suggestions.

\appendix
\section{Electronic Database of Objects in $\rho$ Oph Core A}
The $\rho$~Oph A region has been studied extensively in most
regions of the spectrum and a large amount of observational
data  exist. In order to make comparisons between
the new {\it Chandra} and {\em VLA} data and existing data,
we used IDL to construct a database of the cataloged infrared
\citep[][2MASS\footnote{2MASS 2002, 2d Incremental Release, Point Source Catalog.}]{Bar97,Bon01,All02},
X-ray \citep{Cas95,Gro00} and radio sources \citep{Sti88,Leo91} in the
{\it Chandra} FOV.  We matched sources which fell within the
combined positional uncertainty published for each source.  We visually checked
every match using the {\it ds9} tool in CIAO
and against the alternative associations
determined by the author of each catalog.  Similarly, the measured photometry
of each source was also checked for consistency.  We omitted two sources from 
the 2MASS catalog as they had no counterparts and appeared to be reflection 
nebulae in the 2MASS images.

We used the findings of \citet{Sim95}, who performed a search for binaries 
by lunar occultation.  Their study included the portion of $\rho$~Oph 
discussed in this paper.  Three binary systems, Oph~S1, DoAr~24E, and Elias~30,
are in the {\it Chandra} FOV.  In our database, all information previously
associated with Oph~S1 is assigned to Oph~S1~A. We
assign the same V-magnitude absorption ($A_{\rm V}$) and column density
($N_{\rm H}$) to both Oph~S1~A and Oph~S1~B.
We associate all {\it Chandra} X-ray emission with Oph~S1~B (see \S7.3).
Because of the close 20 mas separation of Oph~S1~A and Oph~S1~B, some caution
is needed when attributing emission to either of the two components.
Any further determinations of the origin of cataloged emission for Oph~S1 are
at the discretion of the user of the data base.  We also
associate the {\it Chandra} X-ray emission in the region of DoAr~24E with the
binary companion, DoAr~24E~B.  We use the IR photometry determined by
\citet{All02} for this source.  All other cataloged information is assigned to
DoAr~24E~A in this database.  The user should be aware that the IR photometry
measurements for DoAr~24E~A are likely combined in the IR catalogs.  Elias~30
is separated in \citet{Bar97}, and we use their catalog for the IR photometry.
We use the K-band magnitudes derived by \citet{Sim95} for each of these six
sources.

A total of 318 previously cataloged objects exist in the {\it Chandra}
ACIS-I FOV.  In our observations we add 27 new objects detected with
{\it Chandra}, one of which (J162607.3-242530)
was simultaneously detected with the {\it VLA} (see \S4.2).  
We matched these 345 sources with known 
brown dwarf (BD) candidates \citep{Neu99,Wil99}, spectral types 
\citep{Mar98,Wil99,Wil01}, and classification designations based on studies 
of spectral energy distributions (SEDs), using classifications in 
order of preference from 
\citet{Wil01,LR99,AM94,Gre94,Wil89}.

The \citet[][2MASS]{Sti88,Leo91,Cas95,Str95,Bar97,Mar98,Wil99,Neu99,LR99,Gro00,Bon01,Wil01,All02}
catalogs were used to construct a database of 345 sources in the {\it Chandra} FOV.
The IR photometry was preferenced
using \citet{Bar97} measurements first, then \citet{All02}, then 2MASS,
and \citet{Bon01} last.  For sources with cataloged JHK photometry we used,
\begin{equation}
   f_{\lambda} = F_{\lambda_{\rm Vega}}\Delta\lambda\times 10^{-0.4\lambda}
\end{equation}
to determine the absorbed flux in each band.  From these values, 
and the associated absorbed X-ray flux  $f_{\rm X}$ (Table 1), we also calculated the ratio of 
$f_{\rm X}/f_{\lambda}$ for each band.

The $A_{\rm V}$ for each source was determined
by \citet{Wil01}, \citet{Wil99}, and \citet{Gro00},
in order of preference.
$A_V$ determinations based only on JHK photometry \citep[e.g.,][]{Bon01}
were not used except in estimating upper limits (\S7.1) because these 
methods {\em assume} a dereddened color to
extract the color excess and $A_V$. This can lead to substantial uncertainties
in $A_V$.  For example, \citet{Bon01} used $(J-H)_0=0.85$, close to the locus
of T Tauri stars of \citet{Mey97}.  
We note though that for WTTS and CTTS, $(J-H)_0$ is in the range 0.1--1.1. 
A $0\fm5$ error in $(J-H)_0$ corresponds to a substantial  error of $4\fm5$ in $A_{\rm V}$.
To determine the $N_{\rm H}$ to each source with a
measured $A_{\rm V}$ we used,
\begin{equation}
   N_{\rm H}= A_{\rm V}\cdot 1.57\times10^{21} {\rm~cm}^{-2},
\end{equation}
\citep{Vuo03}.
Brown dwarf candidates were gleaned from
\citet{Neu99} and \citet{Wil99}.
The database of information on the 345 YSOs in the {\it Chandra} FOV is
publicly available\footnote{ftp://astro.wcupa.edu/pub/mgagne/roph}
in a single IDL save file.  Other supplemental materials are available
at this URL.

\section{A Non-parametric Method for Estimating 
           Spectral Parameters}

In the traditional method for estimating spectral parameters from X-ray CCD data,
the list of photon energies is binned to create an X-ray spectrum. The 
spectrum is fit by forward folding a model spectrum through an appropriate
set of spectral and effective-area response files. Model parameters are
usually derived by minimizing some statistic like $\chi^2$ or the Cash
C statistic. One can compute the statistic over a
grid of free parameters to estimate confidence limits for
each parameter. This method of estimating spectral parameters is
computationally efficient and often quite robust when each spectral bin
contains a signicantly large number of counts. However, as the number 
of counts per bin becomes small, the fit results become uncertain because
(a) the statistic is not a reliable 
goodness-of-fit statistic for small N, and (b) the binning process is 
arbitrary: there is no established method for choosing bin centers,
bin widths and minimum counts per bin. 

Some methods have been devised to address the first problem 
related to small N statistics. For example, \citet{Geh86}
derived a  method for estimating upper and lower limits
for small N. Also, \citet{Nou89} discussed various minimization
techniques using $\chi^2$ and the C statistic. \citet{Mig99} suggest 
a more robust statistic for small $N$: $\chi^2_\gamma$. However,
these methods do not address the second problem of how to bin
the event data.

The above problems - particularly the second - 
can be avoided by operating on unbinned photon
event lists, rather than binning the data according to photon
energy. Using this strategy, one can construct empirical
(or cumulative) distribution functions (EDF) and then take
advantage of a number of non-parametric goodness-of-fit statistics 
that use EDFs (e.g. \citet{Ste74,BF96}). We briefly 
describe below a procedure that we have developed for estimating
spectral parameters ($N_{\rm H}$, $\log T$) using unbinned photon data. 
These parameters are then used to estimate the unabsorbed-to-absorbed
flux ratio from which we derive the unabsorbed X-ray luminosity 
$L_{\rm X}$ from $f_{\rm X}$ and distance.
We used this procedure with unbinned photon event lists for sources
in $\rho$ Oph A to derive the spectral parameters in Table 4 
and Figs. 2--8, along with their 1$\sigma$ uncertainties.
As noted in Table~4, these errors do not include systematic
effects such as time variability, calibration uncertainties
in the response files used in the MARX simulations, and uncertainties
associated with the choice of spectral models.
For this paper, we used the Wisconsin absorption model \citep{Mor83}
and a single-temperature VAPEC emission model \citep{Smi01} with solar abundances
except for Fe set at 0.5 times solar.
If, for example, the X-ray emission were time-variable
or involved multiple temperature components or had different
elemental abundances, then $\log T$ would lie outside the stated range.
Distance uncertainties would lead to much larger errors
in $L_{\rm X}$.

XSPEC was used to generate photon spectra over a grid of 
column densities ($20.30 \leq \log N_{\rm H} \leq 23.30$ in increments of 0.04) and
plasma temperatures ($6.30 \leq \log T \leq 8.35$ in increments of 0.05).
These spectra were then input to the MARX ray-tracing software to 
generate simulated event files for 75 sources distributed over the
ACIS-I detector (similar to the situation encountered in
$\rho$ Oph). We collected  16,000--40,000 counts for each 
($\log N_{\rm H}$, $\log T$) pair, from which we extracted event lists
for thousands of simulated sources of varying brightness in
the range 10-1000 counts. We then compared the EDFs of these
low-count sources with bright samples containing 8192 counts. 
These comparisons were made using both the Cramer-von Mises (CvM) statistic
and the two-sample Kolmogorov-Smirnov (KS) statistic.
The choice of 8192 counts for the bright samples was somewhat
arbitrary, the main consideration being that the bright source
samples contain enough counts to totally dominate the noise in
simulated spectra.

In order to test the robustness of the method, we tested the method's
ability to recover the input model from a fake event list.
For example, Figure~\ref{f18} shows confidence contours for a fake source with
input $\log N_{\rm H} = 22.18$ and $\log T = 7.40$. The cross indicates
the location of the input values.
In one realization (upper panel of Fig.~\ref{f18}), 25 source counts
(and the appropriate number of background counts) are extracted from a larger list
corresponding to a bright source and its empirical distribution function (EDF)
is constructed.  The 25-count EDF is then compared to the bright-source EDF for
all ($\log N_{\rm H}$, $\log T$) pairs by computing the CvM statistic.
The dot indicates the $\log N_{\rm H}$, $\log T$ with the lowest CvM
statistic ${\rm CvM}_{\rm min}$ for this realization. The contours indicate 
${\rm CvM} - {\rm CvM}_{\rm min} = \Delta {\rm CvM} = 0.174$, 0.284 and 0.347, corresponding 
respectively to 68\%, 90\% and 95\% confidence contours for the CvM statistic \citep{Ste74}.
The bottom panel of Fig.~\ref{f18} is a similar plot for a 60-count source.
Visual inspection of these plots and many like them show that the
best-fit parameters (dot) are generally close to
the input values (cross) and lie within the 68\% contours approximately
68\% of the time.

To quantify the uncertainty in the best-fit parameters using this method,
200 simulations were realized for each ($\log T$, $\log N_{\rm H}$) pair.
The standard deviation of the 200 output best-fit $\log T$ and $\log N_{\rm H}$
values was computed for each input ($\log T$, $\log N_{\rm H}$).
This was done for 10, 25, 60, 160, and 400-count sources with background addition.
This was also done for the two-sample KS statistic.
Some results of this analysis for the CvM statistic are shown in Table~7.
The first three columns show the input model:  $\log N_{\rm H}$, $\log T$ and FCF,
the unabsorbed-to-absorbed flux ratio.
The next column is source counts. The next four columns show the output results:
mean and standard deviation of the best-fit $\log N_{\rm H}$ and $\log T$ values.
These standard deviations are shown in Fig.~\ref{f18} as dashed boxes around the best-fit
value for that realization.
The last two columns in Table~7 show the mean and 68\%-confidence upper and lower bounds
on the FCF used to calculate $L_{\rm X}$.

Table~7 allows us to make the following observations: for moderate column
densities ($\log N_{\rm H} = 21.50$ and 22.18) and high temperatures
($\log T = 7.4$ and 7.8), $\log N_{\rm H}$, $\log T$ and FCF can be estimated
with some precision with as few as 25 counts. If $\log N_{\rm H}$ is very
high or $\log T$ is low, then the parameters cannot be reliably estimated,
even with 400 counts. That is, photons from a luminous, cool, highly absorbed
source will be difficult to distinguish from a less luminous, hotter, less absorbed
source. Hence the large uncertainities in FCF for large $N_{\rm H}$ and low $T$.

The ability to estimate spectral parameters can be further improved by
constraining $\log N_{\rm H}$ or $\log T$. A similar analysis in which $\log N_{\rm H}$
is known $\pm 10\%$ is shown in Table~8. It is clear that if $A_V$ data
can be used to constrain $N_{\rm H}$, then $\log T$, FCF and hence $L_{\rm X}$
can be reliably estimated in most cases with as few as 25 counts.
Finally, we note that although both the CvM and KS statistic worked well,
the $\Delta{\rm CvM}$ confidence intervals of \citep{Ste74} were more accurate than
those for the KS statistic for low-count sources. With more than 100 counts, both
statistics found very similar errors.
Thus, in estimating $\log T$ and $L_{\rm X}$ in Table~4 and in Figs. 2-8, we used
the CvM statistic and published $A_V$ data when available.

To estimate spectral parameters from real data, photon-event 
(evt2.fits) and ancillary-response (.arf) files are extracted for the source
and the source EDF is computed. Based on the exposure time and the size of the
source region, the number of 0.5-7.0~keV background counts in the
source region is estimated. The source EDF is then compared to the full set
of simulated event lists.
Each ($\log N_{\rm H}$, $\log T$) pair on our VAPEC grid has a simulated
event file. Based on the number of simulated events,
the number of real source events, and the estimated
number of background counts, the appropriate number of background events
are randomly drawn from a large background event list and 
{\em added} to the simulated events. The simulated background-added
EDF is computed and compared to the real source EDF. This way,
a CvM (or KS) statistic is computed at every point on the ($\log N_{\rm H}$, $\log T$)
grid. A full grid search requires approximately 7 seconds on a dual-processor 2.8-GHz 
Xeon machine running IDL version 6.0 under Linux kernel 2.4.

When $\log N_{\rm H}$ is estimated from $A_V$, then the grid search is restricted to
values of $\log N_{\rm H}$ within 10\% of the estimated value. This 
reduces the error in $\log T$, FCF, and $L_{\rm X}$.
The best-fit values in Table~4 represent the grid point with the lowest CvM statistic.
When generating light curves like those in Figs. 2--8, the procedure is
performed at each time step.

As a reality check, we compared the fit results and confidence contours with
those from conventional XSPEC fitting. For sources with
N $>$ 100 counts, the results were nearly identical, provided our VAPEC
grid had sufficient resolution. We found that the confidence contours
for simulated sources with 25 -- 100 counts were somewhat smaller. 
This demonstrates that 
the spectral information content of the unbinned EDF is comparable to
or greater than that of binned spectra. 

In summary, the advantage of fitting binned background-subtracted
spectra in XSPEC or Sherpa is that the minimization process is
more efficient than deriving confidence contours from XSPEC
and MARX simulations. The non-parametric
method we describe provides spectral information for faint
sources that may be more statistically valid because
the photon energies are not binned.

\begin{deluxetable}{lrrrrrrcl}
\tablewidth{0pt}
\tabletypesize{\scriptsize}
\tablecolumns{9}
\small
\tablecaption{{\it Chandra} X-ray sources in $\rho$~Oph Core A.}
\tablehead{
\colhead{GDS}               & \colhead{R.A.}              &
\colhead{Decl.}             & \colhead{Counts}            &
\colhead{$\sigma$}          & \colhead{KS}                &
\colhead{$\langle{E}\rangle$}                             &
\colhead{$\log{f_{\rm X}}$} & \colhead{IR/Radio\tablenotemark{b}} \\
\colhead{Designation}       & \colhead{(J2000)}           &
\colhead{(J2000)}           & \colhead{(cts)}             &
\colhead{(cts)}             & \colhead{}                  &
\colhead{(keV)}             &
\colhead{(erg~cm$^{-2}$s$^{-1}$)}                         &
\colhead{Designation} \\
\colhead{ (1)} &
\colhead{ (2)} &
\colhead{ (3)} &
\colhead{ (4)} &
\colhead{ (5)} &
\colhead{ (6)} &
\colhead{ (7)} &
\colhead{ (8)} &
\colhead{ (9)}
}
\startdata
J162555.3-242536& 16 25 55.40& $-24$ 25 36.5&    16.7&     6.8&  0.65&  3.56& -14.24&              \nodata\\
J162603.0-242336\tablenotemark{a}& 16 26 03.03& $-24$ 23 36.1& 80776.5&   285.4& 16.81&  2.41& -10.98&              DoAr 21\\
J162605.7-242755& 16 26 05.72& $-24$ 27 56.0&    19.9&     6.3&  0.50&  3.21& -14.48&              \nodata\\
J162607.1-242724& 16 26 07.13& $-24$ 27 24.3&   731.6&    28.3&  2.51&  3.12& -12.87&  BKLT J162607-242725\\
J162607.3-242530\tablenotemark{a}& 16 26 07.36& $-24$ 25 30.9&    24.9&     6.8&  0.59&  3.97& -14.14&              \nodata\\
J162607.6-242741& 16 26 07.63& $-24$ 27 41.4&   880.7&    30.9&  3.89&  2.88& -12.85&  BKLT J162607-242742\\
J162610.3-242054& 16 26 10.35& $-24$ 20 55.0&   184.7&    14.8&  3.66&  3.38& -13.23&               GSS 26\\
J162612.5-242848& 16 26 12.51& $-24$ 28 48.5&    76.5&    10.3&  0.67&  3.60& -13.73&              \nodata\\
J162613.3-242823& 16 26 13.32& $-24$ 28 23.3&    15.8&     5.7&  0.55&  3.44& -14.42&              \nodata\\
J162615.7-242749& 16 26 15.75& $-24$ 27 49.4&    58.1&     9.0&  0.59&  3.70& -13.88&              \nodata\\
J162615.8-241922& 16 26 15.83& $-24$ 19 22.2&   165.7&    14.0&  1.34&  2.42& -13.71&              SKS 1-7\\
J162616.3-241841& 16 26 16.35& $-24$ 18 41.9&    12.4&     4.7&  0.84&  4.25& -14.47&              \nodata\\
J162616.8-242223& 16 26 16.87& $-24$ 22 23.0&  2078.2&    46.7&  7.74&  2.47& -12.59&              Oph S28\\
J162617.0-242021& 16 26 17.08& $-24$ 20 21.5&  3772.8&    62.5&  4.23&  1.61& -12.58&              DoAr 24\\
J162617.1-241240& 16 26 17.15& $-24$ 12 40.4&   126.2&    13.5&  1.46&  2.02& -13.89&  BKLT J162617-241241\\
J162617.4-241809& 16 26 17.49& $-24$ 18 09.0&    39.7&     7.6&  0.93&  3.73& -14.08&              \nodata\\
J162617.9-241518& 16 26 17.97& $-24$ 15 18.6&    41.1&     8.1&  0.82&  3.21& -14.12&              \nodata\\
J162618.8-242819& 16 26 18.88& $-24$ 28 19.8&   381.7&    20.7&  1.31&  2.54& -13.29&               VSSG 1\\
J162621.5-242600& 16 26 21.54& $-24$ 26 01.0&    10.8&     4.7&  1.24&  2.15& -15.09&                 GY 5\\
J162622.2-242447& 16 26 22.23& $-24$ 24 48.0&     8.8&     4.3&  0.53&  2.59& -15.01&             SKS 3-12\\
J162622.4-242252\tablenotemark{a}& 16 26 22.41& $-24$ 22 52.8&  3493.1&    60.1&  2.50&  3.04& -12.25&             GSS 30\#2\\
J162623.3-242059& 16 26 23.39& $-24$ 20 59.6&   992.7&    32.6&  3.04&  2.21& -12.99&           DoAr 24E A\\
J162623.6-242439& 16 26 23.61& $-24$ 24 39.6&   445.6&    22.2&  3.52&  4.44& -12.90&                GY 21\\
J162624.0-242448& 16 26 24.07& $-24$ 24 48.1&  2625.2&    52.3& 24.17&  3.45& -12.24&               Oph S2\\
J162624.0-241613\tablenotemark{a}& 16 26 24.09& $-24$ 16 13.3&  1923.8&    45.0&  3.04&  2.60& -12.56&             Elias 24\\
J162624.1-241518& 16 26 24.17& $-24$ 15 18.7&    20.4&     7.1&  0.88&  3.21& -14.25&              \nodata\\
J162625.2-242323& 16 26 25.24& $-24$ 23 24.0&   378.0&    20.5&  4.71&  3.52& -13.14&                GY 31\\
J162625.2-242444& 16 26 25.30& $-24$ 24 44.7&    21.0&     5.8&  1.03&  2.12& -14.65&                GY 29\\
J162625.7-241427& 16 26 25.72& $-24$ 14 27.3&   105.7&    12.1&  3.16&  1.82& -14.10&  BKLT J162625-241430\\
J162627.4-242418& 16 26 27.48& $-24$ 24 18.1&    24.9&     6.2&  2.77&  5.18& -14.02&              \nodata\\
J162627.8-242642& 16 26 27.83& $-24$ 26 42.6&    10.5&     4.4&  0.83&  2.00& -15.00&                GY 37\\
J162627.8-242359\tablenotemark{a}& 16 26 27.88& $-24$ 23 59.1&    12.6&     4.7&  1.91&  5.24& -14.29&               LFAM 6\tablenotemark{c}\\
J162628.5-241540& 16 26 28.51& $-24$ 15 40.4&    17.2&     5.4&  0.53&  2.62& -14.62&  BKLT J162628-241543\\
J162629.7-241905& 16 26 29.70& $-24$ 19 05.3&   316.8&    18.9&  1.02&  2.87& -13.33&             SKS 1-19\\
J162630.5-242256\tablenotemark{a}& 16 26 30.55& $-24$ 22 56.8&   449.4&    22.3&  3.72&  3.07& -13.13&                GY 51\\
J162630.6-242023& 16 26 30.62& $-24$ 20 23.3&    22.3&     5.9&  0.68&  4.30& -14.23&              \nodata\\
J162630.8-242108& 16 26 30.84& $-24$ 21 08.9&    10.0&     4.6&  0.94&  3.65& -14.58&              \nodata\\
J162630.9-243106& 16 26 30.91& $-24$ 31 06.6&    15.6&     5.9&  0.82&  2.87& -14.46&                GY 54\\
J162631.3-241833\tablenotemark{a}& 16 26 31.32& $-24$ 18 33.3&    11.5&     4.8&  0.42&  3.59& -14.48&              LFAM 10\\
J162631.3-242530& 16 26 31.37& $-24$ 25 30.3&    36.0&     7.1&  1.26&  2.45& -14.40&                GY 59\\
J162631.5-242318& 16 26 31.52& $-24$ 23 18.4&    21.6&     5.8&  0.54&  3.97& -14.33&              \nodata\\
J162633.2-241951& 16 26 33.25& $-24$ 19 51.3&    13.4&     4.8&  0.77&  3.62& -14.48&              \nodata\\
J162634.1-242328& 16 26 34.19& $-24$ 23 28.2&  2583.1&    51.9&  1.61&  2.52& -12.51&             Oph S1 B\\
J162634.5-242228& 16 26 34.51& $-24$ 22 28.7&    39.4&     7.4&  0.96&  3.90& -14.08&              \nodata\\
J162636.8-241933& 16 26 36.81& $-24$ 19 33.1&    14.0&     5.0&  0.62&  3.62& -14.60&              \nodata\\
J162636.8-241551& 16 26 36.83& $-24$ 15 51.2&   106.2&    11.9&  0.76&  3.49& -13.64&  BKLT J162636-241554\\
J162636.8-241900& 16 26 36.86& $-24$ 19 00.1&     8.2&     4.1&  0.42&  3.00& -14.96&  BKLT J162636-241902\\
J162637.1-241601& 16 26 37.12& $-24$ 16 01.1&    64.0&     9.6&  0.76&  1.80& -14.26&  BKLT J162637-241602\\
J162637.9-241942& 16 26 37.90& $-24$ 19 42.7&     9.4&     4.3&  0.71&  3.45& -14.83&              \nodata\\
J162640.4-242714& 16 26 40.48& $-24$ 27 14.4&   162.4&    13.8&  5.02&  3.69& -13.48&                GY 91\\
J162641.8-241603& 16 26 41.88& $-24$ 16 03.3&    18.3&     6.0&  0.94&  3.54& -14.35&              \nodata\\
J162642.3-242625\tablenotemark{a}& 16 26 42.39& $-24$ 26 25.9&    45.4&     7.9&  0.75&  3.18& -14.13&               GY 101\\
J162642.5-242631& 16 26 42.58& $-24$ 26 31.7&   117.2&    11.9&  0.75&  3.49& -13.66&               GY 103\\
J162642.8-242029& 16 26 42.89& $-24$ 20 29.9&  1225.2&    36.1&  2.72&  2.54& -12.80&               GY 110\\
J162642.9-242259& 16 26 42.92& $-24$ 22 59.0&    11.4&     4.6&  1.28&  2.71& -14.89&               GY 109\\
J162643.4-242434& 16 26 43.41& $-24$ 24 34.9&    19.6&     5.6&  1.05&  3.88& -14.42&              \nodata\\
J162643.8-241632\tablenotemark{a}& 16 26 43.80& $-24$ 16 32.9&   565.7&    25.0&  1.08&  2.35& -13.18&              VSSG 11\\
J162643.9-242717& 16 26 43.92& $-24$ 27 17.8&    11.9&     4.7&  0.65&  4.52& -14.35&              \nodata\\
J162645.0-242307& 16 26 45.06& $-24$ 23 07.7&    71.4&     9.5&  0.80&  3.42& -13.90&               GY 116\\
J162646.8-241907& 16 26 46.81& $-24$ 19 07.9&    66.4&     9.3&  1.58&  2.30& -14.12&  BKLT J162646-241910\\
J162648.3-242834& 16 26 48.33& $-24$ 28 34.2&    31.5&     7.1&  1.47&  3.22& -14.16&  BKLT J162648-242836\\
J162648.6-242839& 16 26 48.61& $-24$ 28 39.8&    24.5&     6.5&  0.47&  3.51& -14.29&               GY 128\\
J162649.1-242143& 16 26 49.18& $-24$ 21 43.8&    18.9&     5.6&  0.59&  3.71& -14.33&              \nodata\\
J162649.2-242002& 16 26 49.27& $-24$ 20 02.7&  1348.3&    37.8&  4.42&  2.53& -12.78&               GY 135\\
J162651.9-243039& 16 26 51.93& $-24$ 30 39.6&    12.6&     5.7&  0.63&  3.23& -14.57&               GY 144\\
J162652.4-243133& 16 26 52.44& $-24$ 31 33.0&    30.4&     8.0&  0.63&  4.12& -13.92&              \nodata\\
J162653.4-243236& 16 26 53.49& $-24$ 32 36.7&    31.9&     7.5&  1.85&  3.10& -14.19&               GY 146\\
J162654.0-242323& 16 26 54.03& $-24$ 23 23.8&    11.6&     4.8&  0.64&  3.54& -14.55&              \nodata\\
J162654.0-242840& 16 26 54.10& $-24$ 28 40.4&    11.0&     5.0&  0.50&  3.93& -14.58&              \nodata\\
J162654.4-241554& 16 26 54.47& $-24$ 15 54.5&    16.9&     6.1&  0.89&  2.63& -14.56&              \nodata\\
J162654.4-242620& 16 26 54.48& $-24$ 26 20.6&   428.2&    21.8&  1.79&  2.64& -13.25&               GY 153\\
J162655.0-242229& 16 26 55.00& $-24$ 22 29.6&   285.2&    18.0&  1.69&  2.79& -13.40&               GY 156\\
J162658.3-242130& 16 26 58.38& $-24$ 21 30.2&    41.7&     7.8&  1.11&  3.12& -14.11&               GY 171\\
J162658.6-241834& 16 26 58.65& $-24$ 18 34.5&    18.7&     5.7&  0.90&  2.46& -14.65&  BKLT J162658-241836\\
J162658.9-242215& 16 26 58.90& $-24$ 22 15.2&    11.5&     4.8&  0.57&  3.34& -14.70&              \nodata\\
J162700.0-242640\tablenotemark{a}& 16 27 00.07& $-24$ 26 40.4&   175.5&    14.5&  0.94&  4.16& -13.32&              LFAM 21\\
J162700.5-241624& 16 27 00.53& $-24$ 16 24.9&    58.7&     9.0&  2.41&  1.94& -14.30&  BKLT J162700-241627\\
J162701.2-242913& 16 27 01.28& $-24$ 29 13.6&    12.9&     7.3&  0.72&  3.51& -14.48&              \nodata\\
J162703.0-242614& 16 27 03.00& $-24$ 26 14.4&    44.1&     8.1&  2.36&  3.81& -13.99&               GY 188\\
J162704.1-242828& 16 27 04.11& $-24$ 28 28.2&    39.8&     8.1&  1.80&  3.28& -14.11&               GY 192\\
J162704.6-242715& 16 27 04.62& $-24$ 27 15.4&  4614.4&    69.0& 23.33&  3.37& -12.04&               GY 195\\
J162705.2-242007& 16 27 05.28& $-24$ 20 07.4&   201.6&    15.5&  1.81&  2.61& -13.53&               GY 199\\
J162706.0-242618& 16 27 06.01& $-24$ 26 18.5&   228.6&    16.4&  5.82&  3.17& -13.37&               GY 203\\
J162710.3-241912& 16 27 10.32& $-24$ 19 12.6&   911.4&    31.5&  1.20&  1.59& -13.22&           Elias 30 A\\
J162710.3-241918& 16 27 10.35& $-24$ 19 18.5&   110.7&    11.7&  0.98&  1.67& -14.11&           Elias 30 B\\
J162713.7-241816& 16 27 13.73& $-24$ 18 16.7&    42.3&     8.4&  1.52&  2.42& -14.31&              VSSG 24\\
J162722.9-241800& 16 27 22.93& $-24$ 18 00.1&   486.2&    23.6&  2.08&  2.65& -13.05&              VSSG 22\\
\enddata
\tablecomments{Col. (1) J2000 IAU designation, cols. (2) - (3) {\it Chandra} X-ray position 
with astrometric correction applied, col. (4) net counts (background subtracted),
col. (5) 1$\sigma$ error in net counts \citep{Geh86}, col. (6)  Kolmogorov-Smirnov (KS) 
variability statistic (variable sources have KS $>$ 1), col. (7) mean photon energy
from event file, col. (8) calculated absorbed flux in 0.5 - 7 keV band (\S~3.1), col. (9) 
probable counterpart.}
\tablenotetext{a}{Simultaneously detected with the VLA. See Table 3.}
\tablenotetext{b}{DoAr=\citet{DA59}; GSS=\citet{Gra73}; VSSG=\citet{Vrb75}; 
               Elias=\citet{Eli78}; LFAM=\citet{Leo91}; GY=\citet{GY92}; 
               SKS=\citet{Str95}; BKLT=\citet{Bar97}.}
\tablenotetext{c}{The identification as LFAM~6 is uncertain due to an offset
               of $\approx2.4\arcsec$ between the X-ray coordinates and
               precessed LFAM coordinates.}
\end{deluxetable}

\clearpage
\begin{deluxetable}{cccclc} 
\tablewidth{0pc} 
\tablecaption{Summary of VLA Observations of $\rho$~Oph Core~A} 
\tabletypesize{\scriptsize}
\tablehead{ 
\colhead{Frequency}                                  & 
\colhead{Time Range\tablenotemark{a}}                & 
\colhead{Pointings}                                  & 
\colhead{On-source Time\tablenotemark{b}}            & 
\colhead{Flux cal./$\phi$-cal\tablenotemark{c}}      & 
\colhead{Beam FWHM\tablenotemark{d}}                 \\ 
\colhead{(GHz)}       & 
\colhead{(IAT)}       & 
\colhead{ }           & 
\colhead{(min)}       & 
\colhead{(J2000)}     & 
\colhead{($\arcsec$)} } 
\startdata 
 1.42  & 07:41 - 09:16 & 1                  & 20 & 0137$+$331 / 1626$-$298 & 33.0 $\times$ 15.9 \\ 
 4.86  & 04:25 - 11:38 & 9\tablenotemark{e} & 245& 0137$+$331 / 1626$-$298 & 9.5 $\times$ 5.0 \\ 
 8.46  & 05:13 - 08:58 & 4                  & 52 & 0137$+$331 / 1626$-$298 & 5.7 $\times$ 2.6 \\ 
\enddata 
\tablenotetext{a}{Observed on 16 May 2000.} 
\tablenotetext{b}{Total on-source time was composed of 2 separate scans at 
	1.42~GHz, 19 scans at 4.86~GHz, and 5 scans at 8.46~GHz.} 
\tablenotetext{c}{$0137+331=3C48$. At low elevations when $1626-298$ was not 
	visible, phase calibration was achieved using $1554-270$ (4.86~and 
	8.46~GHz) and $1700-261$ (4.86~GHz).} 
\tablenotetext{d}{Typical synthesized beam in cleaned maps; VLA in 
	C-configuration with  26 operational antennas (no subarrays). } 
\tablenotetext{e}{One pointing centered near Oph~S1~A and remaining pointings offset 
	by $\sim$one-half primary beam width ($4.5\arcmin$) in a grid around Oph~S1~A.} 
\end{deluxetable} 


\clearpage
\begin{deluxetable}{lllrrclc} 
\tablecolumns{8} 
\small 
\tablewidth{0pc} 
\tablecaption{VLA 6~cm Radio Sources in $\rho$~Oph Core~A \tablenotemark{a}} 
\tabletypesize{\footnotesize}
\tablehead{ 
\colhead{GDS}                       &
\colhead{R.A.}                      & \colhead{Decl.}                     & 
\colhead{S$_{6~{\rm cm}}^{(peak)}$} & \colhead{S$_{6~{\rm cm}}^{(total)}$}& 
\colhead{RMS Noise}                 & \colhead{Ident.\tablenotemark{b}}    &
\colhead{Class\tablenotemark{c}}    \\ 
\colhead{}                          & 
\colhead{(J2000)}                   & \colhead{}                          & 
\colhead{($\mu$Jy)}                 & \colhead{($\mu$Jy)}                 & 
\colhead{($\mu$Jy/beam)}            & \colhead{    }                      & 
\colhead{}                          } 
\small 
\startdata 
J162556.1-243014*\tablenotemark{d}& 16 25 56.10 & $-$24 30 14.2 & 436    & 545   & 36 & WLY~2-11&\nodata\\ 
J162603.0-242336*                 & 16 26 03.06 & $-$24 23 36.6 & 11995  & 12404 & 43 & SFAM~7  & III \\ 
J162607.3-242530                  & 16 26 07.45 & $-$24 25 31.8 & 410    & 657   & 43 & \nodata &\nodata\\ 
J162610.5-242853                  & 16 26 10.51 & $-$24 28 53.8 & 275    & 211   & 37 & \nodata &\nodata\\ 
J162611.0-242908                  & 16 26 11.04 & $-$24 29 08.2 & 452    & 430   & 34 & \nodata &\nodata\\ 
J162615.5-243428*\tablenotemark{d}& 16 26 15.53 & $-$24 34 28.6 & 4090   & 7475  & 72 & \nodata &\nodata\\ 
J162621.7-242251                  & 16 26 21.73 & $-$24 22 51.4 & 265    & 355   & 43 & LFAM~1  & I   \\ 
J162622.4-242252*                 & 16 26 22.39 & $-$24 22 53.4 & 2665(v)  & 2620  & 57 & LFAM~2  & III \\ 
J162624.0-241613*                 & 16 26 24.01 & $-$24 16 12.3 & 190    & 200   & 32 & WSB~31  & II  \\ 
J162625.6-242429                  & 16 26 25.68 & $-$24 24 29.8 & 280    & 282   & 43 & LFAM~4  &\nodata\\ 
J162626.3-242431                  & 16 26 26.38 & $-$24 24 31.0 & 330    & 325   & 50 & LFAM~5  & 0   \\ 
J162627.8-242359                  & 16 26 27.79 & $-$24 24 02.2 & 201    & 296   & 35 & \nodata &\nodata\\ 
J162629.5-242317                  & 16 26 29.54 & $-$24 23 17.8 & 240    & 274   & 39 & LFAM~7  &\nodata\\ 
J162630.5-242256*                 & 16 26 30.51 & $-$24 22 57.4 & 1061   & 1054  & 39 & LFAM~9  & IIF \\ 
J162631.3-241833                  & 16 26 31.28 & $-$24 18 32.8 & 280    & 369   & 38 & LFAM~10 &\nodata\\ 
J162632.7-242245*                 & 16 26 32.71 & $-$24 22 45.4 & 200    & 250   & 39 & \nodata &\nodata\\ 
J162633.4-241216\tablenotemark{d} & 16 26 33.47 & $-$24 12 16.0 & 1840   & 2020  & 86 & SFAM~12 &\nodata\\ 
J162633.5-242448                  & 16 26 33.58 & $-$24 24 49.0 & 219    & 207   & 39 & \nodata &\nodata\\ 
J162634.2-242328*                 & 16 26 34.20 & $-$24 23 28.6 & 7100(v)  & 11455 & 36 & LFAM~11 &\nodata\\ 
J162635.3-242405                  & 16 26 35.34 & $-$24 24 05.8 & 758    & 720   & 36 & LFAM~13 &\nodata\\ 
J162639.0-243052                  & 16 26 39.02 & $-$24 30 52.6 & 300    & 288   & 36 & \nodata &\nodata\\ 
J162642.3-242625                  & 16 26 42.42 & $-$24 26 26.8 & 3364   & 3560  & 37 & LFAM~15 & III \\ 
J162643.8-241632*                 & 16 26 43.77 & $-$24 16 33.2 & 2648   & 2700  & 50 & SFAM~15 & III \\ 
J162646.3-242001                  & 16 26 46.32 & $-$24 20 02.0 & 250    & 270   & 38 & LFAM~17 &\nodata\\ 
J162646.4-241935                  & 16 26 46.40 & $-$24 19 35.6 & 223    & 230   & 38 & \nodata &\nodata\\ 
J162652.8-241953                  & 16 26 52.89 & $-$24 19 53.8 & 288    & 430   & 44 & \nodata &\nodata\\ 
J162658.7-242651                  & 16 26 58.76 & $-$24 26 52.0 & 1720   & 2940  & 51 & LFAM~19 &\nodata\\ 
J162700.0-242640*                 & 16 26 59.99 & $-$24 26 40.2 & 55987  & 57010 & 51 & LFAM~21 &\nodata\\ 
J162700.0-242209                  & 16 27 00.09 & $-$24 22 09.4 & 209    & 340   & 39 & \nodata &\nodata\\ 
J162702.1-241928                  & 16 27 02.11 & $-$24 19 28.6 & 585    & 549   & 39 & \nodata &\nodata\\ 
J162702.3-242724                  & 16 27 02.37 & $-$24 27 24.4 & 280    & 360   & 42 & \nodata &\nodata\\ 
\enddata 
\tablenotetext{a}{An asterisk following the source number indicates that 
	additional notes follow below.  VLA data (C-configuration) are from cleaned maps and 
	fluxes are corrected for primary beam attenuation.  Total flux is the 
	average of the values measured using IMFIT (Gaussian fit) and TVSTAT 
	(pixel summation). A (v) following the total flux indicates that
        the flux is variable. RMS noise ($\pm1\sigma$) is measured in immediate 
	vicinity of source. Typical 6~cm radio position uncertainties are 
	$\pm0.05$~s in RA and $\pm0\arcsec.7$ in Dec.} 
\tablenotetext{b}{LFAM = \citealt{Leo91}; SFAM = \citealt{Sti88}; WLY = \citet{Wil89};
WSB = \citet{Wil87}} 
\tablenotetext{c}{SED classification designations in order of preference are:
	\citet{Wil01,Wil99,LR99,Str95,Gre94,AM94}.} 
\tablenotetext{d}{Outside the {\it Chandra} FOV.}
\end{deluxetable}
\clearpage

\footnotesize{
\noindent Comments on individual sources: \\ 
\noindent J162556.1-243014 - New radio detection. Probable counterpart is infrared source no. 11
in Table 2 of \citet{Wil89}.  \\
\noindent J162603.0-242335 - Emission at 6~cm is left circularly polarized (LCP) with a detection
	significance ${\rm S/N}=16$ in Stokes V maps and a fractional circular 
	polarization computed from peak Stokes V and I fluxes of 
	$\pi_{c}=-0.059$ (LCP). \\ 
\noindent J162615.5-243428 - Possible double radio source, but source structure is uncertain due
to beam elongation at low source elevation. \\ 
\noindent J162622.4-242252 - Peak flux increased by a factor of $\sim$5 
	during the time interval 0629 - 0931 UT. \\
\noindent J162624.0-241613 - New radio detection. Probable counterpart is  H$\alpha$ emission-line
star  no. 31 in Table 1 of \citet{Wil87}, which is cross-listed as  Elias-24 \citep{Eli78}.  \\ 
\noindent J162630.5-242256 - The peak flux density at 3.6 cm is comparable to that at 6 cm,
                   suggesting a flat spectral energy distribution and possible nonthermal 
                   radio emission. \\
\noindent J162632.7-242245 - Marginal detection (${\rm S/N}=5.1$). \\ 
\noindent J162634.1-242328 - Stokes I emission is extended.  Emission at 6~cm
        is left circularly 
	polarized (LCP) with a detection significance ${\rm S/N}=9$ in Stokes V maps 
	and a fractional circular polarization $\pi_{c}=-0.046$ (LCP).
        Stokes V maps with Oph~S1 located near phase center give peak 
fluxes V(6~cm)$=-330$~($\pm$38)~$\mu$Jy and V(3.5~cm)$=-390$~($\pm$30)~$\mu$Jy, 
where the minus sign denotes  LCP.  The peak flux 
ratios V/I give fractional circular polarizations $\pi_{c}$(6~cm)$=-0.046$ (LCP) and 
$\pi_{c}$(3.5~cm)$=-0.062$ (LCP).  The Stokes V map at 20~cm has an rms noise of 
65~$\mu$Jy/beam, giving a 3$\sigma$ upper limit V(20~cm)~$\leq195$~$\mu$Jy 
and $\pi_{c}$(20~cm) $\leq0.0035$.  The 6-cm Stokes V flux given above is 
consistent with the value V(6~cm)$=-280$~($\pm$50)~$\mu$Jy obtained with the 
VLA on 1987 Aug 6 \citep{And88}. \\
\noindent J162643.8-241632 - The infrared counterpart is offset by $3.\arcsec2$ from the radio 
	peak position making the identification uncertain. \\ 
\noindent J162700.0-242640 - Emission at 6~cm is right circularly polarized with a detection 
	significance ${\rm S/N}=34$ in Stokes V maps and a fractional circular 
	polarization computed from peak Stokes V and I fluxes of 
	$\pi_{c}=+0.034$.  Corresponds to \#20 (BZ6) in Table II of 
	\citet{Sti88}, who argue that it is extragalactic.} 

\clearpage
\begin{deluxetable}{lllrrrccrcrr}
\rotate
\tablewidth{0pt}
\tabletypesize{\scriptsize}
\tablecolumns{12}
\small
\tablecaption{X-ray sources in $\rho$~Oph Core A with JHK Photometry}
\tablehead{
\colhead{GDS}               & \colhead{IR/Optical}        &
\colhead{LFAM/}             & \colhead{J}                 &
\colhead{H}                 & \colhead{K}                 &
\colhead{$\log{{f_{\rm X}/f_{\rm K}}}$}                   &
\colhead{SED\tablenotemark{a}} & \colhead{$A_{\rm V}$}    &
\colhead{$N_{\rm H}$}                                     &
\colhead{$\log T$}                                     &
\colhead{$\log{L_{\rm X}}$} \\
\colhead{Designation}       & \colhead{Designation}       &
\colhead{SFAM}              & \colhead{}                  &
\colhead{}                  & \colhead{}                  &
\colhead{}                  &
\colhead{}                  & \colhead{}                  &
\colhead{($10^{22}{\rm cm}^{-2})$} &
\colhead{(K)}   &
\colhead{(ergs~s$^{-1}$)}   }
\startdata
J162603.0-242336 &              DoAr 21 &   SFAM 7 &     8.01 &     6.75 &     6.16 & -1.75 &      III  &     6.6 &  $ 1.15^{+ 0.02}_{-0.02}$  & $ 7.55^{+ 0.02}_{-0.02}$ & $31.82^{+ 0.02}_{-0.02}$ \\
J162607.1-242724 &  BKLT J162607-242725 &  \nodata &    15.49 &    12.50 &    10.48 & -1.91 &  \nodata  &    19.5 & ($ 3.47                 $) & $ 7.45^{+ 0.02}_{-0.02}$ & $30.18^{+ 0.02}_{-0.02}$ \\
J162607.6-242741 &  BKLT J162607-242742 &  \nodata &    14.73 &    11.86 &    10.27 & -1.97 &  \nodata  &    20.6 & ($ 2.88                 $) & $ 7.40^{+ 0.02}_{-0.05}$ & $30.20^{+ 0.07}_{-0.02}$ \\
J162610.3-242054 &               GSS 26 &  \nodata &    14.97 &    11.63 &     9.38 & -2.71 &      IIF  &    22.6 & ($ 4.17                 $) & $ 7.55^{+ 0.30}_{-0.05}$ & $29.79^{+ 0.03}_{-0.17}$ \\
J162615.8-241922 &              SKS 1-7 &  \nodata &    13.95 &    11.44 &     9.98 & -2.95 &      III  &    17.4 & ($ 2.40                 $) & $ 7.25^{+ 0.05}_{-0.10}$ & $29.45^{+ 0.22}_{-0.06}$ \\
J162616.8-242223 &              Oph S28 &  LFAMp  1 &    11.11 &     9.23 &     8.19 & -2.54 &       II  &    11.0 & ($ 1.66                 $) & $ 7.40^{+ 0.02}_{-0.02}$ & $30.35^{+ 0.02}_{-0.02}$ \\
J162617.0-242021 &              DoAr 24 &  \nodata &     9.72 &     8.67 &     8.09 & -2.57 &       II  &     1.8 &  $ 0.24^{+ 0.02}_{-0.02}$  & $ 7.35^{+ 0.02}_{-0.05}$ & $30.08^{+ 0.02}_{-0.02}$ \\
J162617.1-241240 &  BKLT J162617-241241 &  \nodata &    13.11 &    11.47 &    10.65 & -2.86 &  \nodata  & \nodata &  $ 1.26^{+ 0.56}_{-0.71}$  & $ 7.15^{+ 0.30}_{-0.15}$ & $29.22^{+ 0.37}_{-0.37}$ \\
J162618.8-242819 &               VSSG 1 &  \nodata &    13.49 &    10.76 &     8.68 & -3.05 &       II  &    17.1 &  $ 2.40^{+ 0.02}_{-0.02}$  & $ 7.25^{+ 0.05}_{-0.02}$ & $29.87^{+ 0.02}_{-0.06}$ \\
J162621.5-242600 &                 GY 5 &  \nodata &    12.70 &    11.57 &    10.91 & -3.96 &      III\ &     4.5 & ($ 0.79                 $) & $ 7.50^{+ 0.85}_{-0.25}$ & $27.67^{+ 0.12}_{-0.13}$ \\
J162622.2-242447 &             SKS 3-12 &  \nodata & $>$17.00 &    15.22 &    12.97 & -3.05 &  \nodata  & \nodata &  $ 8.71^{+11.24}_{-8.05}$  & $ 6.85^{+ 1.50}_{-0.50}$ & $29.73^{+ 3.94}_{-2.10}$ \\
J162622.4-242252 &            GSS 30\#2 &   LFAM 2 &    14.22 &    11.47 &     9.60 & -1.64 &      III  &    16.0\tablenotemark{c} &  $ 2.88^{+ 0.02}_{-0.02}$  & $ 7.50^{+ 0.05}_{-0.02}$ & $30.72^{+ 0.02}_{-0.03}$ \\
J162623.3-242059 &           DoAr 24E A &  \nodata &     8.98 &     7.53 &     6.90 & -3.46 &       II  &     6.1 & ($ 0.87                 $) & $ 7.50^{+ 0.02}_{-0.05}$ & $29.79^{+ 0.03}_{-0.02}$ \\
J162623.6-242439 &                GY 21 &   LFAM 3 &    14.08 &    11.63 &     9.94 & -2.15 &      IIF  &    14.0 &  $ 2.63^{+ 0.02}_{-0.02}$  & $ 8.35^{+ 0.02}_{-0.02}$ & $29.87^{+ 0.02}_{-0.02}$ \\
J162624.0-242448 &               Oph S2 &  \nodata &    10.83 &     8.68 &     7.20 & -2.59 &       II  &    11.8 &  $ 3.47^{+ 0.70}_{-0.02}$  & $ 7.70^{+ 0.02}_{-0.15}$ & $30.67^{+ 0.11}_{-0.02}$ \\
J162624.0-241613 &             Elias 24 &  \nodata &    10.18 &     8.23 &     6.77 & -3.08 &       II  &    10.0 & ($ 1.38                 $) & $ 7.60^{+ 0.02}_{-0.05}$ & $30.25^{+ 0.03}_{-0.02}$ \\
J162625.2-242323 &                GY 31 &  \nodata & $>$17.00 & $>$15.50 &    13.24 & -1.08 &      III\ &    56.0 & ($ 7.94                 $) & $ 7.30^{+ 0.02}_{-0.10}$ & $30.31^{+ 0.28}_{-0.02}$ \\
J162625.2-242444 &                GY 29 &  \nodata &    15.92 &    12.97 &    10.86 & -3.53 &      III  &    19.1 & ($ 3.47                 $) & $ 7.00^{+ 0.20}_{-0.15}$ & $29.12^{+ 0.34}_{-0.52}$ \\
J162625.7-241427 &  BKLT J162625-241430 &  \nodata &    13.04 &    12.02 &    11.48 & -2.74 &  \nodata  & \nodata &  $ 0.32^{+ 0.55}_{-0.29}$  & $ 7.40^{+ 0.50}_{-0.25}$ & $28.57^{+ 0.32}_{-0.15}$ \\
J162627.8-242642 &                GY 37 &  \nodata &    14.25 &    12.94 &    11.99 & -3.43 &      III\ &     6.3 & ($ 1.05                 $) & $ 7.15^{+ 0.40}_{-0.35}$ & $28.05^{+ 0.52}_{-0.29}$ \\
J162628.5-241540 &  BKLT J162628-241543 &  \nodata &    15.42 &    12.43 &    10.77 & -3.54 &  \nodata  &    21.7 & ($ 2.88                 $) & $ 7.30^{+ 0.25}_{-0.25}$ & $28.52^{+ 0.55}_{-0.19}$ \\
J162629.7-241905 &             SKS 1-19 &   LFAM 8 &    16.61 &    13.24 &    11.19 & -2.08 &      III  &    23.0 & ($ 3.16                 $) & $ 7.35^{+ 0.05}_{-0.05}$ & $29.79^{+ 0.08}_{-0.05}$ \\
J162630.5-242256 &                GY 51 &   LFAM 9 &    16.69 &    13.46 &    10.72 & -2.07 &      IIF  &    22.0 & ($ 3.47                 $) & $ 7.40^{+ 0.10}_{-0.05}$ & $29.96^{+ 0.08}_{-0.11}$ \\
J162630.9-243106 &                GY 54 &  \nodata &    14.79 &    12.36 &    10.80 & -3.37 &  \nodata  &    16.6 & ($ 2.19                 $) & $ 7.40^{+ 0.35}_{-0.35}$ & $28.53^{+ 0.59}_{-0.16}$ \\
J162631.3-242530 &                GY 59 &  \nodata &    14.75 &    12.89 &    11.68 & -2.95 &      III\ &    11.0 & ($ 1.51                 $) & $ 7.45^{+ 0.20}_{-0.25}$ & $28.49^{+ 0.28}_{-0.08}$ \\
J162634.1-242328 &             Oph S1 B &  \nodata &    -9.00 &    -9.00 &     8.30 & -2.42 &           &    10.0 & ($ 1.82                 $) & $ 7.40^{+ 0.02}_{-0.02}$ & $30.44^{+ 0.02}_{-0.02}$ \\
J162636.8-241551 &  BKLT J162636-241554 &  \nodata &    12.13 &    10.44 &     9.38 & -3.12 &  \nodata  &     7.6 &  $ 1.38^{+ 0.02}_{-0.02}$  & $ 8.35^{+ 0.02}_{-0.02}$ & $29.06^{+ 0.02}_{-0.02}$ \\
J162636.8-241900 &  BKLT J162636-241902 &  \nodata &    16.07 &    13.93 &    12.31 & -3.27 &  \nodata  &    14.0 & ($ 2.63                 $) & $ 7.60^{+ 0.75}_{-0.30}$ & $27.95^{+ 0.21}_{-0.17}$ \\
J162637.1-241601 &  BKLT J162637-241602 &  \nodata &    12.49 &    11.37 &    10.73 & -3.19 &  \nodata  & \nodata &  $ 0.18^{+ 1.08}_{-0.16}$  & $ 7.35^{+ 0.40}_{-0.60}$ & $28.38^{+ 1.07}_{-0.11}$ \\
J162640.4-242714 &                GY 91 &  \nodata & $>$17.00 &    16.40 &    12.51 & -1.71 &  \nodata  & \nodata &  $ 7.24^{+ 5.34}_{-3.44}$  & $ 7.45^{+ 0.70}_{-0.30}$ & $29.75^{+ 0.77}_{-0.40}$ \\
J162642.3-242625 &               GY 101 &  LFAM 15 & $>$17.00 &    16.34 &    12.58 & -2.33 &      III  &    50.0 & ($ 6.61                 $) & $ 7.20^{+ 0.05}_{-0.20}$ & $29.45^{+ 0.70}_{-0.10}$ \\
J162642.5-242631 &               GY 103 &  \nodata & $>$17.00 &    15.82 &    12.52 & -1.88 &  \nodata  &    48.5 & ($ 6.61                 $) & $ 7.35^{+ 0.10}_{-0.15}$ & $29.67^{+ 0.37}_{-0.11}$ \\
J162642.8-242029 &               GY 110 &  \nodata &    10.71 &     8.93 &     8.00 & -2.83 &       II  &     8.5 & ($ 1.51                 $) & $ 7.50^{+ 0.02}_{-0.02}$ & $30.06^{+ 0.02}_{-0.02}$ \\
J162642.9-242259 &               GY 109 &  LFAM 16 &    15.37 &    12.91 &    11.39 & -3.56 &  \nodata  &    14.6 & ($ 2.63                 $) & $ 7.40^{+ 0.75}_{-0.15}$ & $28.14^{+ 0.16}_{-0.28}$ \\
J162643.8-241632 &              VSSG 11 &  SFAM 15 &    13.12 &    10.85 &     9.58 & -2.57 &      III  &    14.0 & ($ 2.00                 $) & $ 7.25^{+ 0.05}_{-0.05}$ & $29.94^{+ 0.09}_{-0.06}$ \\
J162645.0-242307 &               GY 116 &  LFAMp 3 &    13.14 &    10.50 &     8.88 & -3.58 &       II  &    17.0 & ($ 3.16                 $) & $ 7.75^{+ 0.60}_{-0.15}$ & $28.98^{+ 0.06}_{-0.12}$ \\
J162646.8-241907 &  BKLT J162646-241910 &  \nodata &    14.05 &    11.67 &    10.34 & -3.21 &  \nodata  & \nodata &  $ 1.51^{+ 1.12}_{-0.91}$  & $ 7.35^{+ 0.55}_{-0.25}$ & $28.83^{+ 0.49}_{-0.29}$ \\
J162648.3-242834 &  BKLT J162648-242836 &  \nodata & $>$17.00 &    15.73 &    12.48 & -2.40 &  \nodata  & \nodata &  $ 1.66^{+ 3.84}_{-0.61}$  & $ 8.20^{+ 0.15}_{-1.00}$ & $28.56^{+ 0.78}_{-0.05}$ \\
J162648.6-242839 &               GY 128 &  \nodata & $>$17.00 &    14.05 &    10.99 & -3.12 &  \nodata  &    38.6 & ($ 7.24                 $) & $ 7.30^{+ 0.35}_{-0.10}$ & $29.13^{+ 0.20}_{-0.40}$ \\
J162649.2-242002 &               GY 135 &  LFAM 18 &    12.15 &     9.82 &     8.62 & -2.56 &      III  &    16.0 &  $ 2.19^{+ 0.02}_{-0.02}$  & $ 7.35^{+ 0.02}_{-0.05}$ & $30.25^{+ 0.04}_{-0.02}$ \\
J162651.9-243039 &               GY 144 &  \nodata & $>$17.00 &    15.75 &    13.46 & -2.42 &  \nodata  &    26.8 & ($ 3.80                 $) & $ 7.55^{+ 0.80}_{-0.40}$ & $28.43^{+ 0.59}_{-0.19}$ \\
J162653.4-243236 &               GY 146 &  \nodata & $>$17.00 & $>$15.50 &    13.18 & -2.15 &  \nodata  &    43.5 & ($ 6.03                 $) & $ 7.20^{+ 0.15}_{-0.20}$ & $29.35^{+ 0.68}_{-0.25}$ \\
J162654.4-242620 &               GY 153 &  \nodata &    14.72 &    11.70 &     9.88 & -2.53 &      III  &    19.7 & ($ 2.63                 $) & $ 7.30^{+ 0.05}_{-0.05}$ & $29.87^{+ 0.12}_{-0.05}$ \\
J162655.0-242229 &               GY 156 &  \nodata &    15.04 &    11.97 &    10.19 & -2.55 &      III  &    22.5 & ($ 3.47                 $) & $ 7.30^{+ 0.10}_{-0.10}$ & $29.79^{+ 0.21}_{-0.13}$ \\
J162658.3-242130 &               GY 171 &  \nodata &    16.12 &    13.38 &    11.74 & -2.65 &  \nodata  &    19.9 & ($ 3.47                 $) & $ 7.45^{+ 0.45}_{-0.10}$ & $28.94^{+ 0.09}_{-0.24}$ \\
J162658.6-241834 &  BKLT J162658-241836 &  \nodata &    15.47 &    13.06 &    11.54 & -3.26 &  \nodata  &    14.2 & ($ 2.63                 $) & $ 7.20^{+ 0.35}_{-0.10}$ & $28.60^{+ 0.18}_{-0.37}$ \\
J162700.5-241624 &  BKLT J162700-241627 &  \nodata &    13.46 &    11.86 &    10.95 & -3.15 &  \nodata  & \nodata &  $ 0.50^{+ 1.01}_{-0.43}$  & $ 7.40^{+ 0.75}_{-0.45}$ & $28.44^{+ 0.74}_{-0.20}$ \\
J162703.0-242614 &               GY 188 &  \nodata & $>$17.00 & $>$15.50 &    12.52 & -2.21 &  \nodata  &    20.5 & ($ 3.80                 $) & $ 8.35^{+ 0.02}_{-0.35}$ & $28.83^{+ 0.03}_{-0.02}$ \\
J162704.1-242828 &               GY 192 &  \nodata &    16.22 &    13.07 &    10.81 & -3.01 &  \nodata  &    20.9 & ($ 3.80                 $) & $ 7.45^{+ 0.55}_{-0.10}$ & $28.97^{+ 0.09}_{-0.26}$ \\
J162704.6-242715 &               GY 195 &  \nodata &    16.67 &    13.55 &    11.30 & -0.75 &  \nodata  &    20.6 &  $ 3.80^{+ 0.02}_{-0.33}$  & $ 7.60^{+ 0.05}_{-0.02}$ & $30.94^{+ 0.02}_{-0.04}$ \\
J162705.2-242007 &               GY 199 &  \nodata &    12.48 &    10.41 &     9.27 & -3.05 &  \nodata  & \nodata &  $ 2.00^{+ 0.40}_{-0.85}$  & $ 7.40^{+ 0.35}_{-0.10}$ & $29.44^{+ 0.13}_{-0.22}$ \\
J162706.0-242618 &               GY 203 &  \nodata & $>$17.00 &    13.57 &    11.05 & -2.18 &  \nodata  &    36.5 & ($ 5.01                 $) & $ 7.30^{+ 0.05}_{-0.10}$ & $29.93^{+ 0.24}_{-0.06}$ \\
J162710.3-241912 &           Elias 30 A &  \nodata &     8.56 &     7.33 &     6.30 & -3.93 &      IID  &     3.5 &  $ 0.55^{+ 0.05}_{-0.05}$  & $ 7.20^{+ 0.05}_{-0.02}$ & $29.62^{+ 0.02}_{-0.04}$ \\
J162710.3-241918 &           Elias 30 B &  \nodata &    11.27 &    10.00 &     9.25 & -3.64 &  \nodata  &     3.5 & ($ 0.50                 $) & $ 7.30^{+ 0.10}_{-0.10}$ & $28.66^{+ 0.09}_{-0.05}$ \\
J162713.7-241816 &              VSSG 24 &  \nodata &    12.30 &    10.27 &     9.32 & -3.81 &  \nodata  &    10.7 & ($ 1.82                 $) & $ 7.30^{+ 0.25}_{-0.10}$ & $28.72^{+ 0.14}_{-0.19}$ \\
J162722.9-241800 &              VSSG 22 &  \nodata &    13.27 &    10.79 &     9.41 & -2.52 &      III  &    16.0 & ($ 2.88                 $) & $ 7.30^{+ 0.05}_{-0.02}$ & $30.09^{+ 0.02}_{-0.07}$ \\
\enddata
\tablecomments{The quoted errors are internal errors only based on the non-parametric method described in Appendix B.
Additional errors due to systematic effects such as distance and $A_{\rm V}$ uncertainties are not included.}
\tablecomments{All JHK photometry from \citet{Bar97} except Oph~S1~B, K=8.3 \citep{Sim95}.
$L_{\rm X}$ is the unabsorbed 0.5-7.0 keV luminosity assuming $d=165$~pc.
Parentheses indicate $N_{\rm H}$ estimated from $A_V$ assuming $1.57\times10^{21}{\rm cm}^{-2}$ 
per mag \citep{Vuo03}.}
\tablenotetext{a}{D = double peaked spectrum; F = flat spectrum.}
\tablenotetext{b}{Brown dwarf candidate.}
\tablenotetext{c}{Near-IR measurements give $A_V$ $\approx$ 27 mag \citep{Cas85}}.
\end{deluxetable}

\clearpage
\begin{deluxetable}{llll}
\tablewidth{0pt}
\tabletypesize{\scriptsize}
\tablecolumns{4}
\small
\tablecaption{X-ray Upper Limits for Class 0-III Objects in $\rho$~Oph}
\tablehead{
\colhead{Name}         &
\colhead{Class}        &
\colhead{log N$_{H}$}  &
\colhead{log L$_{X}$ (0.5 - 7 keV)}  \\
\colhead{     }          &
\colhead{     }          &
\colhead{(cm$^{-2}$)}    &
\colhead{(ergs s$^{-1}$)} 
}
\startdata
LFAM~5     & 0        & 22.48\tablenotemark{a} & $\leq$27.81 \\
GSS 30-1   & I        & 22.78\tablenotemark{b} & $\leq$28.32 \\
GSS 30-3   & I        & 22.48\tablenotemark{a} & $\leq$28.05 \\
GY~81      & II(flat) & 22.46                  & $\leq$28.00 \\
GY~84      & III\tablenotemark{c}      & 22.40 & $\leq$27.71 \\
GY~154     & II       & 22.59                  & $\leq$28.27 \\
SKS~1-11   & III\tablenotemark{a}      & 22.48 & $\leq$27.92 \\
\enddata
\tablenotetext{a}{Column density is an assumed value that is typical of sources in
$\rho$ Oph core-A.}
\tablenotetext{b}{Based on A$_V$ = 38 \citep{Cas85} and the conversion of
                  \citet{Vuo03}.}
\tablenotetext{c}{Classification uncertain.}
\end{deluxetable}

\clearpage
\begin{deluxetable}{llll}
\tablewidth{0pt}
\tabletypesize{\footnotesize}
\tablecaption{Spectral Fit Results  for Oph S1\tablenotemark{a}}
\tablehead{
\colhead{Parameter}        &
\colhead{PL }        &
\colhead{1T    }        &
\colhead{2T    }    }
\startdata
N$_{\rm H}$ (10$^{22}$ cm$^{-2}$) & 2.16 [1.99 - 2.36] & 1.55 [1.48 - 1.68] & 1.90 [1.50 - 2.45]  \\
kT$_1$ (keV)                      & ...                & 2.38 [2.17 - 2.60] & 0.46 [0.29 - 1.95]    \\
kT$_2$ (keV)                      & ...                & ...                & 2.32 [2.03 - 2.69]  \\
Photon index                      & 3.00 [2.84 - 3.19] & ...                & ...                \\
Fe                                & ...                & 0.10 [0.00 - 0.33] & 0.24 [0.00 - 0.57]     \\
EM$_{2}$/EM$_{1}$                 & ...                & ...                & 1.79                   \\
$\chi^2$/dof                      & 147.0/135          & 141.2/134          & 137.3/132              \\
$\chi^2_{red}$                    & 1.09               & 1.05               & 1.04                 \\
f$_{X}$ (10$^{-13}$ ergs cm$^{-2}$ s$^{-1}$)   & 3.00  & 2.99               & 3.01     \\          
log L$_{X}$ (ergs s$^{-1}$)       & 30.75              & 30.37              & 30.59     
\enddata
\tablenotetext{a}{PL = power-law model. 1T and 2T denote
VAPEC optically thin plasma models with one and two temperature
components. Fe abundance is 
relative to solar  and abundances for
all other elements were held fixed at their solar values (Anders \& Grevesse 1989). 
f$_{X}$ is the absorbed flux (0.5 - 7 keV) measured from model fit. 
Unabsorbed luminosity L$_{X}$ (0.5 - 7 keV) assumes d = 165 pc.
Brackets enclose 90\% confidence ranges.
Spectra are corrected for charge transfer inefficiency (CTI) using CIAO v. 2.3 and CALDB
v. 2.20 and rebinned to a minimum of 15 counts per bin.}
\end{deluxetable}

\clearpage
\begin{deluxetable}{lcrrcccccrrr}
\tablewidth{0pt}
\tabletypesize{\scriptsize}
\tablecolumns{11}
\tablecaption{MARX VAPEC Simulations: $N_{\rm H}$ unknown}
\tablehead{
\multicolumn{3}{c}{Input Model} &
\multicolumn{1}{c}{} &
\multicolumn{7}{c}{Output Parameters} \\
\cline{1-3} \cline{5-11} \\
\multicolumn{1}{c}{$\log N_{\rm H}$} &
\multicolumn{1}{c}{$\log T$} &
\multicolumn{1}{c}{$FCF$} &
\multicolumn{1}{c}{$N$} &
\multicolumn{1}{c}{$\log\overline{N_{\rm H}}$} &
\multicolumn{1}{c}{$\sigma_{\overline{N_{\rm H}}}$} &
\multicolumn{1}{c}{$\log\overline{T}$} &
\multicolumn{1}{c}{$\sigma_{\overline{T}}$} &
\multicolumn{1}{r}{$\overline{\rm FCF}$} &
\multicolumn{2}{c}{FCF 68\% c.l.} \\
\multicolumn{1}{c}{(cm$^{-2}$)} &
\multicolumn{1}{c}{(K)} &
\multicolumn{1}{c}{} &
\multicolumn{1}{c}{(cts)} &
\multicolumn{2}{c}{(cm$^{-2}$)} &
\multicolumn{2}{c}{(K)} &
\multicolumn{3}{c}{}  }
\startdata
21.50 & 7.0 &     2.04 &   10 & 21.26 &  0.65 &  6.92 &  0.33 &       1.68 &       1.06 &      53.28 \\
21.50 & 7.4 &     1.47 &   10 & 21.60 &  0.58 &  7.32 &  0.51 &       1.66 &       1.12 &      14.74 \\
21.50 & 7.8 &     1.28 &   10 & 21.62 &  0.60 &  7.65 &  0.56 &       1.37 &       1.10 &       6.36 \\
22.18 & 7.0 &     7.62 &   10 & 22.19 &  0.41 &  6.92 &  0.35 &      11.26 &       2.42 &     286.65 \\
22.18 & 7.4 &     2.53 &   10 & 22.27 &  0.36 &  7.36 &  0.46 &       3.64 &       1.52 &      24.16 \\
22.18 & 7.8 &     1.79 &   10 & 22.31 &  0.35 &  7.70 &  0.51 &       2.04 &       1.51 &      10.08 \\
22.90 & 7.0 &    51.41 &   10 & 22.92 &  0.26 &  6.99 &  0.35 &      87.53 &       6.52 &    4493.47 \\
22.90 & 7.4 &     6.33 &   10 & 22.97 &  0.25 &  7.45 &  0.46 &       7.22 &       2.57 &     207.73 \\
22.90 & 7.8 &     3.31 &   10 & 23.05 &  0.23 &  7.73 &  0.44 &       4.52 &       2.57 &      26.98 \\
21.50 & 7.0 &     2.04 &   25 & 21.29 &  0.50 &  6.97 &  0.21 &       1.69 &       1.13 &       4.11 \\
21.50 & 7.4 &     1.47 &   25 & 21.45 &  0.51 &  7.37 &  0.31 &       1.43 &       1.12 &       4.52 \\
21.50 & 7.8 &     1.28 &   25 & 21.56 &  0.39 &  7.79 &  0.41 &       1.33 &       1.14 &       1.96 \\
22.18 & 7.0 &     7.62 &   25 & 22.15 &  0.21 &  7.00 &  0.22 &       7.62 &       3.36 &      19.39 \\
22.18 & 7.4 &     2.53 &   25 & 22.17 &  0.22 &  7.46 &  0.33 &       2.47 &       1.60 &       5.89 \\
22.18 & 7.8 &     1.79 &   25 & 22.21 &  0.18 &  7.81 &  0.37 &       1.94 &       1.46 &       2.86 \\
22.90 & 7.0 &    51.41 &   25 & 22.91 &  0.17 &  6.98 &  0.20 &      58.52 &      14.20 &     541.86 \\
22.90 & 7.4 &     6.33 &   25 & 22.90 &  0.22 &  7.49 &  0.39 &       5.27 &       2.51 &      58.16 \\
22.90 & 7.8 &     3.31 &   25 & 23.01 &  0.19 &  7.73 &  0.40 &       4.03 &       2.57 &      18.50 \\
21.50 & 7.0 &     2.04 &   60 & 21.33 &  0.42 &  6.98 &  0.14 &       1.86 &       1.18 &       3.71 \\
21.50 & 7.4 &     1.47 &   60 & 21.49 &  0.30 &  7.40 &  0.22 &       1.43 &       1.22 &       1.92 \\
21.50 & 7.8 &     1.28 &   60 & 21.48 &  0.30 &  7.81 &  0.28 &       1.30 &       1.15 &       1.53 \\
22.18 & 7.0 &     7.62 &   60 & 22.16 &  0.11 &  7.00 &  0.12 &       7.62 &       4.20 &      13.98 \\
22.18 & 7.4 &     2.53 &   60 & 22.16 &  0.13 &  7.46 &  0.19 &       2.32 &       1.87 &       3.64 \\
22.18 & 7.8 &     1.79 &   60 & 22.19 &  0.13 &  7.86 &  0.31 &       1.81 &       1.52 &       2.40 \\
22.90 & 7.0 &    51.41 &   60 & 22.88 &  0.12 &  7.01 &  0.14 &      45.27 &      13.04 &     199.65 \\
22.90 & 7.4 &     6.33 &   60 & 22.87 &  0.16 &  7.49 &  0.29 &       5.39 &       2.81 &      16.81 \\
22.90 & 7.8 &     3.31 &   60 & 22.96 &  0.13 &  7.82 &  0.32 &       3.63 &       2.71 &       6.33 \\
21.50 & 7.0 &     2.04 &  160 & 21.40 &  0.27 &  7.00 &  0.08 &       1.85 &       1.42 &       2.60 \\
21.50 & 7.4 &     1.47 &  160 & 21.48 &  0.16 &  7.42 &  0.11 &       1.43 &       1.30 &       1.63 \\
21.50 & 7.8 &     1.28 &  160 & 21.48 &  0.15 &  7.86 &  0.19 &       1.28 &       1.17 &       1.38 \\
22.18 & 7.0 &     7.62 &  160 & 22.17 &  0.06 &  7.00 &  0.08 &       7.62 &       4.86 &      12.56 \\
22.18 & 7.4 &     2.53 &  160 & 22.16 &  0.08 &  7.43 &  0.13 &       2.43 &       2.01 &       3.12 \\
22.18 & 7.8 &     1.79 &  160 & 22.16 &  0.08 &  7.90 &  0.23 &       1.71 &       1.55 &       2.04 \\
22.90 & 7.0 &    51.41 &  160 & 22.88 &  0.08 &  7.01 &  0.10 &      45.27 &      21.09 &     142.62 \\
22.90 & 7.4 &     6.33 &  160 & 22.87 &  0.10 &  7.46 &  0.15 &       5.27 &       3.77 &       8.79 \\
22.90 & 7.8 &     3.31 &  160 & 22.93 &  0.07 &  7.85 &  0.23 &       3.31 &       2.71 &       4.39 \\
21.50 & 7.0 &     2.04 &  400 & 21.44 &  0.15 &  7.01 &  0.04 &       1.94 &       1.69 &       2.29 \\
21.50 & 7.4 &     1.47 &  400 & 21.45 &  0.10 &  7.43 &  0.07 &       1.40 &       1.30 &       1.55 \\
21.50 & 7.8 &     1.28 &  400 & 21.47 &  0.11 &  7.85 &  0.16 &       1.28 &       1.19 &       1.33 \\
22.18 & 7.0 &     7.62 &  400 & 22.17 &  0.03 &  7.01 &  0.05 &       7.62 &       5.85 &       9.25 \\
22.18 & 7.4 &     2.53 &  400 & 22.16 &  0.06 &  7.42 &  0.08 &       2.43 &       2.09 &       2.87 \\
22.18 & 7.8 &     1.79 &  400 & 22.15 &  0.06 &  7.90 &  0.18 &       1.71 &       1.55 &       1.89 \\
22.90 & 7.0 &    51.41 &  400 & 22.88 &  0.06 &  7.02 &  0.08 &      51.41 &      21.09 &      70.98 \\
22.90 & 7.4 &     6.33 &  400 & 22.87 &  0.08 &  7.46 &  0.11 &       4.98 &       3.90 &       8.54 \\
22.90 & 7.8 &     3.31 &  400 & 22.92 &  0.06 &  7.86 &  0.21 &       3.31 &       2.73 &       4.03 \\
\enddata
\end{deluxetable}

\clearpage
\begin{deluxetable}{lcrcccccrrr}
\tablewidth{0pt}
\tabletypesize{\scriptsize}
\tablecolumns{11}
\tablecaption{MARX VAPEC Simulations: 10\% $N_{\rm H}$ error}
\tablehead{
\multicolumn{3}{c}{Input Model} &
\multicolumn{1}{c}{} &
\multicolumn{7}{c}{Output Parameters} \\
\cline{1-3} \cline{5-11} \\
\multicolumn{1}{c}{$\log N_{\rm H}$} &
\multicolumn{1}{c}{$\log T$} &
\multicolumn{1}{c}{$FCF$} &
\multicolumn{1}{c}{$N$} &
\multicolumn{1}{c}{$\log\overline{N_{\rm H}}$} &
\multicolumn{1}{c}{$\sigma_{\overline{N_{\rm H}}}$} &
\multicolumn{1}{c}{$\log\overline{T}$} &
\multicolumn{1}{c}{$\sigma_{\overline{T}}$} &
\multicolumn{1}{r}{$\overline{\rm FCF}$} &
\multicolumn{2}{c}{FCF 68\% c.l.} \\
\multicolumn{1}{c}{(cm$^{-2}$)} &
\multicolumn{1}{c}{(K)} &
\multicolumn{1}{c}{} &
\multicolumn{1}{c}{(cts)} &
\multicolumn{2}{c}{(cm$^{-2}$)} &
\multicolumn{2}{c}{(K)} &
\multicolumn{3}{c}{}  }
\startdata
21.50 & 7.0 &     2.04 &   10 & 21.49 &  0.04 &  7.01 &  0.17 &       2.04 &       1.68 &       2.34 \\
21.50 & 7.4 &     1.47 &   10 & 21.50 &  0.04 &  7.48 &  0.31 &       1.47 &       1.30 &       1.66 \\
21.50 & 7.8 &     1.28 &   10 & 21.51 &  0.04 &  7.80 &  0.39 &       1.31 &       1.22 &       1.47 \\
22.18 & 7.0 &     7.62 &   10 & 22.19 &  0.04 &  6.98 &  0.22 &       8.35 &       4.00 &      18.88 \\
22.18 & 7.4 &     2.53 &   10 & 22.19 &  0.04 &  7.47 &  0.30 &       2.63 &       1.89 &       4.00 \\
22.18 & 7.8 &     1.79 &   10 & 22.19 &  0.03 &  7.81 &  0.36 &       1.80 &       1.62 &       2.47 \\
22.90 & 7.0 &    51.41 &   10 & 22.90 &  0.04 &  7.00 &  0.15 &      58.52 &      17.19 &     169.44 \\
22.90 & 7.4 &     6.33 &   10 & 22.90 &  0.04 &  7.54 &  0.37 &       5.94 &       2.95 &      12.83 \\
22.90 & 7.8 &     3.31 &   10 & 22.92 &  0.03 &  7.88 &  0.39 &       3.31 &       2.79 &       5.94 \\
21.50 & 7.0 &     2.04 &   25 & 21.49 &  0.03 &  7.01 &  0.08 &       2.04 &       1.85 &       2.19 \\
21.50 & 7.4 &     1.47 &   25 & 21.49 &  0.04 &  7.45 &  0.21 &       1.43 &       1.33 &       1.59 \\
21.50 & 7.8 &     1.28 &   25 & 21.51 &  0.04 &  7.84 &  0.30 &       1.29 &       1.22 &       1.38 \\
22.18 & 7.0 &     7.62 &   25 & 22.18 &  0.04 &  6.96 &  0.14 &       8.35 &       4.86 &      15.21 \\
22.18 & 7.4 &     2.53 &   25 & 22.18 &  0.04 &  7.43 &  0.17 &       2.47 &       2.07 &       3.36 \\
22.18 & 7.8 &     1.79 &   25 & 22.18 &  0.04 &  7.90 &  0.29 &       1.75 &       1.61 &       2.07 \\
22.90 & 7.0 &    51.41 &   25 & 22.90 &  0.04 &  6.99 &  0.09 &      58.52 &      29.05 &     122.10 \\
22.90 & 7.4 &     6.33 &   25 & 22.90 &  0.04 &  7.43 &  0.18 &       6.75 &       3.96 &      10.16 \\
22.90 & 7.8 &     3.31 &   25 & 22.91 &  0.03 &  7.92 &  0.33 &       3.22 &       2.79 &       4.39 \\
21.50 & 7.0 &     2.04 &   60 & 21.50 &  0.03 &  7.00 &  0.05 &       2.04 &       1.94 &       2.19 \\
21.50 & 7.4 &     1.47 &   60 & 21.49 &  0.04 &  7.41 &  0.12 &       1.47 &       1.37 &       1.56 \\
21.50 & 7.8 &     1.28 &   60 & 21.50 &  0.04 &  7.83 &  0.24 &       1.30 &       1.22 &       1.36 \\
22.18 & 7.0 &     7.62 &   60 & 22.18 &  0.04 &  7.00 &  0.08 &       7.62 &       5.65 &      10.20 \\
22.18 & 7.4 &     2.53 &   60 & 22.18 &  0.04 &  7.40 &  0.10 &       2.63 &       2.21 &       3.12 \\
22.18 & 7.8 &     1.79 &   60 & 22.18 &  0.04 &  7.87 &  0.23 &       1.75 &       1.62 &       1.96 \\
22.90 & 7.0 &    51.41 &   60 & 22.89 &  0.03 &  7.00 &  0.08 &      51.41 &      32.06 &      90.28 \\
22.90 & 7.4 &     6.33 &   60 & 22.90 &  0.03 &  7.43 &  0.12 &       5.94 &       4.31 &       8.16 \\
22.90 & 7.8 &     3.31 &   60 & 22.90 &  0.03 &  7.96 &  0.28 &       3.16 &       2.73 &       3.65 \\
21.50 & 7.0 &     2.04 &  160 & 21.49 &  0.03 &  7.00 &  0.03 &       2.04 &       1.94 &       2.19 \\
21.50 & 7.4 &     1.47 &  160 & 21.49 &  0.03 &  7.41 &  0.06 &       1.43 &       1.40 &       1.50 \\
21.50 & 7.8 &     1.28 &  160 & 21.49 &  0.03 &  7.82 &  0.14 &       1.28 &       1.24 &       1.32 \\
22.18 & 7.0 &     7.62 &  160 & 22.18 &  0.03 &  6.99 &  0.06 &       7.62 &       5.85 &      10.20 \\
22.18 & 7.4 &     2.53 &  160 & 22.18 &  0.03 &  7.40 &  0.07 &       2.53 &       2.29 &       2.87 \\
22.18 & 7.8 &     1.79 &  160 & 22.17 &  0.03 &  7.88 &  0.18 &       1.73 &       1.62 &       1.95 \\
22.90 & 7.0 &    51.41 &  160 & 22.89 &  0.03 &  7.01 &  0.05 &      45.27 &      32.06 &      70.98 \\
22.90 & 7.4 &     6.33 &  160 & 22.89 &  0.03 &  7.43 &  0.08 &       5.94 &       4.98 &       7.41 \\
22.90 & 7.8 &     3.31 &  160 & 22.89 &  0.03 &  7.98 &  0.24 &       3.16 &       2.71 &       3.31 \\
21.50 & 7.0 &     2.04 &  400 & 21.50 &  0.03 &  6.99 &  0.02 &       2.04 &       1.94 &       2.16 \\
21.50 & 7.4 &     1.47 &  400 & 21.48 &  0.03 &  7.40 &  0.04 &       1.43 &       1.40 &       1.51 \\
21.50 & 7.8 &     1.28 &  400 & 21.49 &  0.03 &  7.82 &  0.10 &       1.28 &       1.24 &       1.30 \\
22.18 & 7.0 &     7.62 &  400 & 22.17 &  0.03 &  7.00 &  0.05 &       7.62 &       5.85 &       9.25 \\
22.18 & 7.4 &     2.53 &  400 & 22.17 &  0.03 &  7.41 &  0.05 &       2.43 &       2.29 &       2.87 \\
22.18 & 7.8 &     1.79 &  400 & 22.16 &  0.03 &  7.88 &  0.10 &       1.68 &       1.63 &       1.84 \\
22.90 & 7.0 &    51.41 &  400 & 22.89 &  0.03 &  7.01 &  0.03 &      51.41 &      32.06 &      51.41 \\
22.90 & 7.4 &     6.33 &  400 & 22.89 &  0.02 &  7.42 &  0.06 &       6.33 &       5.27 &       6.93 \\
22.90 & 7.8 &     3.31 &  400 & 22.89 &  0.03 &  7.98 &  0.23 &       3.16 &       2.62 &       3.31 \\
\enddata
\end{deluxetable}

\clearpage

\figcaption{Full-resolution grayscale ACIS-I image of $\rho$ Ophiuchus cloud A 
(L1688) with a logarithmic
stretch.  Pixel size is 0.492$''$. 
Sources listed in Table 1 are indicated with open circles and
ellipses.
The ACIS-I aimpoint is the magnetic B4 star Oph~S1 at
$16^{\mathrm h}26^{\mathrm m}34^{\mathrm s} -24\arcdeg23\farcm5$.  
The weak-lined T Tauri star DoAr~21 at far right 
($16^{\mathrm h}26^{\mathrm m}03^{\mathrm s} -24\arcdeg23\farcm6$)
is in the decay stage of a large flare throughout the $\sim96$~ks observation.
The trailed image near DoAr~21 is caused by the $40\mu$sec row-to-row charge transfer time.
\label{f1} }

\figcaption{Light curves of $f_{\rm X}$, $\log T$ and  $L_{\rm X}$
         (assuming d = 165 pc)
         for the optically bright WTTS DoAr~21=J162603.0-242336
         (K0, class III). The bin size is varied so as to include 
         approximately the same number of counts per bin (and similarly
         for Figures 3-8). The flare decay is punctuated by
         smaller heating events with coronal temperatures varying from
         $\approx 28-50$~MK.
\label{f2} }

\figcaption{Light curves of $f_{\rm X}$, $\log T$ and $L_{\rm X}$
         for the CTTS Oph~S2=J162624.0-242448  (class II).
         Oph S2 underwent a very rapid temperature spike ($T > 100$~MK)
         at UT 19h with the luminosity peak occuring nearly two hours later.
         The luminosity decayed with an e-folding time of approximately 6.5h
         and a steady decrease in temperature from $>100$~MK to 25~MK.
\label{f3} }


\figcaption{Light curves of $f_{\rm X}$, $\log T$ and $L_{\rm X}$
         for the B4+K-type binary Oph~S1=J162634.1-242328.
         The presence of hot coronal plasma ($T\approx19$~MK) and flare-like
         variability is typical of Class II or Class III TTSs.  It is likely
         that most of Oph~S1's X-ray emission is produced by the low-mass
         K-type companion discovered via lunar occultation by \citet{Sim95},
         not the magnetic B star of \citet{And88}.
\label{f4} }


\figcaption{Light curves of $f_{\rm X}$, $\log T$ and $L_{\rm X}$
         for GY~195=J162704.6-242715. This unclassified star underwent
         the largest flare-amplitude variation seen in our observation: a twenty-fold increase
         in X-ray luminosity over 1h followed by a 2h decay.
         The temperature spikes (which exceed 50~MK) after 13h UT do not suggest rotational 
         modulation of the inital large flare; rather they suggest two to three
         subsequent reheating events.
\label{f5} }


\figcaption{Light curves of $f_{\rm X}$, $\log T$ and $L_{\rm X}$
         for the CTTS Oph~S28=J162616.8-242223 (K6, class II).
         The initial flare has rise and decay times of approximately
         2h and 6h, respectively.
\label{f6} }


\figcaption{Light curves of $f_{\rm X}$, $\log T$ and $L_{\rm X}$
         for the CTTS DoAr~24=J162617.0-242021 (class II).
         This CTTS light curve shows moderate-intensity, relatively
         short-duration flares throughout.
\label{f7} }


\figcaption{Light curves of $f_{\rm X}$, $\log T$ and  $L_{\rm X}$
         for the WTTS GSS~30 IRS-2 = J162622.4-242252 (K8, class III).
         Typical WTTS light curves like this one are similar to those
         of the CTTS DoAr~24 and the unclassified K-type star Oph~S1~B.
         The persistent increases in mean energy may be the result of 
         nearly continuous reheating
         and the relatively short rise and decay times suggest moderate-sized 
         magnetic loops.
\label{f8} }

\figcaption{(a) Kolmogorov-Smirnov (KS) variablility statistic
            versus mean energy ($\langle{\rm E}\rangle$) for
            flat or Class II sources (squares),
            Class III sources (triangles),
            stars with no SED classification (circles),
            and {\it Chandra} sources without NIR counterparts (crosses).
            Brown-dwarf candidates (BDC) are indicated with gray symbols.
            {\it Chandra} sources without NIR counterparts are typically very 
            hard ($\langle{\rm E}\rangle>3.5$, marked by vertical dotted line) 
            and not variable (${\mathrm KS} < 1.0$) strongly suggesting that
	    these are highly absorbed extragalactic sources.  
         (b) count rate versus $\langle{\rm E}\rangle$, and 
         (c) count rate versus KS statistic for all {\it Chandra}
	    sources in Table~1.  A dashed line at ${\rm KS}=1$ in (a) and (c)
            marks the statistical variability threshold.
\label{f9} }

\figcaption{
Near-infrared color-magnitude diagram for stars in the ACIS-I field of view.
Only stars with measured J, H, and K are shown. A right arrow denotes
objects whose J magnitudes are lower limit. 
Stars detected by {\it Chandra} are indicated with filled symbols
and undetected stars with open symbols.
Class~I (diamonds), flat or Class~II (squares), 
and Class~III (triangles) designations are based on the 
spectral energy distributions of \citet{LR99}, \citet{Wil01},
\citet{AM94}, and \citet{Str95}. The known Class 0 source LFAM~5
was not detected at K-band or with {\it Chandra}, and the 
undetected class I source GSS 30-3 (not shown) has no measured J or H magnitudes.
The dotted polygon encloses the region occupied by 
several dozen undetected near-IR sources in the 
{\em Chandra} FoV which lack published SED classifications.
\label{f10} }

\figcaption{
Near-infrared color-color diagram for stars in the ACIS-I FOV. 
Only stars with measured J, H, and K are shown. Symbol
definitions are the same as in the previous figure.
Right and up arrows denote objects whose J and H magnitudes are 
lower limits, respectively. The $A_V=10$ reddening vector
shown is typical for sources in  $\rho$ Oph core A.
The dotted polygon encloses the region occupied by
several dozen undetected near-IR sources in the 
{\em Chandra} FoV which lack published SED classifications.
The unreddened ZAMS is shown as a dash-dot line at lower
left. The two sloping dashed lines mark the approximate
reddening band for normally reddened M0 V ({\em left}) and
A0 V ({\em right}) stars using data from Bessell \&
Brett (1988) and Rieke \& Lebofsky (1985).
\label{f11} }

\figcaption{X-ray to IR flux ratio for all sources with measured K-band 
        magnitude.  The dotted line represents the total data while the solid 
        line indicates bona fide YSOs with SED classifications.
\label{f12} }

\figcaption{Kaplan-Meier estimator for 14 CTTS 
(12 detections and 2 upper limits)
and 17 WTTS (15 detections and 2 upper limits) in the $\rho$ Oph core-A ACIS-I
field. The generalized Wilcoxon test gives a probability $p$ = 0.53 that the
luminosities are drawn from the same parent distribution.
\label{f13} }

\figcaption{$L_{\rm 6~cm}$ radio luminosity versus $L_{\rm X}$ (0.5 - 7 keV)
for class II (CTTS) and class III (WTTS) sources detected simultaneously
with {\it Chandra} and the {\it VLA}. We assume a distance of 165 pc.
The unusual binary Oph~S1 and three class 0/I sources are also labeled.
Only X-ray upper limits were obtained for the class 0/I sources.
GSS30-IRS1 was not detected by {\it Chandra} or the {\it VLA}.
The quiescent X-ray and radio luminosities of coronal active stars
(main-sequence and evolved giants) generally obey the relation
log L$_{\rm X}$ $\leq$ log L$_{\rm r}$ \citep{Gu93} and are located to the 
left of the dashed line. The PMS stars shown here also satisfy
this inequality.
The five X-ray brightest VLA sources underwent flares during the {\it Chandra}
observation. Their $L_{\rm X}$ ranges are indicated with horizontal lines.
The Class III sources appear to be brighter at 6~cm than other types of YSOs.
The location of Oph S1 in this diagram suggests that both the X-ray and radio
emission are produced on the low-mass companion and that Oph S1 B is a
Class III source.
\label{f14} }

\figcaption{Background-subtracted ACIS-I spectrum of the binary
system Oph~S1 (B4 + K). The spectrum is rebinned to a minimum of 15 counts
per bin. Solid line shows a best-fit 2T VAPEC optically thin plasma
model with a variable Fe abundance. Best-fit values are given in
Table 6. 
\label{f15} }

\figcaption{Background-subtracted ACIS-I light curve of the
brown dwarf candidate GY~31. Nearly all the flare emission is in the 2-7 keV
band.  Although the first flare peak was not observed, flare decay times appear
to be less than five hours, typical of moderate flares on other TTS.
\label{f16} }

\figcaption{
Rebinned background-subtracted ACIS-I spectrum of the brown dwarf candidate
GY~31. Solid line is a best-fit 1T VAPEC model with
$\log N_{\rm H}=22.76$~cm$^{-2}$, $kT=2.3$ keV, Fe abundance Fe = 0.3 solar,
and $\chi^2$/dof = 19/26.
\label{f17} }

\figcaption{
$\log T$ versus $\log N_{\rm H}$ contour plots for two fake
sets of photon events: a source with 25 counts (upper panel), another with 
60 counts (lower panel). The input values were $\log N_{\rm H}=22.18$
and $\log T=7.40$ (indicated with crosses). The best-fit values from
the non-parametric method described in Appendix B are
indicated with dots. Also shown are the 68\%, 90\% and 95\% confidence
contours for the Cramer-von Mises statistic. Multiple fake data sets
were used to estimate the standard deviations of the best-fit $\log N_{\rm H}$
and $\log T=7.40$, indicated as a dashed box. This analysis shows that the
non-parametric method can be used to estimate column density, temperature,
and $L_{\rm X}$ with as few as 25 or 60 counts. See Appendix B and
Tables 7 and 8 for details.
\label{f18} }



\clearpage

\begin{figure}
\figurenum{1}
\plotone{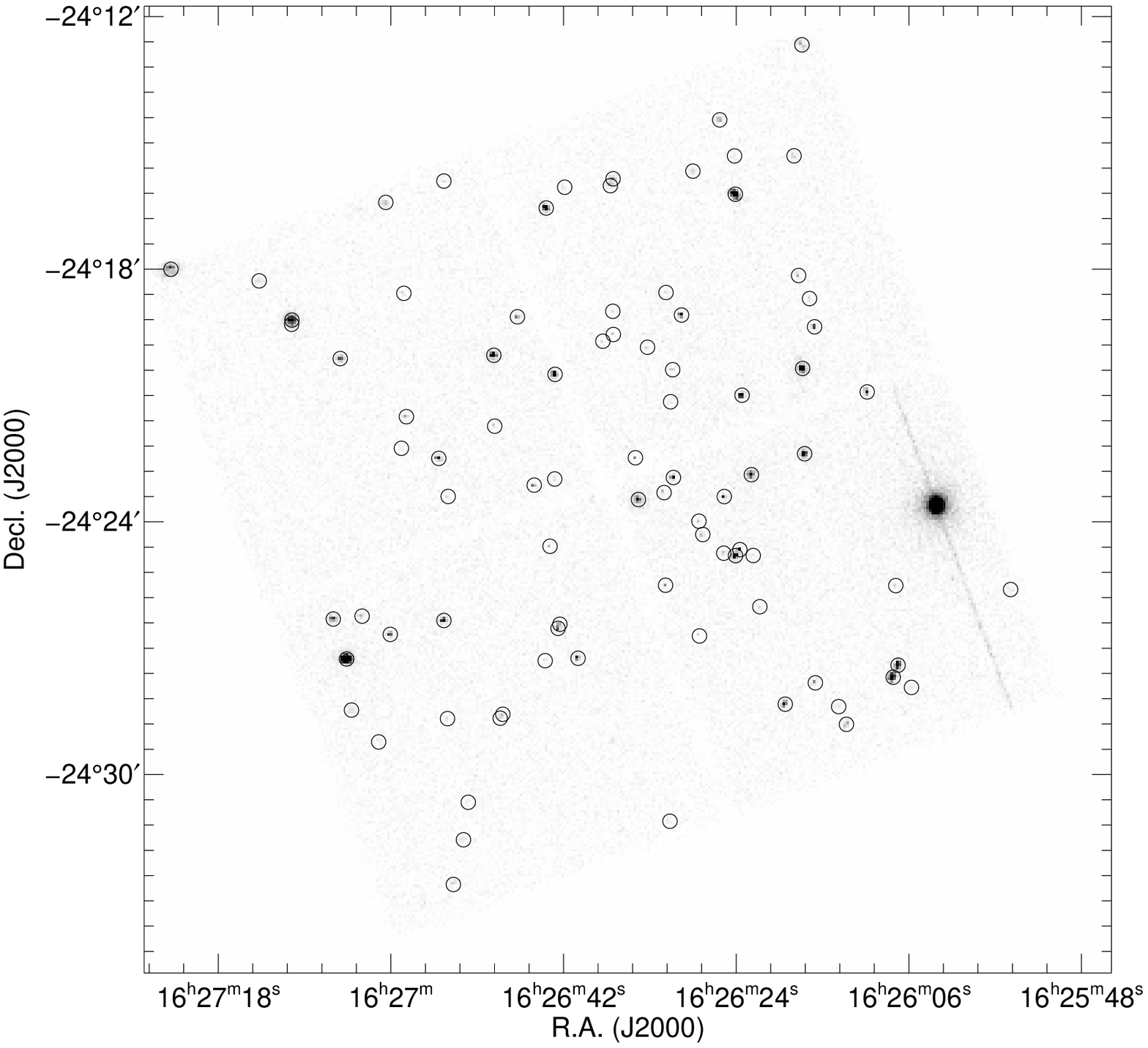}
\caption{}
\end{figure}

\clearpage

\begin{figure}
\figurenum{2}
\plotone{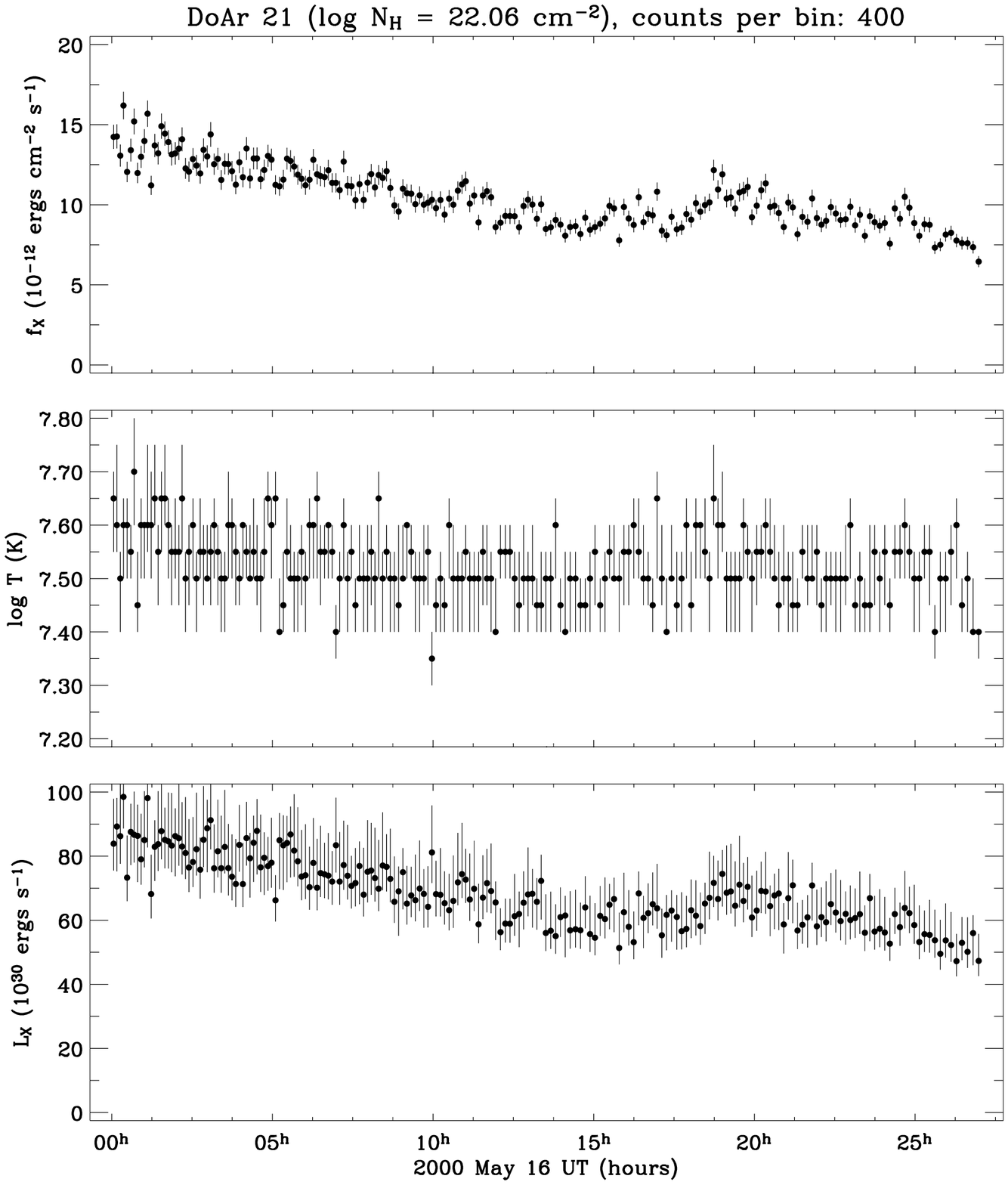}
\caption{}
\end{figure}

\clearpage

\begin{figure}
\figurenum{3}
\plotone{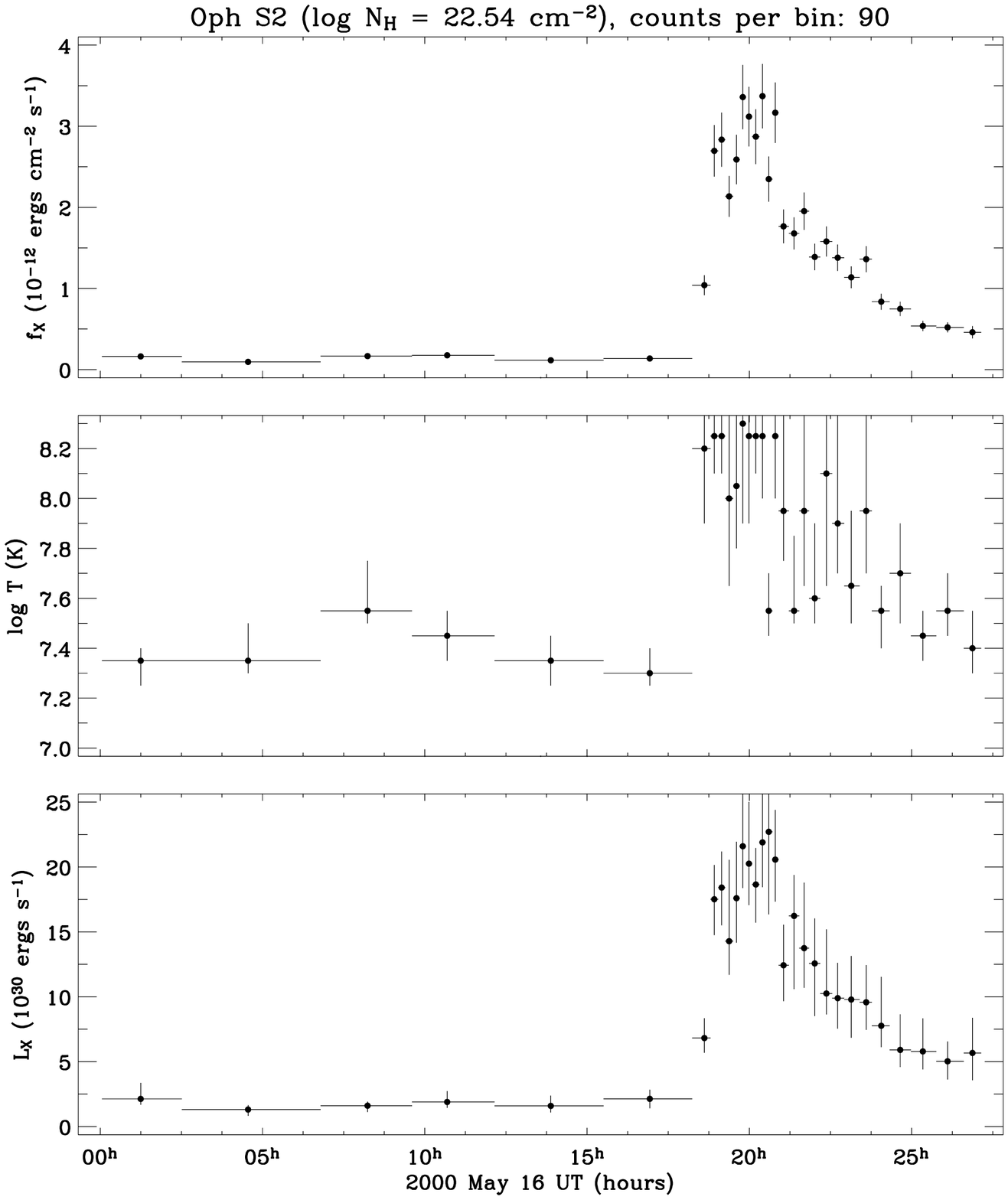}
\caption{}
\end{figure}

\clearpage

\begin{figure}
\figurenum{4}
\plotone{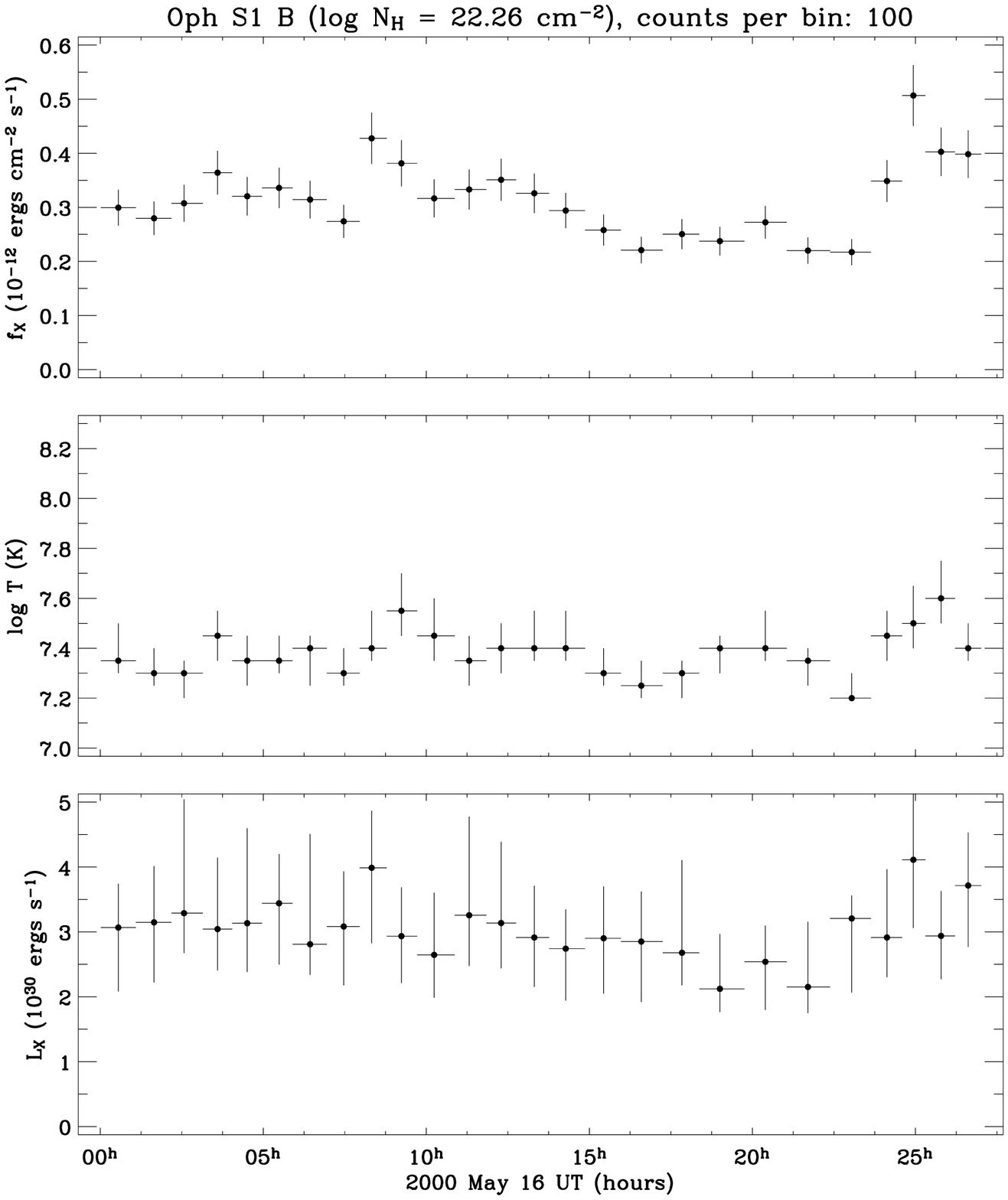}
\caption{}
\end{figure}

\clearpage

\begin{figure}
\figurenum{5}
\plotone{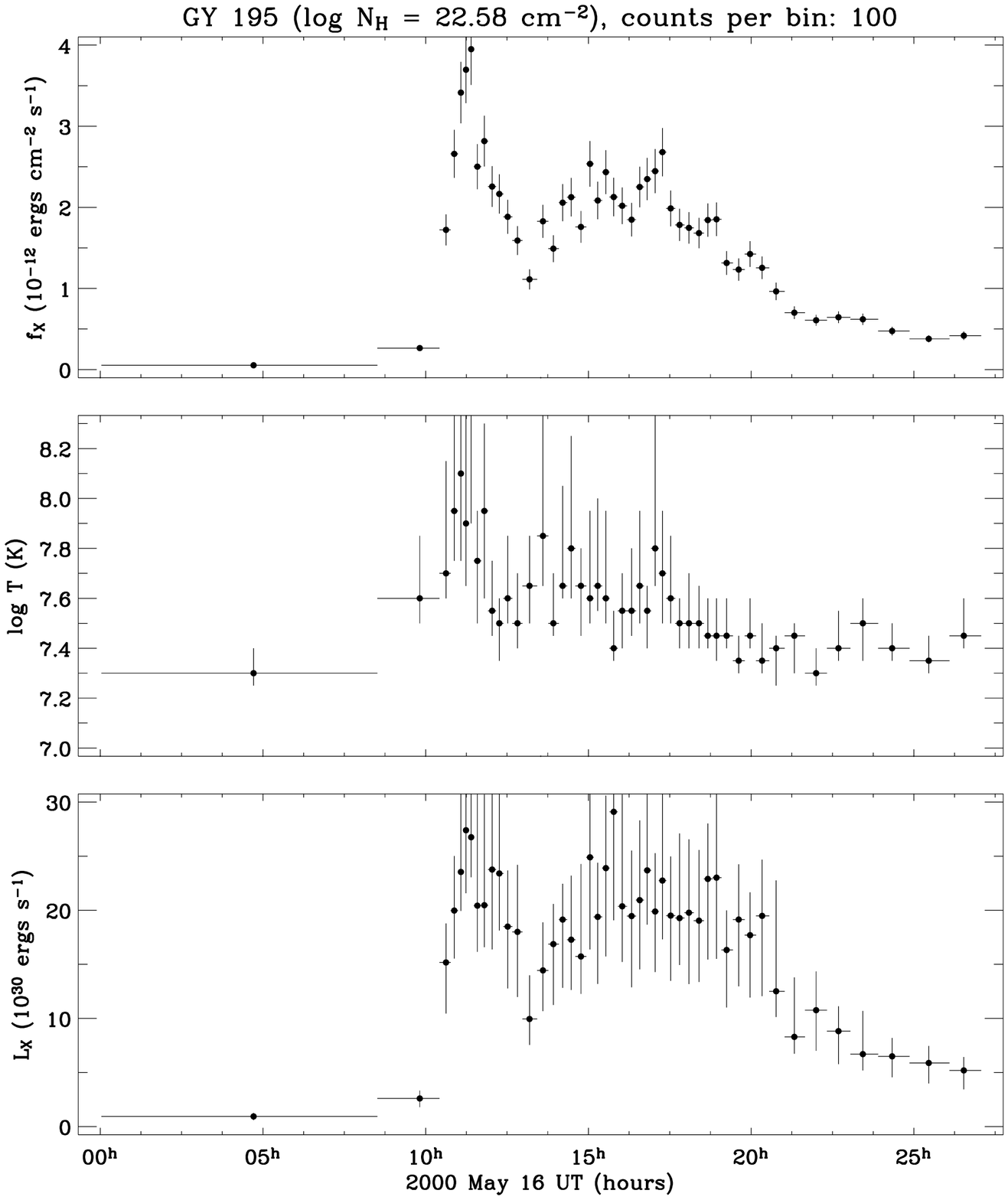}
\caption{}
\end{figure}

\clearpage

\begin{figure}
\figurenum{6}
\plotone{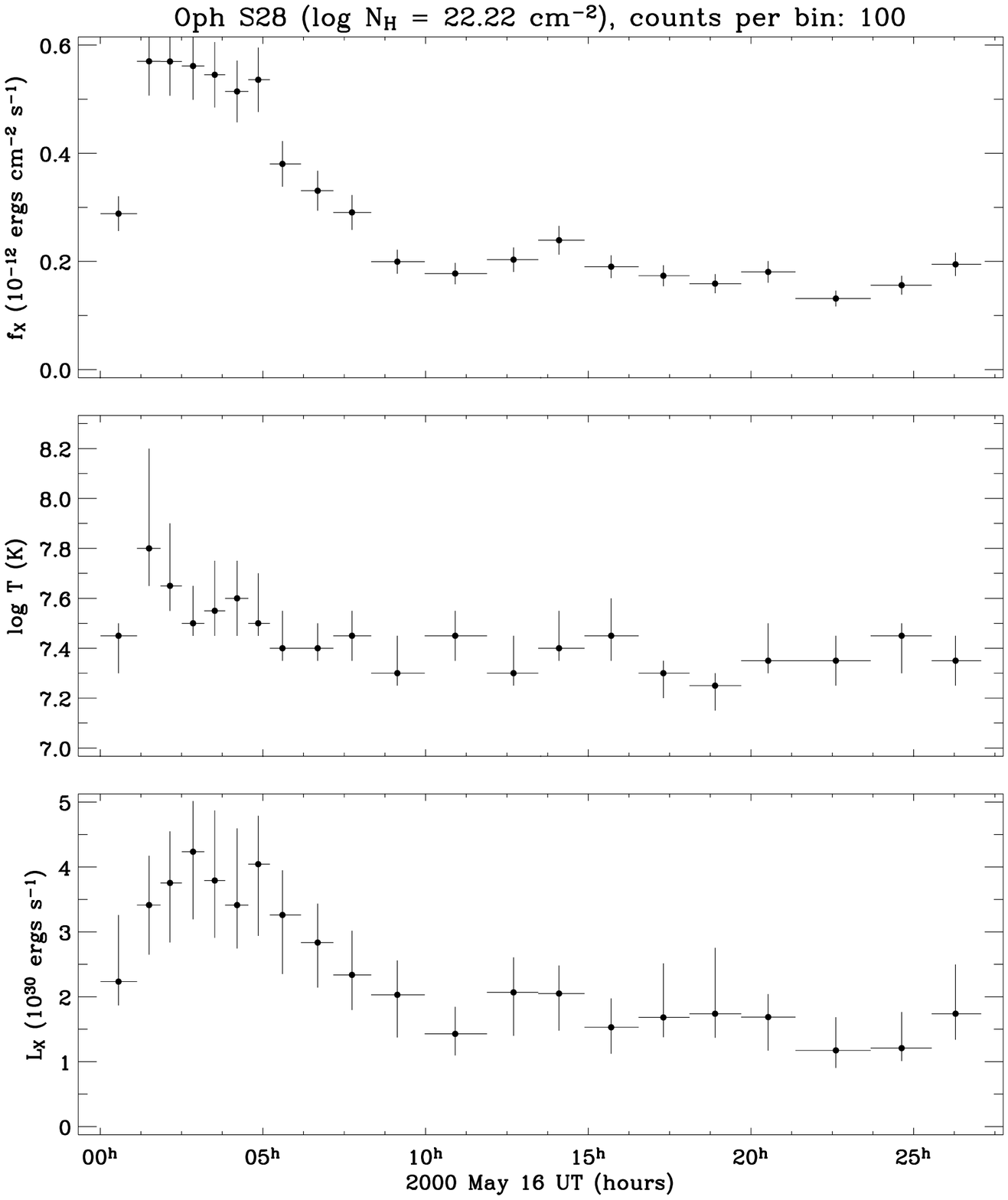}
\caption{}
\end{figure}

\clearpage

\begin{figure}
\figurenum{7}
\plotone{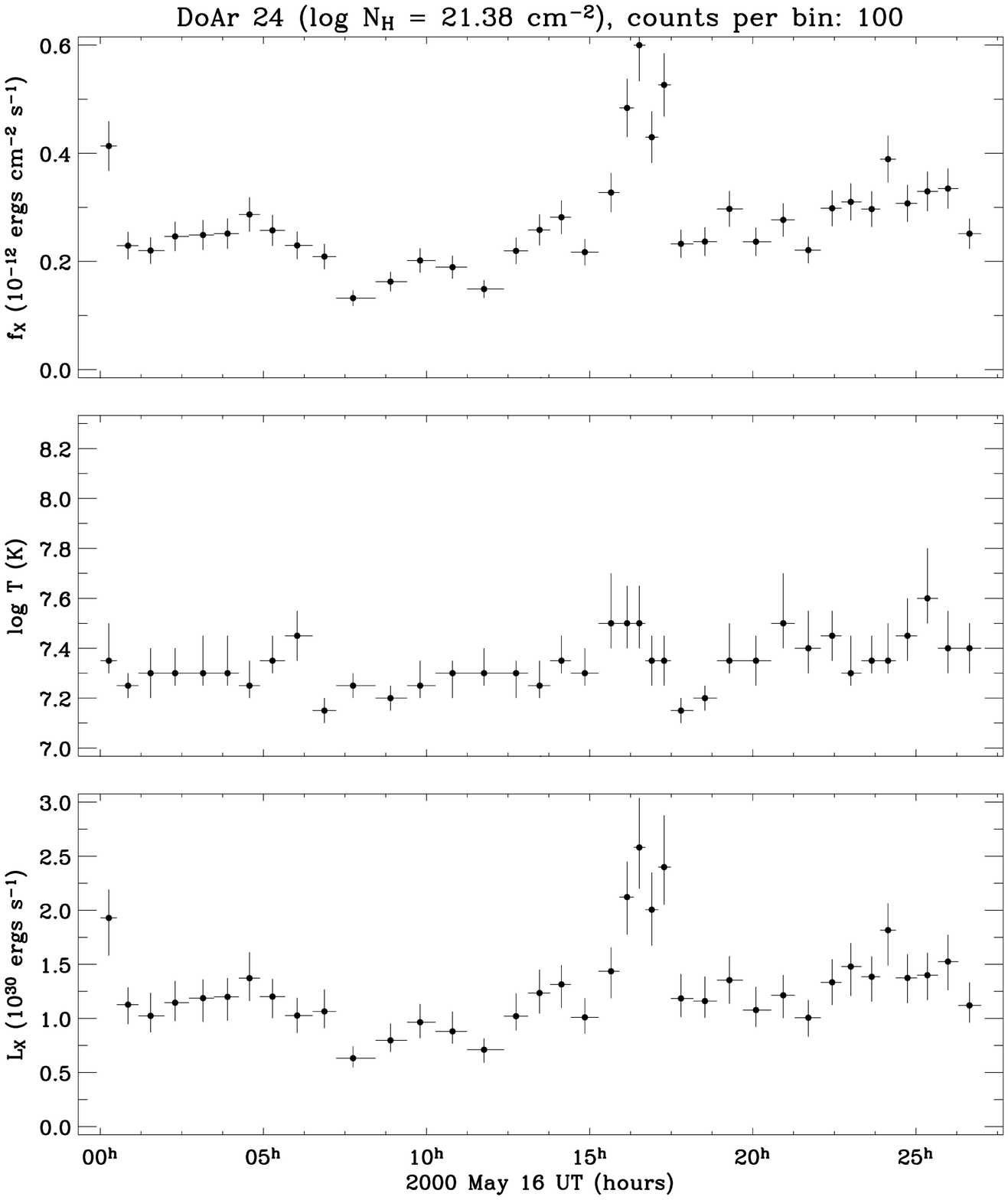}
\caption{}
\end{figure}

\clearpage

\begin{figure}
\figurenum{8}
\plotone{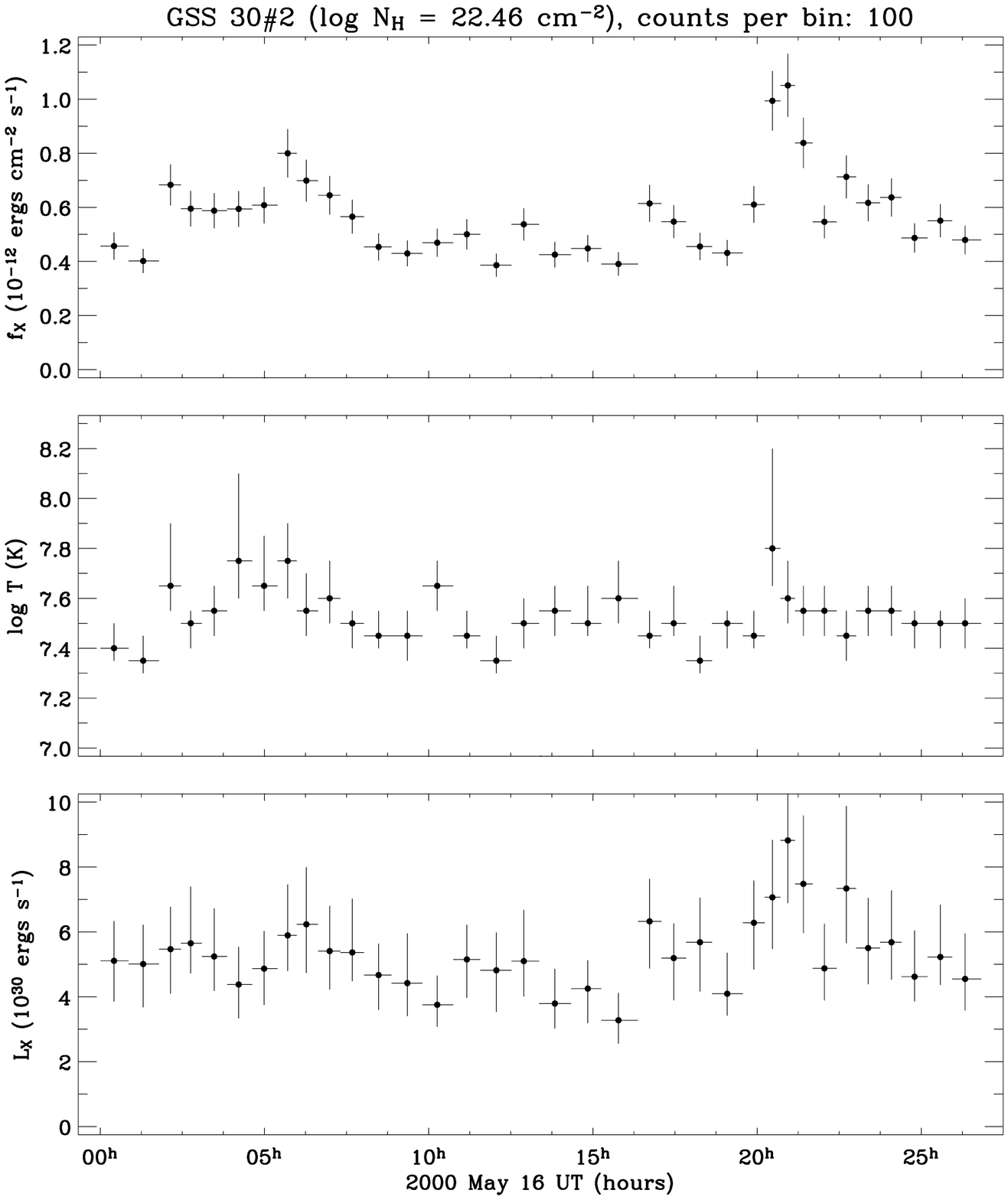}
\caption{}
\end{figure}

\clearpage

\begin{figure}
\figurenum{9a}
\plotone{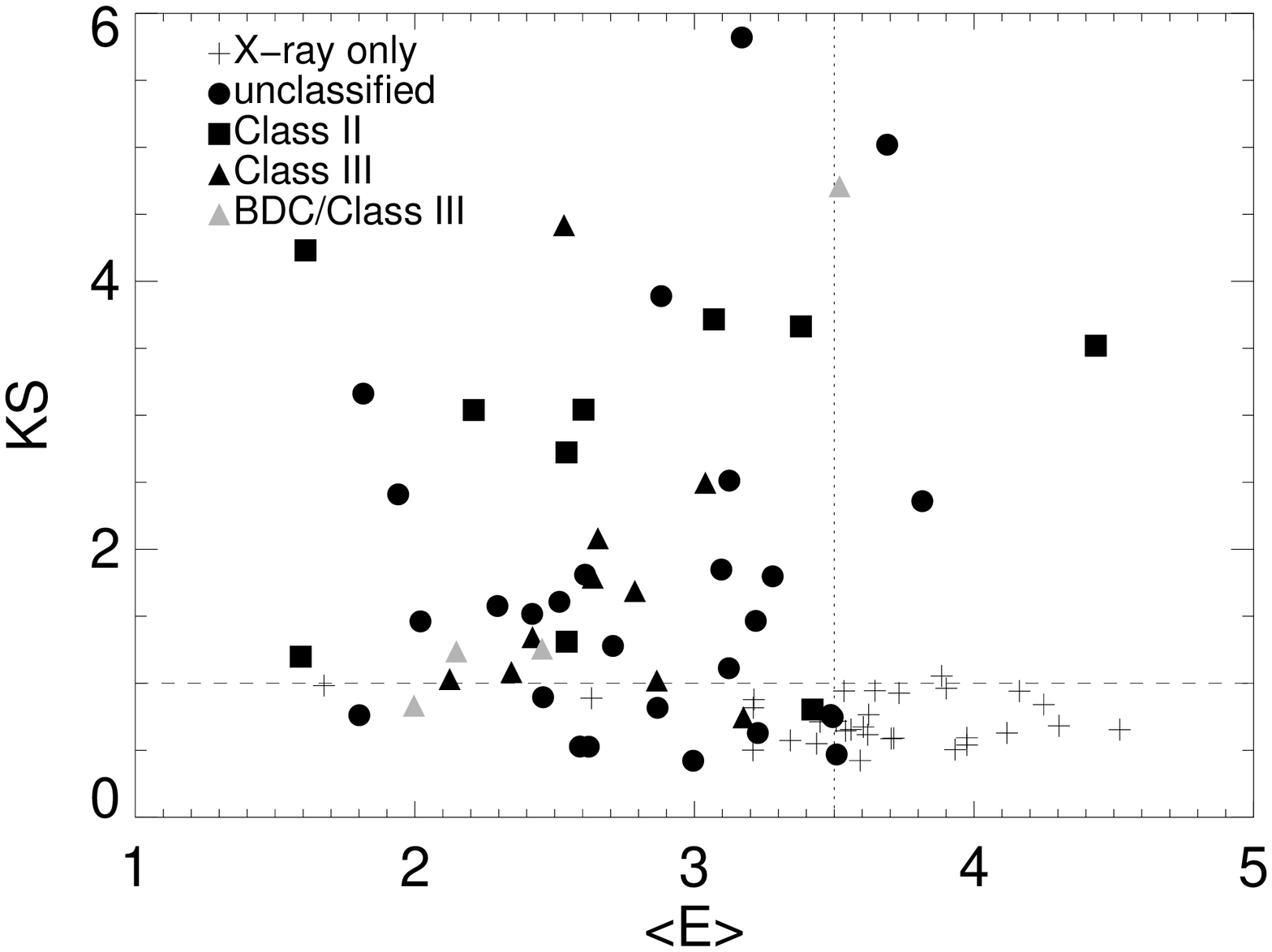}
\caption{}
\end{figure}

\clearpage

\begin{figure}
\figurenum{9b}
\plotone{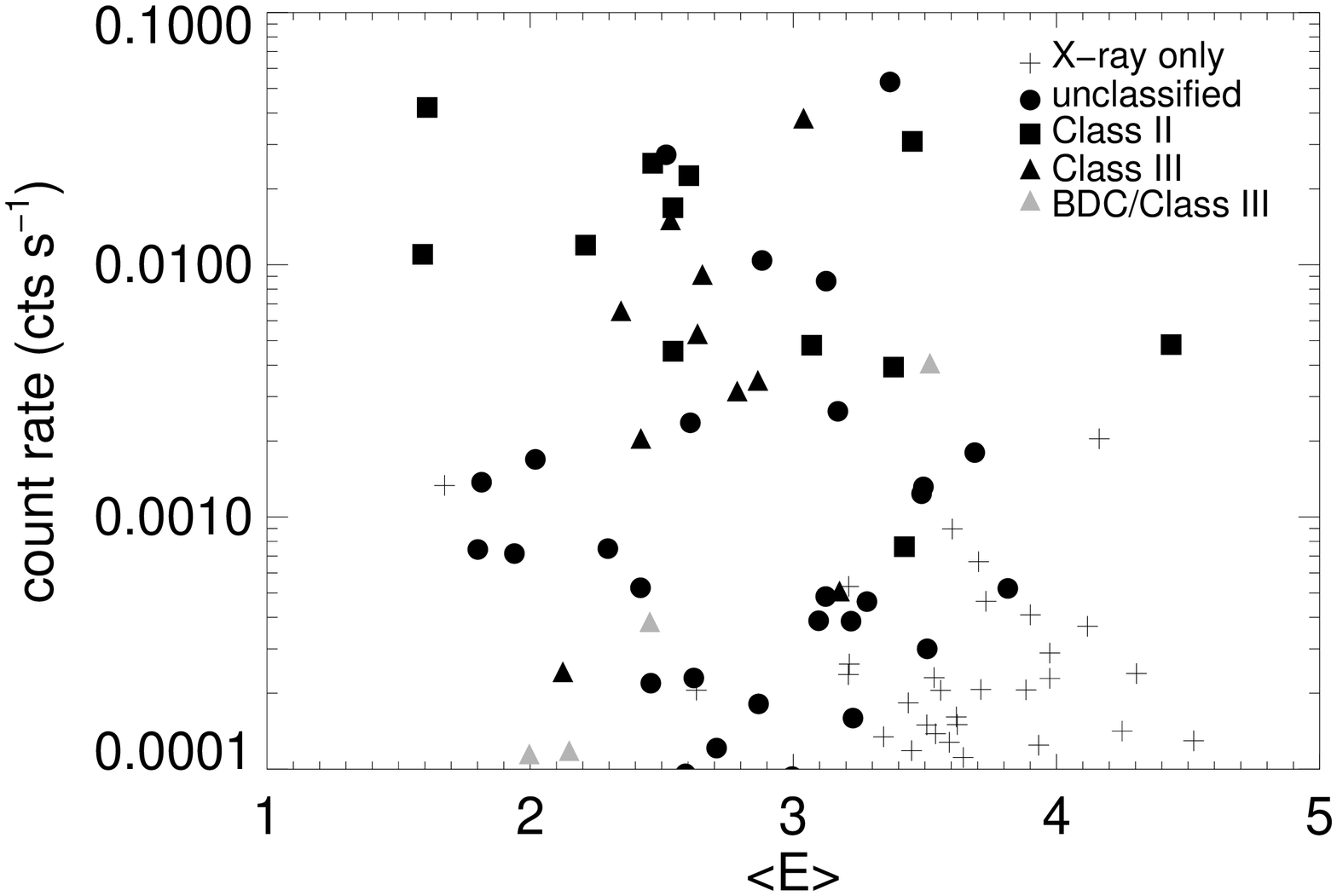}
\caption{}
\end{figure}

\clearpage

\begin{figure}
\figurenum{9c}
\plotone{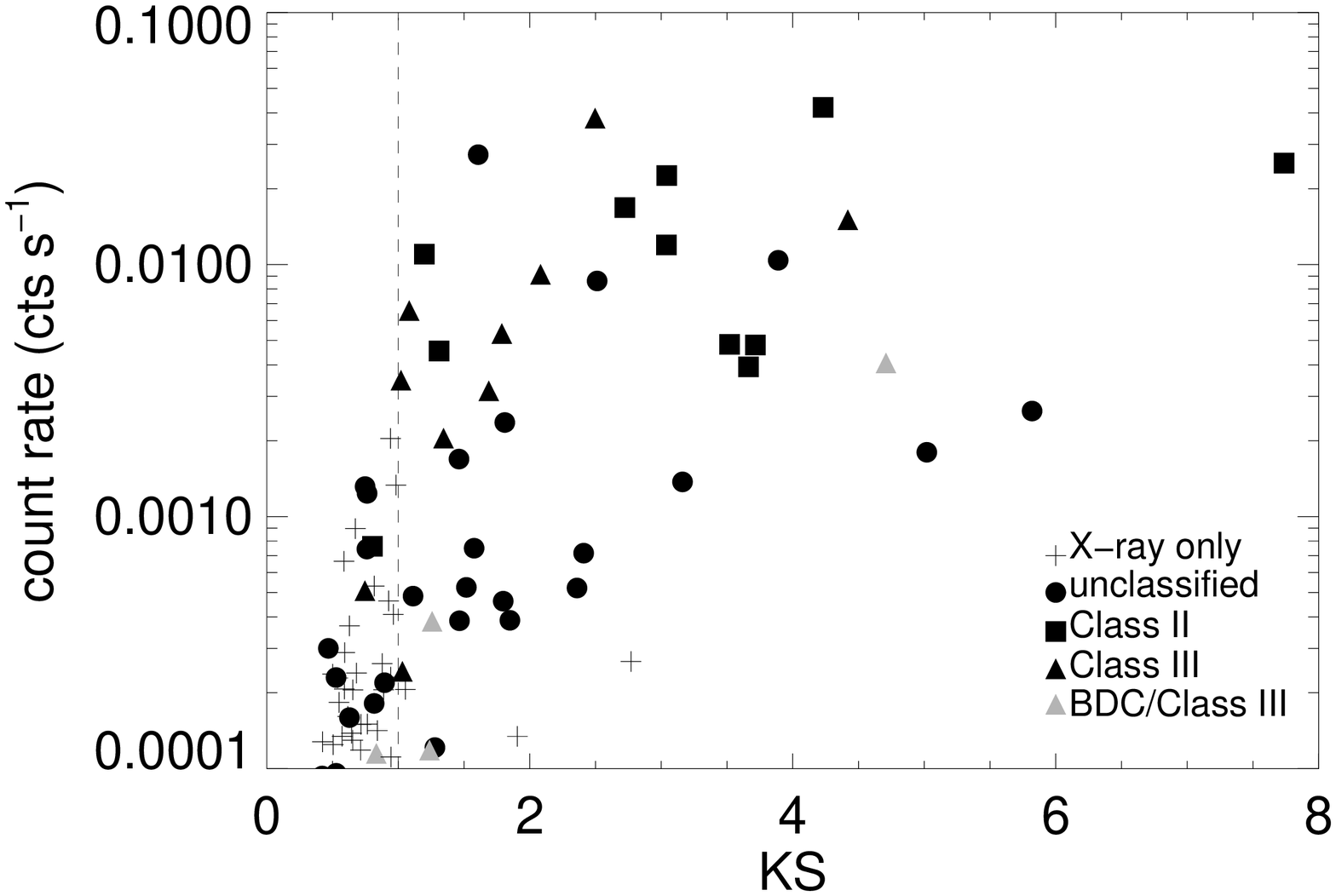}
\caption{}
\end{figure}

\clearpage

\begin{figure}
\figurenum{10}
\plotone{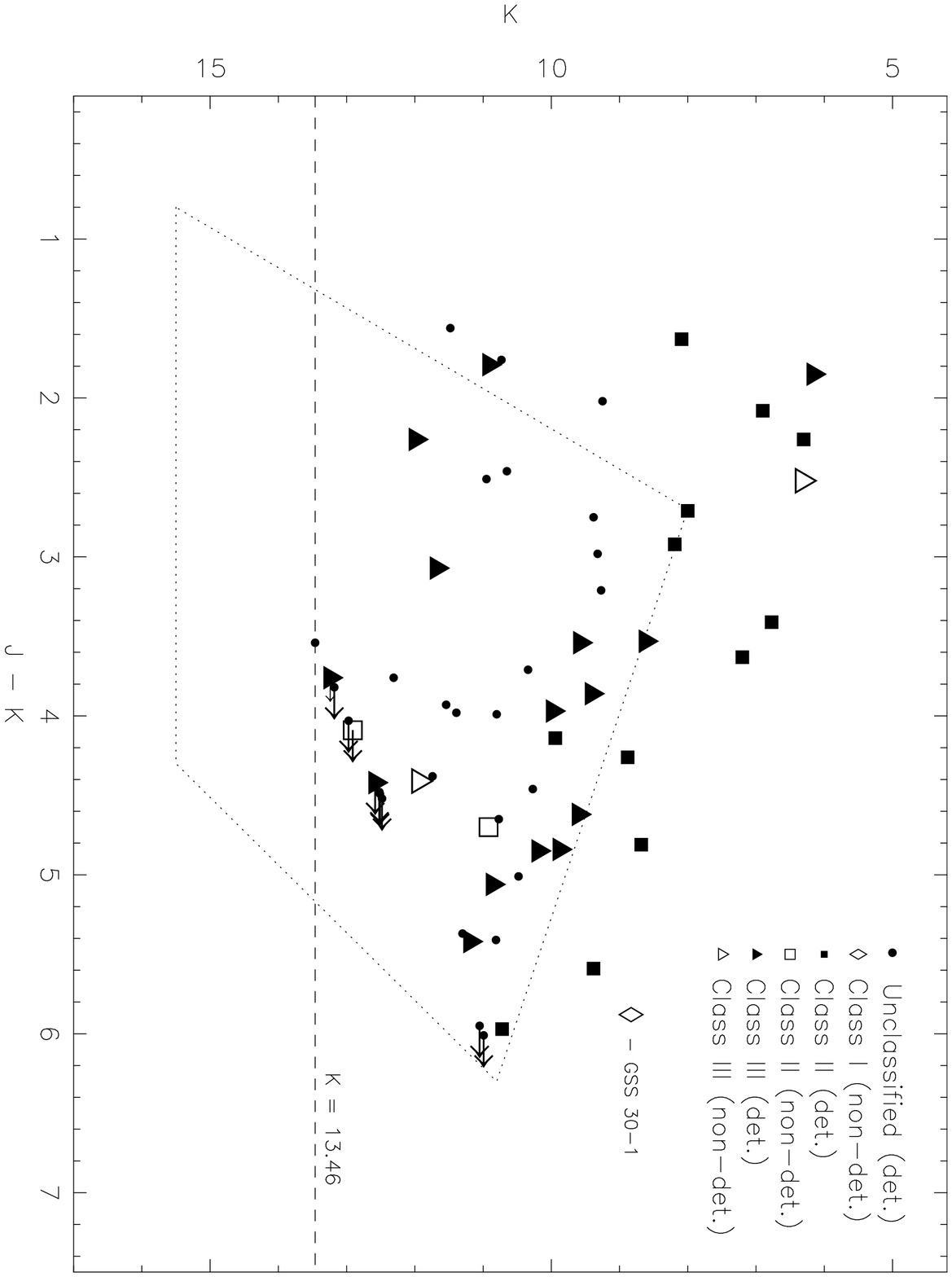}
\caption{}
\end{figure}

\clearpage

\begin{figure}
\figurenum{11}
\plotone{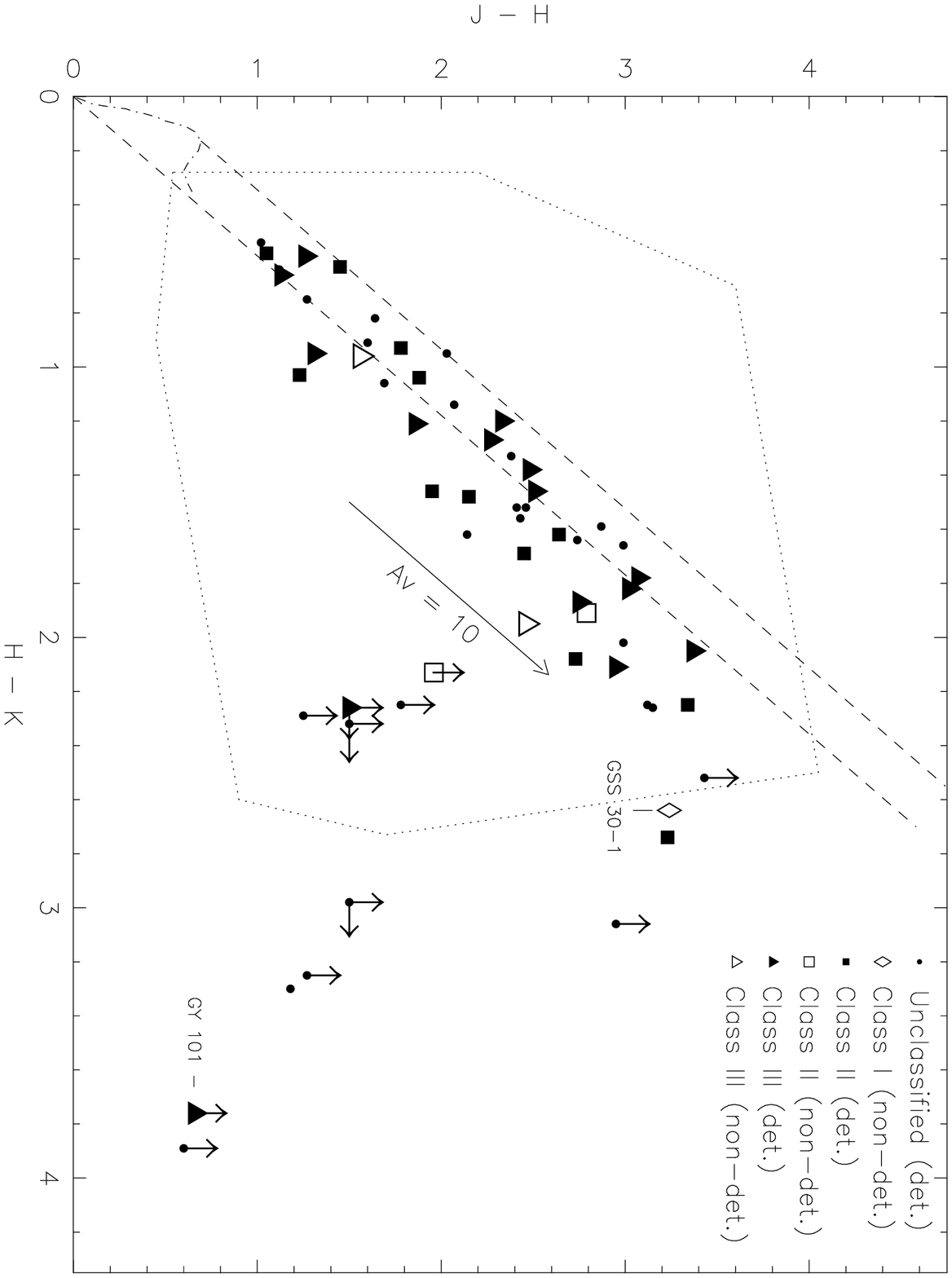}
\caption{}
\end{figure}

\clearpage

\begin{figure}
\figurenum{12}
\plotone{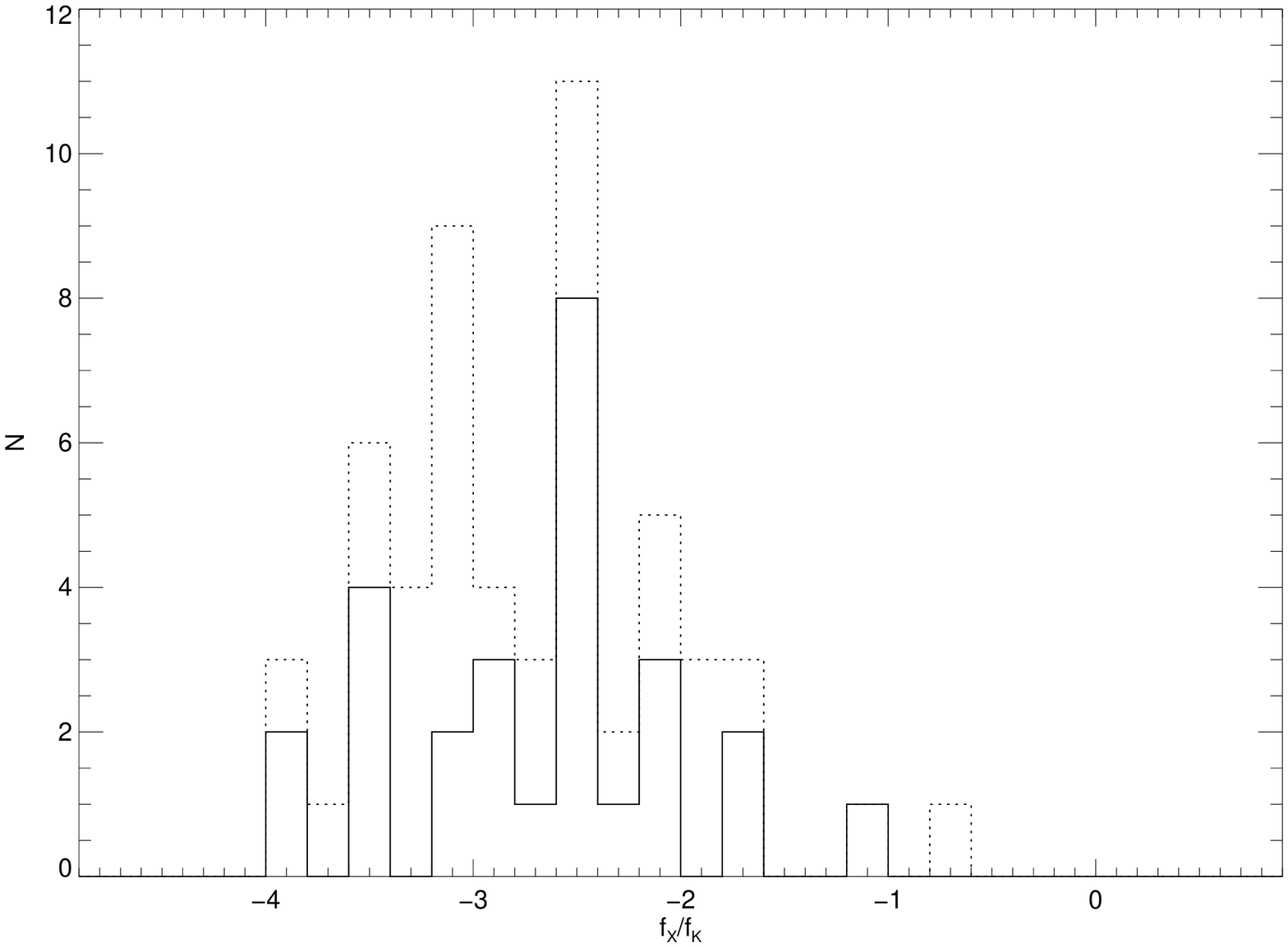}
\caption{}
\end{figure}

\clearpage

\begin{figure}
\figurenum{13}
\plotone{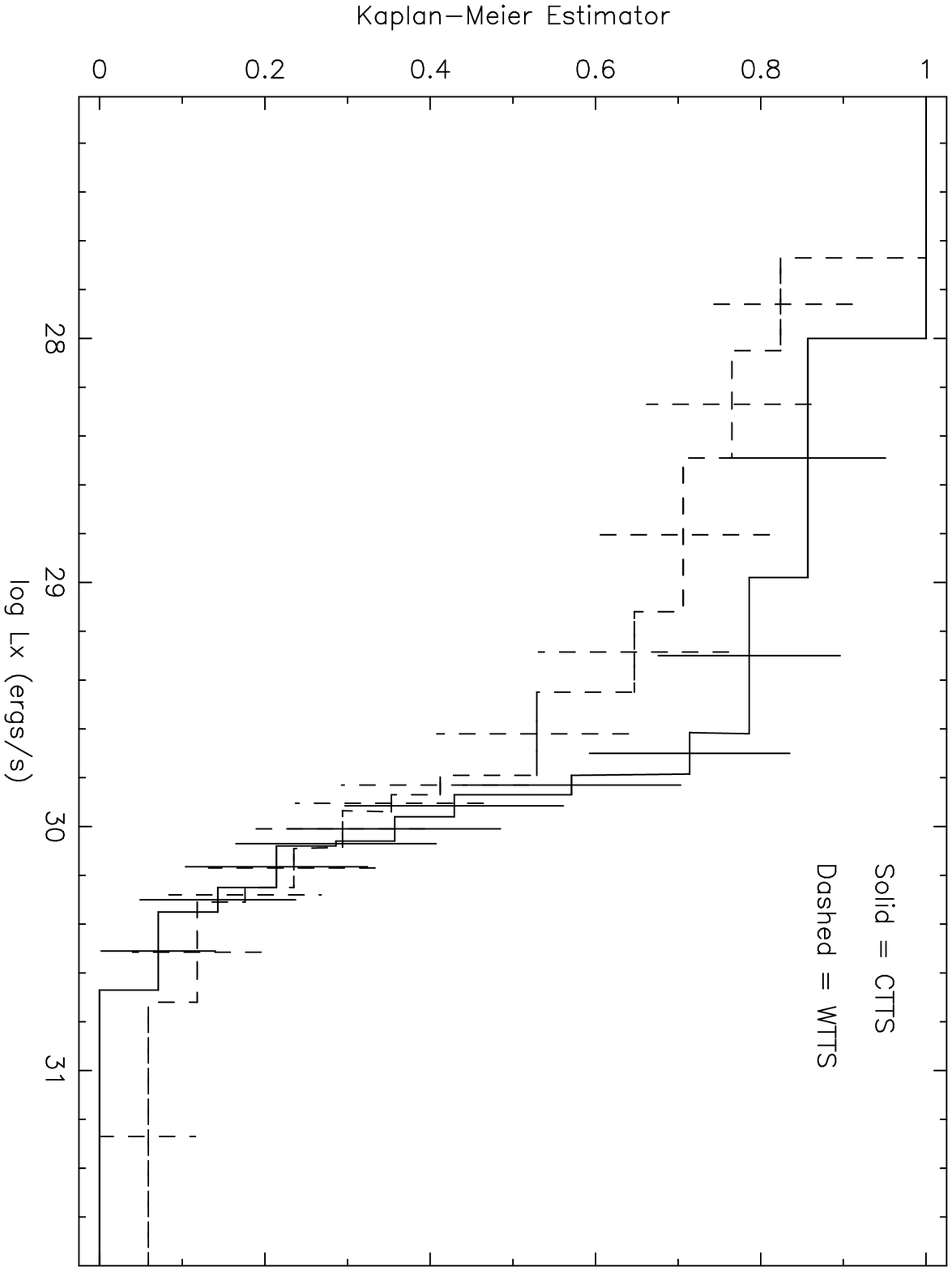}
\caption{}
\end{figure}

\clearpage

\begin{figure}
\figurenum{14}
\plotone{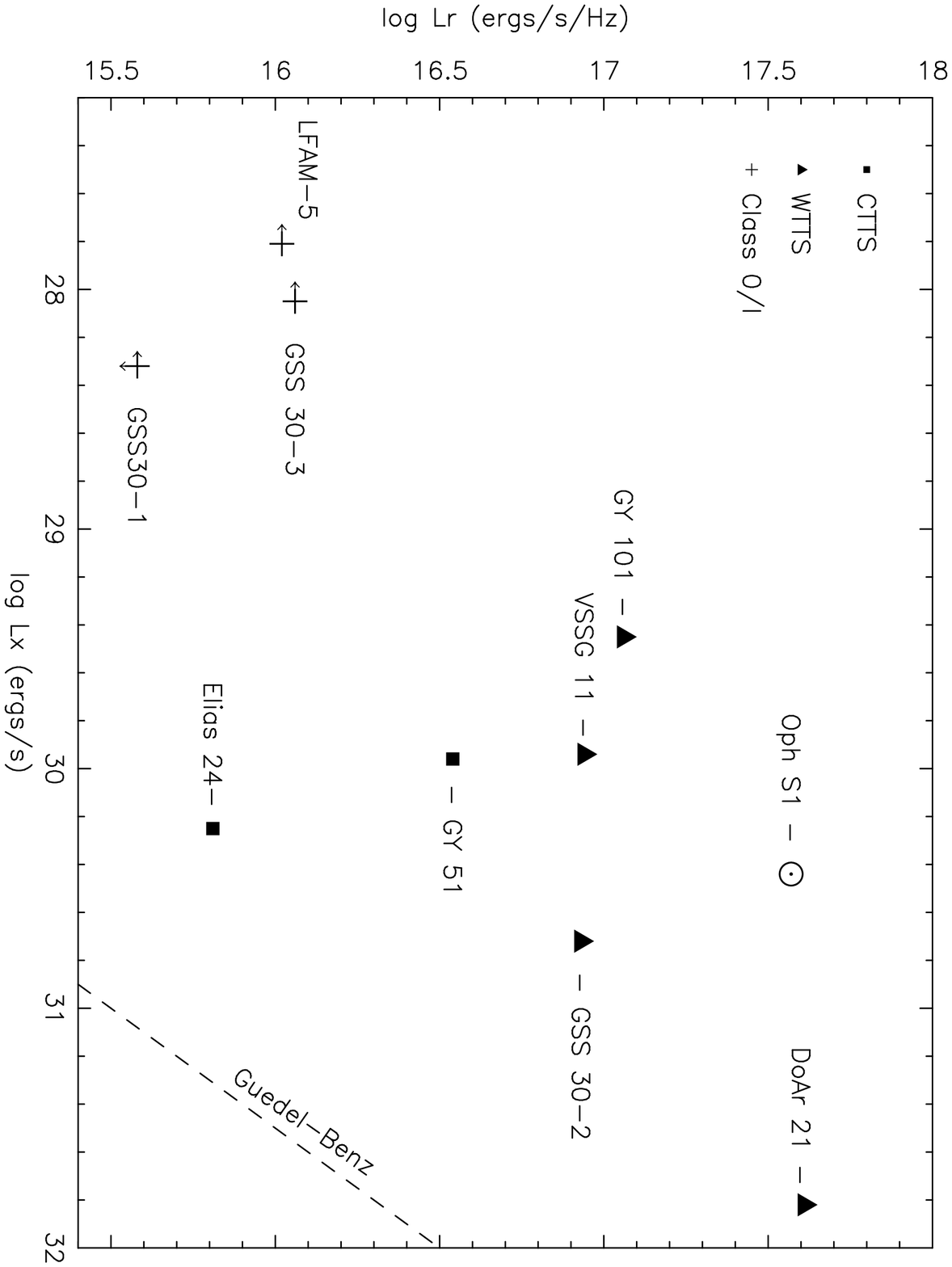}
\caption{}
\end{figure}

\clearpage

\begin{figure}
\figurenum{15}
\plotone{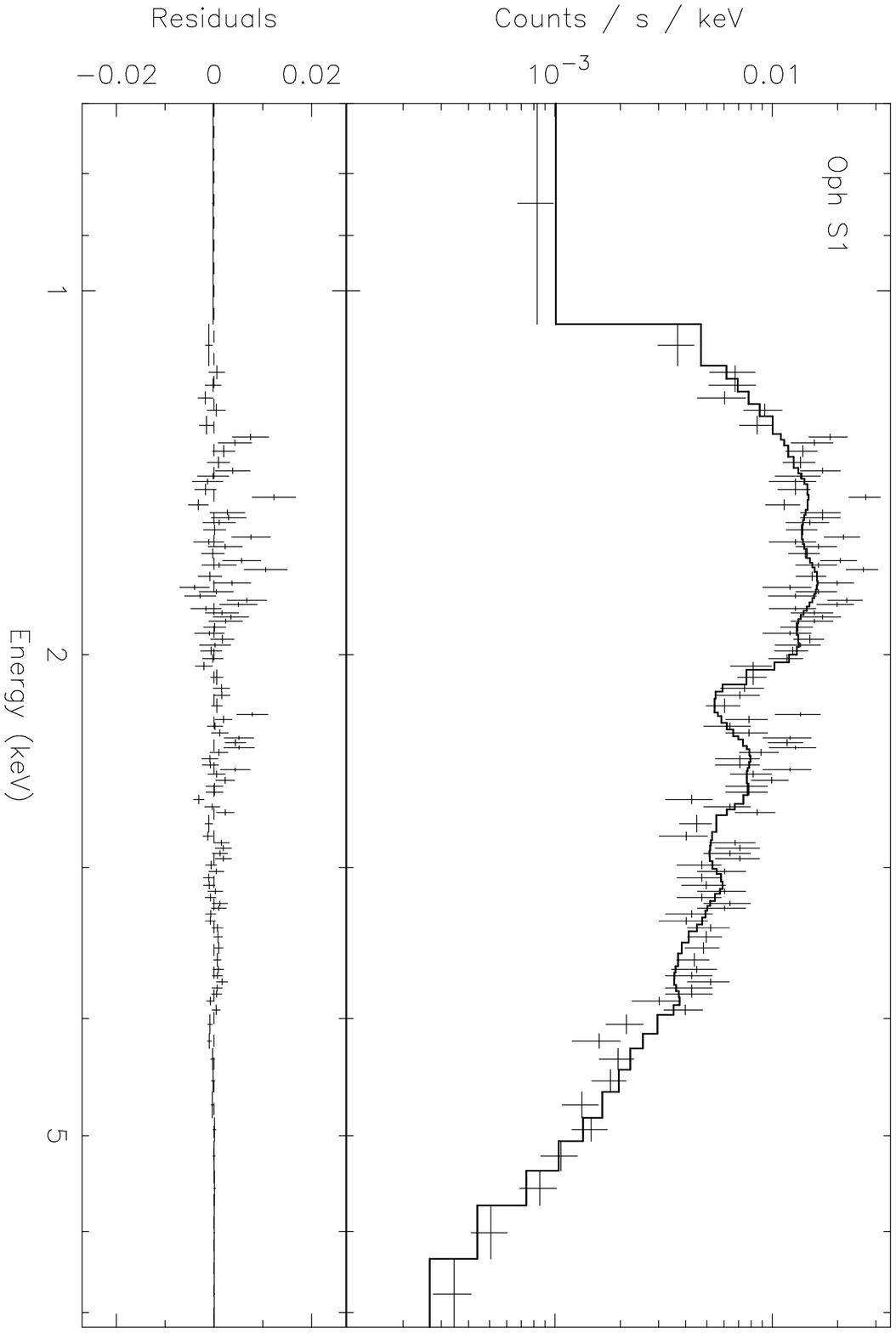}
\caption{}
\end{figure}

\clearpage

\begin{figure}
\figurenum{16}
\plotone{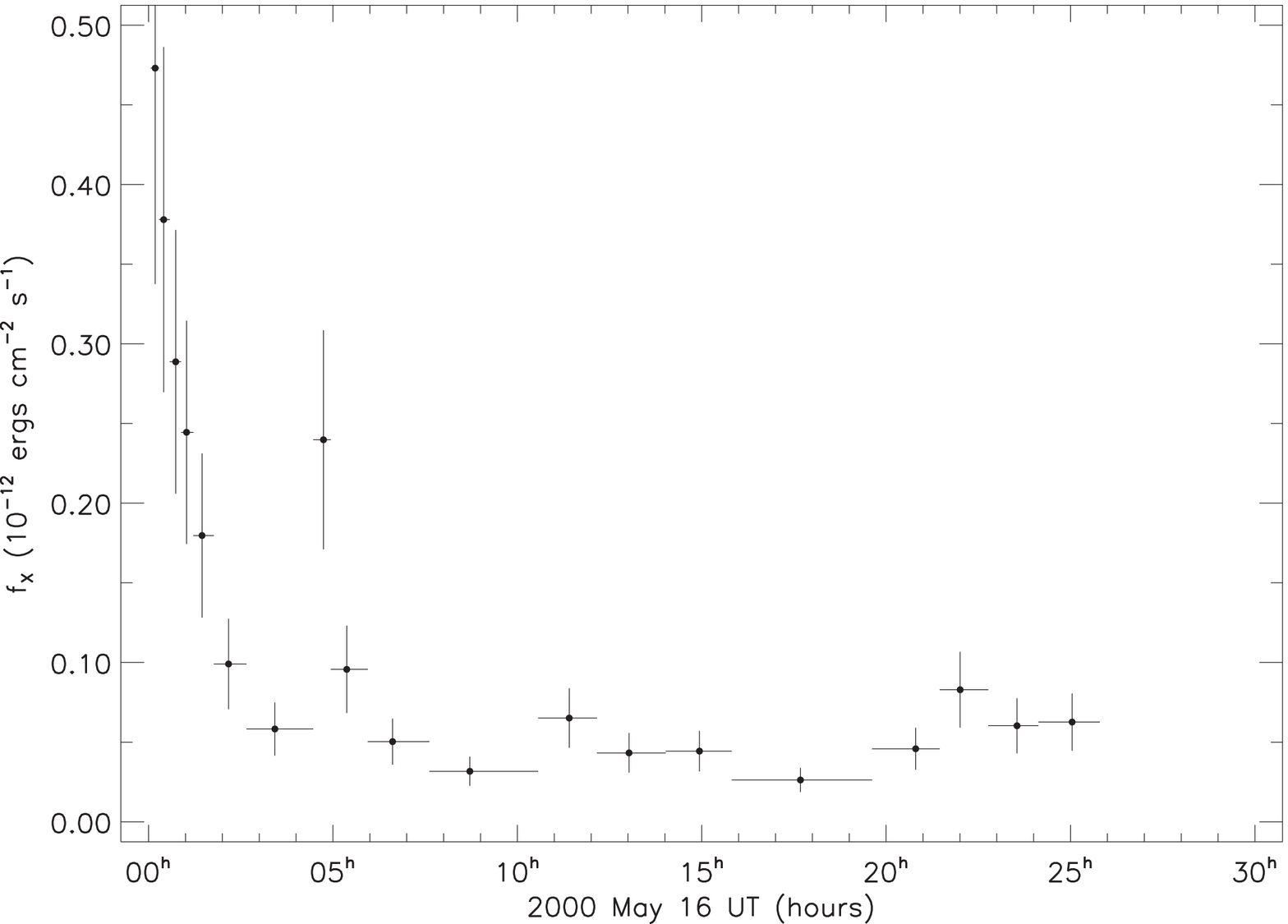}
\caption{}
\end{figure}

\clearpage

\begin{figure}
\figurenum{17}
\plotone{f17.eps}
\caption{}
\end{figure}

\clearpage

\begin{figure}
\figurenum{18}
\plotone{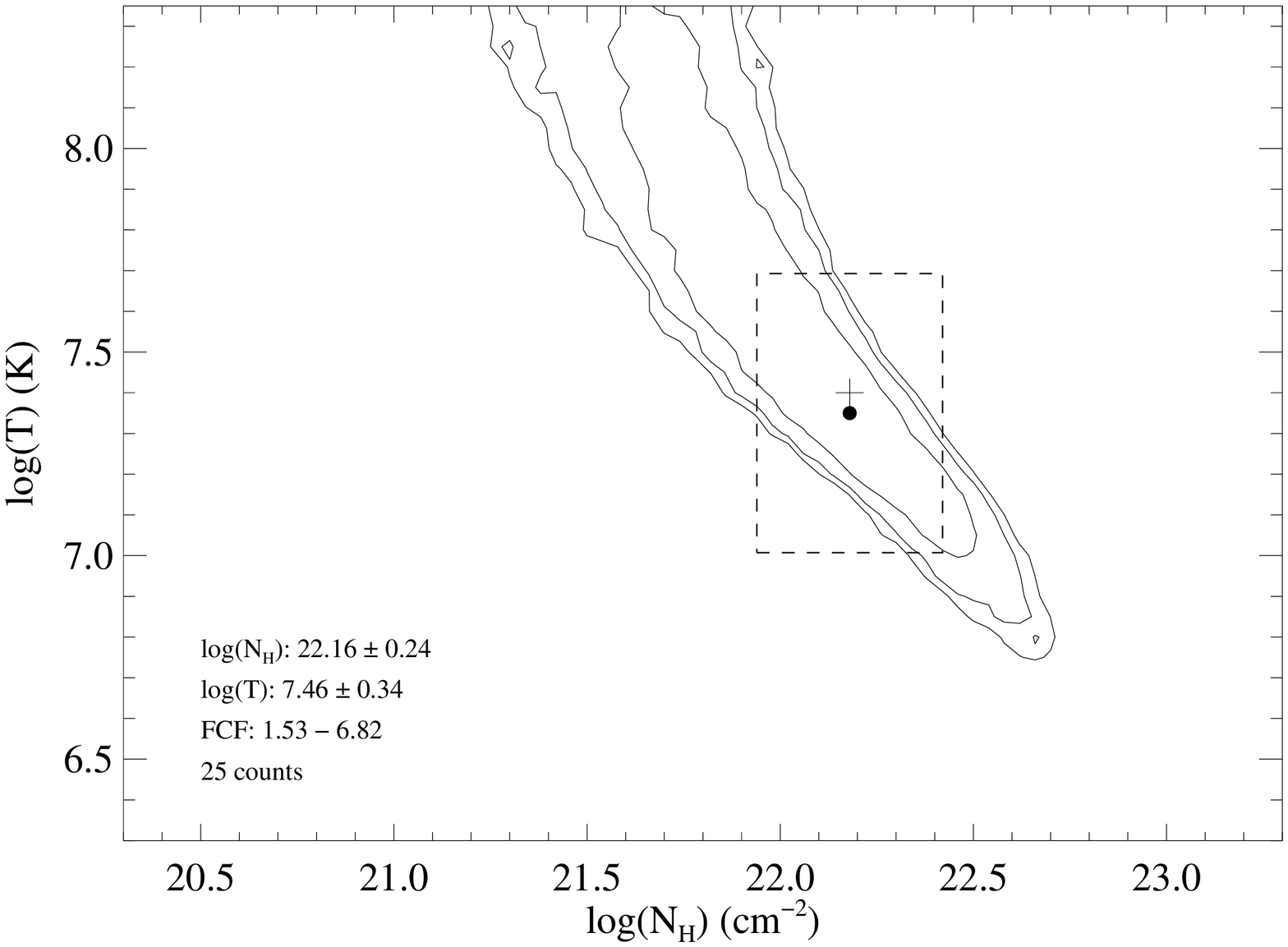}
\plotone{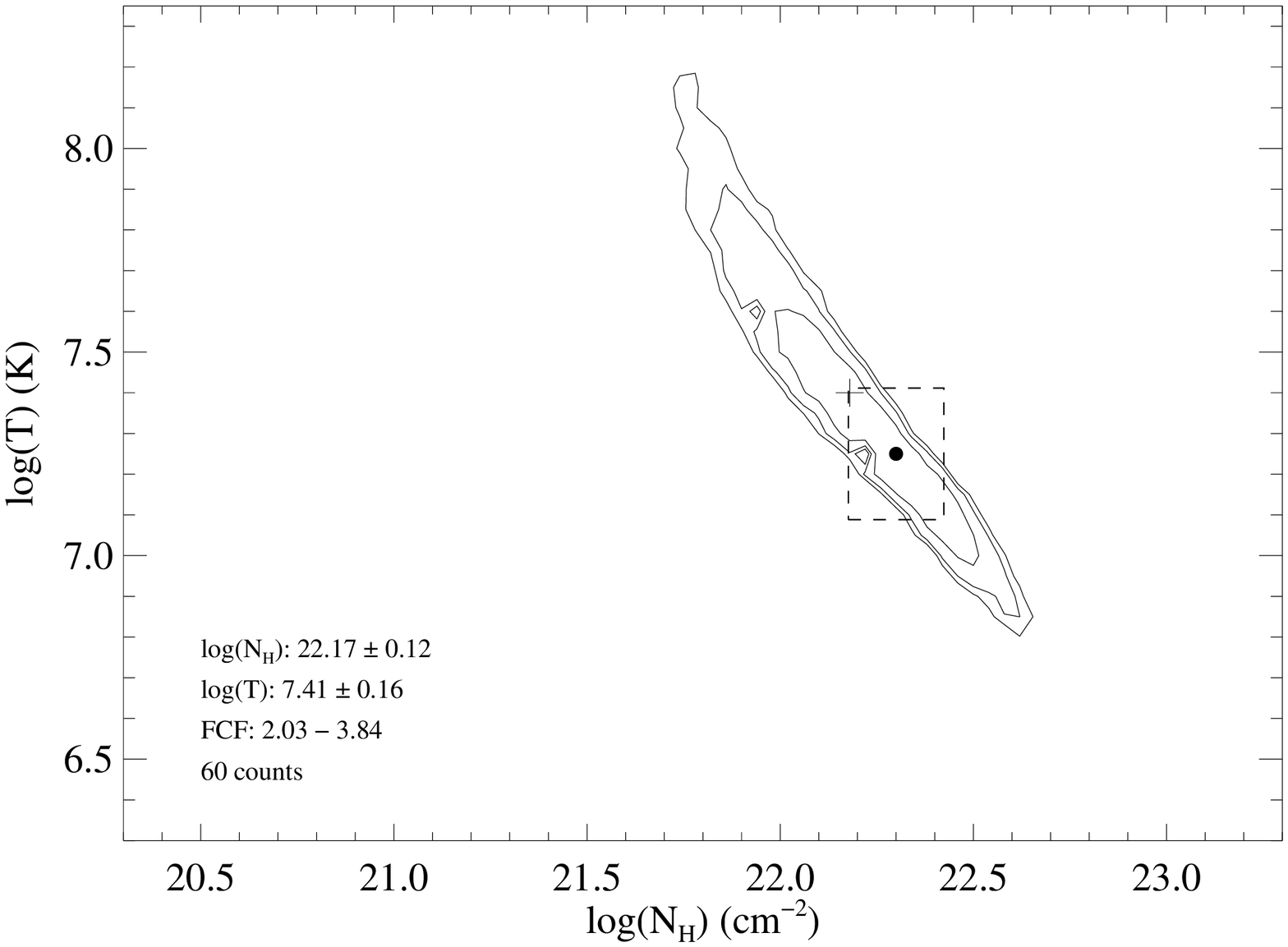}
\caption{}
\end{figure}

\end{document}